\documentclass[twocolumn,superscriptaddress,amsmath,amssymb,longbibliography]{revtex4-1}
\usepackage{graphicx}

\usepackage{color}
\usepackage{natbib}
\usepackage{comment}
\usepackage[normalem]{ulem}
\usepackage{physics}
\usepackage{bbm}
\definecolor{ao(english)}{rgb}{0.0, 0.5, 0.0}
\definecolor{clr}{rgb}{0.0, 0.5, 0.0}
\usepackage{soul} 

\newcommand{\CB}[1]{\textcolor{black} {#1}} 
\newcommand{\AC}[1]{\textcolor{black} {#1}} 

\usepackage{hyperref}
\hypersetup{
	colorlinks = true,
	urlcolor   = blue,
	linkcolor  = blue,
	citecolor  = blue
}

\begin{document}
\title{Feedback-assisted quantum search by continuous-time quantum walks}
\author{Alessandro Candeloro}\email{alessandro.candeloro@unimi.it}
\affiliation{Quantum Technology Lab, Dipartimento di  Fisica {\em Aldo Pontremoli}, Universit\`{a} degli Studi di Milano, I-20133 Milano, Italy}
\author{Claudia Benedetti}\email{claudia.benedetti@unimi.it}
\affiliation{Quantum Technology Lab, Dipartimento di 
Fisica {\em Aldo Pontremoli}, Universit\`{a} degli Studi di Milano, I-20133 Milano, Italy}
\author{Marco G. Genoni}\email{marco.genoni@fisica.unimi.it}
\affiliation{Quantum Technology Lab, Dipartimento di 
Fisica {\em Aldo Pontremoli}, Universit\`{a} degli Studi di Milano, I-20133 Milano, Italy}
\affiliation{INFN, Sezione di Milano, I-20133 Milano, Italy}
\author{Matteo G. A. Paris}\email{matteo.paris@fisica.unimi.it}
\affiliation{Quantum Technology Lab, Dipartimento di 
Fisica {\em Aldo Pontremoli}, Universit\`{a} degli Studi di Milano, I-20133 Milano, Italy}
\affiliation{INFN, Sezione di Milano, I-20133 Milano, Italy}

\begin{abstract}
    We address the quantum search of a target node on a cycle graph by means of a quantum walk assisted by continuous measurement and feedback. Unlike previous spatial search approaches, where the oracle is described as a projector on the target state, we consider a dynamical oracle implemented through a feedback Hamiltonian. The idea is based on continuously monitoring the position of the quantum walker on the graph and then to apply a unitary feedback operation based on the information obtained from measurement. The feedback changes the couplings between the nodes and it is optimized at each time via a numerical procedure. We numerically simulate the stochastic trajectories describing the evolution for graphs of dimensions up to $N=15$, and quantify the performance of the protocol via the average fidelity between the state of the walker and the target node. We  discuss  different constraints on the control strategy. For unbounded controls the protocol is able to quickly localize the walker on the target node. We then discuss how the performance is lowered by posing an upper bound on the control couplings. Finally, we show how a digital feedback protocol seems in general as efficient as the continuous bounded one.
\end{abstract}

\date{\today}
\maketitle

\section{Introduction}
\label{sec:intro}

Quantum walks are used to model the evolution of a quantum particle, or excitation, over a discrete set of positions. If the walker dynamics is continuous in time, i.e. time is a real positive parameter, we talk about continuous-time quantum walk (CTQW) \cite{farhi981,kempe003,venegas12}. In many contexts, quantum walks show a quantum advantage with respect their classical counterpart, by allowing a speed-up in completing certain tasks. Examples are found in quantum computation \cite{childs09,kendon14comp, hines07}, and quantum algorithms \cite{chil02,childs_cleve_03,gamble10,rossi15,cade18,Callison_2019,marsh20,kryukov22,ambainis2003,kadian21,venegas2008quantum}.
Since a  CTQW evolves over a graph, it is strongly related to  applications over networks, including
quantum spatial search \cite{childs04, Janmark14, childs2014, Chakraborty20_anyGraph, paris-oracle-21,benedetti21,portugal2013quantum},
quantum routing \cite{alastair, Chudzicki10,  paganelli13},
quantum transport and state transfer \cite{Christandl05,kay10, tama16}.
The ability to redirect or control information over a graph in an efficient way is essential to develop protocols involving quantum networks and to deal with a large amount of structured data. To this aim, we develop a protocol to  guide the walker toward a target node  on a graph by exploiting the tools of quantum control.

\par
The theory of quantum control \cite{WisemanMilburn} addresses the problem of preparing a quantum system in a desired quantum state or with some desired quantum properties. In quantum feedback-control strategies the quantum system under control is measured (typically continuously in time) and the information acquired is exploited in order to optimize a feedback operation on the system itself. This kind of strategies has been studied in great detail, in particular with the aim of generating quantum states with non-classical properties such as squeezing or entanglement \cite{Wiseman1993,Wiseman1994,Doherty1999,Thomsen2002,SerafozziMancini,Szorkovszky2011,Genoni2013PRA,Genoni2015NJP,Hofer2015,Martin2015,martin2017optimal,Martin:17,Brunelli2019PRL,martin2020quantum,Jiang2020,Zhang2020,Rossi2020PRL,DiGiovanni2021} or to cool optomehcanical systems towards their ground states, with the experimental results recently demonstrated in~\cite{Rossi2018,Magrini2021,Tebbenjohanns2021}.
\par
In this article, we propose a novel approach to quantum search on graphs and establish a new and unexplored line of research combining CTQWs with quantum feedback-control protocols. In our system, the walker is interacting with an environment that is continuously monitored. As a result, the system evolution is a quantum stochastic trajectory and, based on the result of the measurement, a feedback protocol is applied. We will prove that it is possible to drive the walker towards a target state by optimizing the corresponding target fidelity at each step. 
Our method differs from  other methods that have been developed since it allows the walker to be guided continuously to the target node. 
In the  standard spatial search protocol the oracle is described as a projector operator onto the target state. With our approach we modify this paradigm. We consider a \emph{dynamical} oracle encoded in the feedback operation. 
This means that the final projective measurement on the walker, to be performed at any time $t$ after a certain threshold time $t_{th}$, has a high probability of success. In particular, we find that once the walker reaches the target node it remains stuck in it thanks to the feedback operation. This lift the burden of  performing a final measurement at a very specific time, by allowing us to measure the walker position at any time $t\geq t_{th}$.

\par
The manuscript is organized as follows: in Sec. \ref{qwalks} and in Sec. \ref{sec:qcontrol} we give a brief introduction  to continuous-time quantum walks, and  to continuously monitored quantum systems and  unitary (measurement-based) feedback strategies respectively. In Sec. \ref{sec:qtargeting}, we introduce our search scheme and we describe the idea behind our feedback protocol. In Sec. \ref{sec:results} we analyze our results for the different control strategies we have considered, while 
we conclude the manuscript in Sec. \ref{sec:conclusions} with some  remarks and  outlooks.

\section{Quantum walks and spatial search on graphs}
\label{qwalks}
%

A continuous-time quantum walk  describes the continuous motion of a quantum particle over a discrete set of positions. 
Underlying every walk there is a graph $G$, which is described as
 a pair $G=(V, E)$ where $V=\{0,1,\dots,N-1\}$ is the set of vertices and $E$ is the set of undirected edges, i.e. all the pairs of adjacent vertices in $V$. For a CTQW, the  vertices represent the positions that the particle can occupy while 
the edges encode all the possible paths that a walker can move across.
We denote the order of the graph as the number of nodes $N = \vert V\vert$.
We  restrict our discussion to regular graphs, with no loops 
nor multiple edges. 
All this information determines the topology of the graph and is encoded in the adjacency matrix $A$. In the position basis 
$\mathcal{B}_p=\{\ket{k}\}_{k=0}^{N-1}$,
this is given by the matrix elements
\begin{equation}
    \langle i \vert A \vert j \rangle = A_{ij} =
    \begin{cases}
        1 & \textup{if } (i,j)\in E \\
        0 & \textup{if } (i,j)\not\in E
    \end{cases}.
    \label{eq:adjmat}
\end{equation}
For example, in the case of a cycle graph the only non-zero elements  are 
$A_{jk}$ and $A_{kj}$ satisfying the condition ${k=(j+1)\!\!\! \mod N}$, with $j=0,\dots,N-1$.
Associated to $A$ there is a Laplacian matrix, defined as 
$L = D - A$. Here $D$ is a diagonal  matrix, with 
$\langle k \vert D \vert k \rangle=d_k= \sum_{j=0}^{N-1}A_{kj}$  the degree of the $k$th vertex. The Laplacian matrix is  promoted to be the generator of the quantum dynamics and the physical origin behind this choice is the 
correspondence between $L$ and the discretized kinetic operator for regular lattices 
\cite{wong2016laplacian}. 
Hence, the quantum walker Hamiltonian is $H = \gamma L$, where the parameter $\gamma > 0$ is the hopping rate between the nodes and it accounts for the energy scale of the system.
 The generic state of the CTQW at time $t$ is a superposition over the vertices $\ket{\psi(t)}=\sum_{k} a_k\ket{k}$ with  $a_k=\bra{k}e^{-iHt}\ket{\psi_0}$ and $\ket{\psi_0}$ the initial state of the walker.
 Throughout the paper we set $\hbar=1$.
In the case of regular graphs, 
the matrix $D$ is proportional to the identity operator, thus making the evolutions generated by $L$ and $A$ equivalent.
Before proceeding, we  want to remark that using $L$ as the dynamics generator is only one of the possible (infinitely many) choices for a  Hamiltonian. Indeed any Hermitian operator,\CB{ which respects the topology of the graph}, can be used as a legit CTQW Hamiltonian \cite{wong2016laplacian, cqw2,Turner_2021, frigerio21,frigerio22}.
\\
\par
One  application of CTQWs is  the  spatial search algorithm. The goal is to exploit the coherent evolution of a quantum walker to find a marked vertex on a graph faster than its classical counterpart. 
In the quantum spatial search of a marked node $\ket{w}$, the walker evolves under the Hamiltonian:
\begin{equation}
    H_S=\gamma L - \ketbra{w}{w} 
    \label{stardardSearch}
\end{equation}
where $\gamma$ is a   real parameter and the operator $\ketbra{w}{w}$ is the oracle Hamiltonian i.e. a projector onto the target state. 
The walker is usually initialized in the uniform superposition of all nodes \begin{equation}
    \ket{\psi_0}=\frac{1}{\sqrt{N}}\sum_{k=0}^{N-1}\ket{k},
    \label{psi0}
\end{equation}
with no bias toward  the target state. 
The algorithm is successful if the probability of finding the target node $p_w(t_s)=|\bra{w}e^{-iH_S t_s}\ket{\psi_0}|^2 $ is as close as 1 as possible in a time $t_s=\mathcal{O}(\sqrt{N})$.
It was shown  that a $\sqrt{N}$ speedup can be obtained for specific topologies, such as the complete and hypercube graphs and $(d>4)$-dimensional lattices \cite{childs04,farhi98}.
Later studies  proved fast search for different kinds of graphs \cite{Chakraborty16,philipp16,wong16,omar17,wong18,wang20} and
a comprehensive analysis  of the algorithm's performances was carried out in \cite{chakraborty20}, which recovers
previous  graph-dependent results as special cases.
 It is worth mentioning here that considering different oracle operators, such as those which modify the edges connected to the target node, allows to reach
 a search time $t_s=\mathcal{O}(\sqrt{N \ln N})$  in two-dimensional ($d=2$) lattices, by building Hamiltonians that exhibit Dirac points in their dispersion relation  \cite{tanner14, childs2014}. 
 \CB{Since low dimensional lattices, such as the cycle graph, do not sustain fast search with the standard algorithm defined by Hamiltonian \eqref{stardardSearch}, novel strategies must be envisaged to boost the spatial search on these structures. 
 We report in appendix \ref{app:standard-algorithm-cycle} the success probability of the standard search algorithm on the cycle graph  to set a benchmark for our approach.
}

\begin{figure*}
    \centering
    \includegraphics[width=\textwidth]{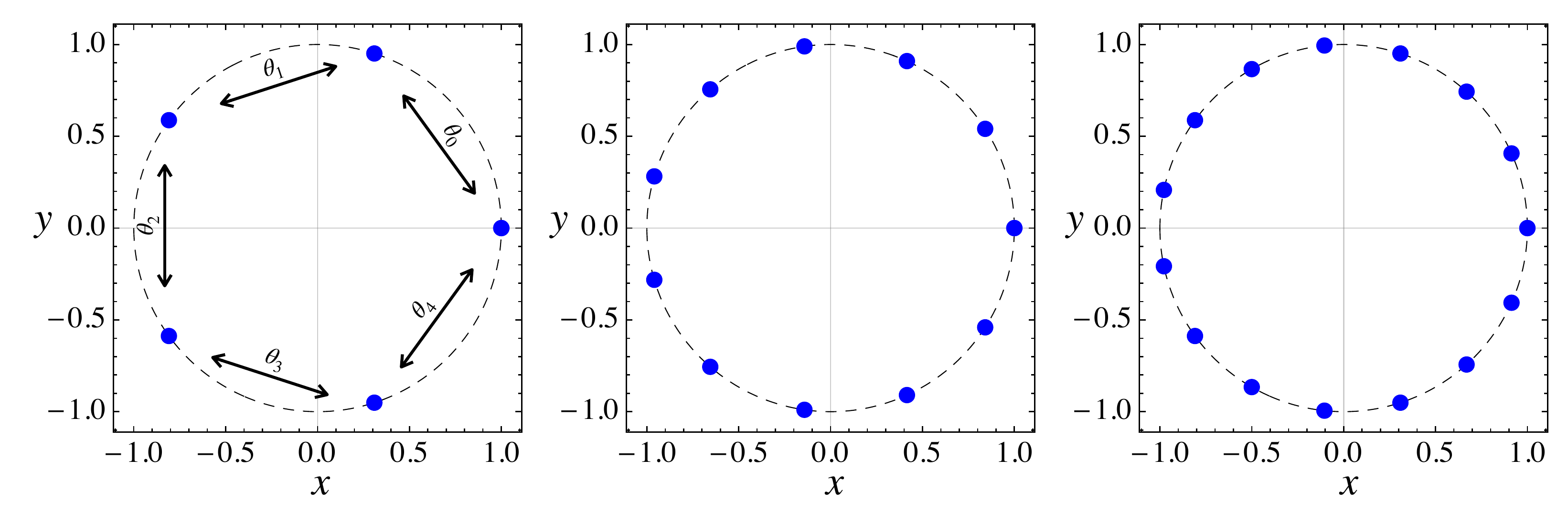}
    \caption{Cycle graph embedded in a plane on a circle of unit radius. Left panel: $N=5$; Central panel: $N=11$; Right panel: $N=15$. 
    }
    \label{fig:cycleembedded}
\end{figure*}

\section{Continuous monitoring and Feedback control}
\label{sec:qcontrol}

\subsection{Continuously monitored quantum systems}
We here provide a very basic introduction to continuously monitored quantum systems. We refer to the following references \cite{WisemanMilburn,SteckJacobs,Brun2002} for a more detailed introduction and for the derivation of the formulas provided in this Section.
We assume that the quantum system under exam interacts with a large Markovian environment described by a train of input bosonic operators $\hat{a}_{j}(t)$ satisfying the canonical commutation relation $[\hat{a}_{j}(t^\prime),\hat{a}^\dagger_{k}(t)]=\delta_{jk} \delta(t-t^\prime)$. The interaction with the system is then given in terms of the time-dependent interaction Hamiltonian 
\begin{align}
\hat{H}^{int} (t) = i \sum_j \sqrt{\kappa_j}(\hat{c}_j\hat{a}^\dagger_{j}(t) \AC{-} \hat{c}_j^\dagger \hat{a}_{j}(t)),
\label{eq:intHamiltonian}
\end{align}
in which $\kappa_j$ represent the coupling strengths, while $\hat{c}_j$ are operators acting on the system Hilbert space (one should also notice that the parameter $t$ in the operators $\hat{a}_j(t)$ is just a label denoting the time at which each operator is interacting with the quantum system via the Hamiltonian). 
We  also assume that the environmental modes $\hat{a}_j(t)$ can be measured continuously in time, just after the interaction, in order to gain information on the state of the system itself. Notice that the interaction with the environment can be either considered already present, and that some degree of control on this environment is achievable in order to perform such a measurement, or that such an interaction can be effectively engineered with the purpose of weakly monitoring the system. Both these approaches are nowadays pursued efficiently in different physical platforms, in particular in circuit QED \cite{Murch:2013aa,Naghiloo:2016aa,Hacohen-Gourgy2016,Ficheux2018,Minev:2019aa} and in optomechanical systems \cite{Wieczorek2015,Rossi:2018aa,Rossi2019,Magrini:2021aa,Tebbenjohanns:2021aa}. It is known that in quantum mechanics a measurement modifies the state of the quantum system that is (directly or indirectly) measured, and that the corresponding {\em conditional state} will depend both on the kind of measurement performed and on the outcome of the measurement. We will focus on continuous homodyne detection of the environmental modes with monitoring efficiencies $\eta_j$, corresponding to a set of continuous photocurrents
\begin{align}
dy_t^{(j)} = \sqrt{\eta_j \kappa_j} \Tr[(\hat{c}_j + \hat{c}^\dagger_j) \varrho^c ] \,dt + dW^{(j)}_t \,, \label{eq:photocurrent}
\end{align}
where $dW^{(j)}_t$ denotes the {\em innovation}, that is the difference between the result of the measurement $dy_t^{(j)}$ and the expected results, and mathematically correspond to independent Wiener increments satisfying $dW^{(j)}_t dW^{(k)}_t= \delta_{jk} dt$. The evolution of the quantum state $\varrho^c(t)$ conditioned on the photocurrents $dy_t^{(j)}$ is then given by the following stochastic master equation (SME)
\begin{align}
   d\varrho^c &= -i[\hat{H}_s,\varrho^{c}(t)]dt +\sum_{j}\kappa_j \mathcal{D}[\hat{c}_j] \varrho^{c}(t) dt \nonumber \\
   &\,\, +\sum_{j}\sqrt{\eta_j \kappa_j} \mathcal{H}[\hat{c}_j] \varrho^{c}(t) \,dW_t^{(j)} \,
   \label{eq:sme}
\end{align}
where $\hat{H}_s$ is the Hamiltonian describing the evolution of the quantum system only, and where we have introduced the two following superoperators
\begin{align}
    \label{eq:superop1}
    \mathcal{D}[\hat{c}]\bullet &= \hat{c}\bullet \hat{c}^\dag - (\hat{c}^\dag\hat{c}\bullet+\bullet\hat{c}^\dag\hat{c})/2 \,, \\
    \label{eq:superop2}
    \mathcal{H}[\hat{c}]\bullet &= \hat{c}\bullet + \bullet \hat{c}^\dag - \Tr[(\hat{c} + \hat{c}^\dag)\bullet]\bullet \,.
\end{align}
The continuous outcomes of the photocurrents $\{dy_t^{(j)}\}$ thus define a particular conditional trajectory for the conditional state of the quantum system. By averaging over all the possible trajectories, i.e. over all the possible outcomes of the measurements, we obtain the evolution of the unconditional state $\varrho^u = \mathbbm{E}_{\sf traj}[\varrho^c]$ that, by exploiting the property $\mathbbm{E}_{\sf traj}[dW_t^{(j)}]=0$, is a Markovian master equation in the Lindblad form
\begin{align}
    \frac{d\varrho^u}{dt} &= -i[\hat{H}_s,\varrho^{u}(t)] +\sum_{j}\kappa_j \mathcal{D}[\hat{c}_j] \varrho^{u}(t). \label{eq:uncondMQE}
\end{align}

The evolution of the conditional states described by the SME (\ref{eq:sme}) can be equivalently described via the formula \cite{rouchon2015arxiv,Rouchon2015} 
\begin{widetext}
\begin{align}
    \varrho^c(t+dt) = \frac{\hat{M}_{{\bf dy}_t} \varrho^c(t) \hat{M}_{{\bf dy}_t}^\dag + \sum_j (1-\eta_j) \hat{c}_j \varrho^c(t) \hat{c}_j^\dag \,dt }{\Tr[\hat{M}_{{\bf dy}_t} \varrho^c(t) \hat{M}_{{\bf dy}_t}^\dag + \sum_j (1-\eta_j) \hat{c}_j \varrho^c(t) \hat{c}_j^\dag \,dt]} \,,
    \label{eq:evcond}
\end{align}
\end{widetext}
where we have introduced the family of Kraus operators
\begin{equation}
    \hat{M}_{{\bf dy}_t} = \mathbb{I} -i \hat{H}_s dt - \sum_{j}\left( \frac{\kappa_j}{2} \hat{c}^\dagger_j\hat{c}_j dt - \sqrt{\eta_j \kappa_j}\, \hat{c}_j\, dy_t^{(j)}\right) \,,
    \label{eq:Kraus}
\end{equation}
with ${\bf dy}_t = \{dy_t^{(1)},...,dy_t^{({\tiny K})}\}$ denoting the vector of the outcomes of the $K$ mesaurement channels.\\
In our protocol we  consider an initial pure state and perfect monitoring efficiency, i.e. $\eta_j = 1$ for all channels. 
Under these assumptions the conditional evolution is described by a stochastic Schr{\"o}dinger equation, or equivalently via the Kraus operators as follows:
\begin{align}
    |\psi^c(t+dt)\rangle = \frac{\hat{M}_{{\bf dy}_t} |\psi^c(t)\rangle}
    {\sqrt{\langle \psi^c(t) | \hat{M}_{{\bf dy}_t}^\dag \hat{M}_{{\bf dy}_t} |\psi^c(t)\rangle}} \,.
    \label{eq:sse}
\end{align}

\subsection{Unitary quantum feedback}
In addition to conditioning the evolution of the quantum state, the outcomes of the measurement performed can  in principle be exploited to further modify the dynamics of the system. 
In this respect, here we  briefly introduce unitary measurement-based quantum feedback. The idea is that, once the measurement outcomes ${\bf dy}_t$ are obtained, one \AC{performs} a unitary operation $\hat{U}_{fb}(t)$ on the quantum state, typically optimized in order to achieve a certain goal as, for example, the preparation of a certain target quantum state. 
This unitary operation may depend only on the last measurement outcomes, or on the whole history of outcomes, and thus on the whole trajectory of the conditional state. In the first instance one talks about Markovian quantum feedback and one can derive a corresponding Markovian feedback master equation \cite{Wiseman1994,WisemanMilburn}. In this work we will focus on the second kind of feedback, and thus our feedback strategy will be optimized by knowing both the last measurement outcomes and the conditional state $|\psi^c(t)\rangle$ (and as a consequence the whole measurement history). In order to obtain the corresponding evolution, we exploit the formulas involving the Kraus operators. 
In particular, if the feedback operation is performed after the measurement, by assuming unit measuring efficiency, initial pure states and no-delay between measurement and feedback, the conditional state at each instant is described via the formula
\begin{align}
       |\psi^{fb}(t+dt)\rangle =  \frac{\hat{U}_{fb} \hat{M}_{{\bf dy}_t} |\psi^{fb}(t)\rangle}
    {\sqrt{\langle \psi^{fb}(t) | {\hat{M}_{{\bf dy}_t}}^\dag \hat{M}_{{\bf dy}_t} |\psi^{fb}(t)\rangle}} \,.
    \label{eq:purefeedback} 
\end{align}
This formula is particularly useful for our numerical approach
where one needs to substitute the time differential $dt$ with a finite but small time increment $\Delta t$, while the Wiener increment $dW_t^{(j)}$ must be replaced by a Gaussian random variable $\Delta W_t^{(j)}$ with zero mean and variance $\Delta t$. The finite increments to the measurement records are
\begin{equation}
   \Delta y_t^{(j)} = \sqrt{\kappa_j} \langle \psi^{c}(t) \vert (\hat{c}^\dagger_j + \hat{c}_j) \vert \psi^{c}(t) \rangle \Delta t + \Delta W_t^{(j)} .
\end{equation}
Due to the finite nature of $\Delta t$, the deterministic identity $\Delta W_t^{(j) 2} = \Delta t$ is no longer satisfied, thus corrections must be considered. This is accomplished by adding an extra term, known as  Euler-Millstein correction \cite{Rouchon2015}, in the Kraus operators that now read 
\begin{align}
    \hat{M}_{{\bf \Delta y}_t} & = \mathbb{I} - i \hat{H}_s \Delta t - \sum_{j} \bigg(\frac{\kappa_j}{2} \hat{c}_j^\dagger \hat{c}_j \Delta t - \nonumber \\
    & \quad -\sqrt{\kappa_j} \hat{c}_j \Delta y_t^{(j)} - \frac{\kappa_j}{2}\hat{c}_j^2({\Delta y_t^{(j)}}^2-\Delta t) \bigg) \,.
\end{align}

\section{Quantum search assisted by feedback}
\label{sec:qtargeting}
Our idea is to continuously monitor the position of a quantum walker during its evolution on a cycle graph, and then to use this information to apply feedback unitary operations as a dynamical oracle with the aim of finding a particular target node. 
The walker is initially prepared in the uniform superposition of all nodes of the graph as in Eq. \eqref{psi0}.
The first step is to describe the continuous monitoring. 
In particular we assume to be able to couple our system to two different environments via the following jump operators 
\begin{align}
    \hat{c}_1 = \hat{x} = \sum_{k=0}^{N-1} \cos(\frac{2\pi k}{N}) \vert k \rangle \langle k \vert\,,\label{eq:jumpoperators1}  \\
	\hat{c}_2 = \hat{y} = \sum_{k=0}^{N-1} \sin(\frac{2\pi k}{N}) \vert k \rangle \langle k \vert \,,
	\label{eq:jumpoperators2}
\end{align}
whose eigenvalues exactly correspond to the coordinates of the position of the $N$ nodes of the graph, corresponding to 
equally spaced points 
on a unit radius ring centered on $(0,0)$ in the $(x,y)$ plane (see Fig. \ref{fig:cycleembedded} for cycle graphs with $N=5, 11, 15$). \AC{By performing continuous homodyne detections one obtains two photocurrents (\ref{eq:photocurrent}) whose average values are indeed proportional to the the expectation values of the operators $\hat{x}$ and $\hat{y}$ on the conditional state $\varrho^c$}.
We show in Appendix \ref{app:uncondME} that the unconditional evolution, corresponding to the master equation (\ref{eq:uncondMQE}) with the cycle graph Hamiltonian $H_S = \gamma L$ and 
jump operators \eqref{eq:jumpoperators1}-\eqref{eq:jumpoperators2}, leads to a symmetric dephasing-like evolution in the position basis, {thus reflecting the translation invariance of the  graph's nodes and} further validating our choice. \AC{In Appendix \ref{app:complexjumpresult} we also mention an alternative choice for the jump operator, i.e. the single non-Hermitian jump operator 
$$
\hat{c}_{0}= \sum_{k=0}^{N-1} e^{i 2\pi k/N} |k\rangle\langle k|\,,
$$
that  satisfies the properties discussed above. We show that the results are comparable with the one obtained via the two Hermitian jump operators $\hat{c}_1$ and $\hat{c}_2$.}
\\
The second  step is to define our feedback strategy. We parametrize the  unitary feedback operator as 
\begin{align}
   &\hat{U}_{fb}( {\boldsymbol \theta})= e^{-i \hat{H}_{fb}( {\boldsymbol \theta}) dt} \,,
\end{align}
where, in general, we  assume to have a finite number of control parameters $ {\boldsymbol \theta}=\{\theta_k\}$ corresponding to a set of control operators $\{\hat{h}_k\}$, such that the feedback Hamiltonian reads
\begin{equation}
\hat{H}_{fb}( {\boldsymbol \theta}) = \sum_k \theta_k \hat{h}_k \,.
\label{eq:feedbHamiltoniansOK}
\end{equation}
The definition of the control operators is indeed crucial. Two natural choices can be considered: the first one is to choose the on-site projectors, i.e. $\hat{h}^{(os)}_k = |k\rangle\langle k|$. However preliminary numerical simulations show that this choice is in general not efficient. 
In order to understand why this is the case, one can for example observe that, being $\{|k\rangle\}$  eigenstates of the control operators above, if the walker during the evolution happens to be in a node $|\bar{k}\rangle$ different from the target, the unitary operation will not be able to change its state and thus the feedback is useless for our purposes. The second natural choice is to consider the hopping operators 
\begin{align}
    \hat{h}^{(hop)}_k = |k\rangle\langle k+1| + |k+1\rangle\langle k| \,, \label{eq:hopping}
\end{align}
with the usual boundary condition $|N\rangle \equiv|0\rangle$.
This set of feedback control operations represents the ability of controlling each coupling between adjacent nodes individually. 
Finally, one needs to decide how to optimize the feedback operation, that is the set of control parameters $ {\boldsymbol \theta}$, in order to find the target node on the graph.
This is typically done by defining a reward function $\Lambda(|\psi^{fb}\rangle)$, that in our case naturally corresponds to the fidelity between the conditional state after the feedback operation $|\psi^{fb}\rangle$ in Eq. (\ref{eq:purefeedback}), and the target state (that we will hereafter denote as $|0\rangle$), i.e.
\begin{equation}
\label{eq:rewardfun}
    \Lambda(|\psi^{fb}(t) \rangle) = \mathcal{F}_{\vert 0 \rangle}(|\psi^{fb}(t) \rangle) = |\langle 0 \vert \psi^{fb}(t)\rangle |^2.
\end{equation}
We will thus choose the parameters ${\boldsymbol \theta}$ as the ones maximizing the fidelity at each step of the trajectory. The same figure of merit will be then used in order to assess the performance of our protocol. We will indeed numerically evaluate
\begin{align}
    \overline{\mathcal{F}}_{\vert 0 \rangle} = \mathbbm{E}_{\sf traj}[\mathcal{F}_{\vert 0 \rangle}(|\psi^{fb}(t) \rangle) ] \,,
\end{align}
that is the fidelity averaged over all the possible trajectories conditioned by the continuous monitoring.
\section{Results}
\label{sec:results}

In the following, we  present our main results, dividing them in three different settings: in Sec. \ref{sec:unbounded} we address the numerical optimization of the feedback operation with unbounded control parameters, that is without posing any bound on the search domain for the parameters $\{\theta_i\}$. Then, in Sec. \ref{sec:multfeedbHnumOpt} we consider numerical optimization of the feedback but with a bounded domain, that from the physical point of view may represent constraints on the physical implementation of the feedback operations. In Sec. \ref{sec:multfeedbHswitch} we study the case of {\em digital feedback} \cite{riste2015}, in which the value of the feedback couplings are not only bounded, but can take values only from a discrete set (one should notice that unlike  previous examples of digital feedback \cite{riste2015}, we still have a continuous measurement output, and only the feedback operations are discrete). 
In all these examples we  consider the system initially prepared in the quantum  state defined in Eq. (\ref{psi0}), corresponding to the uniform superposition over the $N$ nodes of the graph.
\par
The algorithm we used to numerically optimize the feedback couplings is provided by the \texttt{SciPi} library, and in particular the \texttt{scipy.optimize.minimize} function \cite{scipy.opt.min}. The method used for the different strategies are the following: for the unbounded optimization (Sec. \ref{sec:unbounded}) we use the \texttt{BFGS} method; for the bounded optimization (Sec. \ref{sec:multfeedbHnumOpt}) we used the \texttt{L-BFGS-B} method. Differently, the method used for digital feedback (Sec. \ref{sec:multfeedbHswitch}) is a brute force one, i.e. we consider all the possible combinations of the finite discrete values and we select the optimal one.\\
We remark that we have also investigated the scenario with a single feedback Hamiltonian controlling collectively all the couplings via a single parameter $\theta$, i.e. via the Hamiltonian $\hat{H}_{fb} = \theta \sum_k  \hat{h}^{(hop)}_k$. However, as we  show in Appendix \ref{app:singlefeedH}, this kind of feedback is not particularly efficient for our purposes.
\subsection{Numerical optimization with unbounded controls}
\label{sec:unbounded}
We here show the results of our protocol when considering a feedback operation via the control operators introduced in Eq. (\ref{eq:hopping}) and optimized control parameters $\{\theta_k\}$ with unbounded domain. A remark is in order here: the first attempt we have pursued was to follow the approach described by Martin et al. in \cite{Zhang2020,martin2020quantum}, where the feedback operation is assumed to be infinitesimal, i.e. via couplings written as
\begin{align}
    {\boldsymbol \theta}\,dt = A \,{\bf dW} + {\bf B} \,dt \,,
\end{align}
where ${\boldsymbol \theta} = (\theta_1, \dots, \theta_K)^{\sf T}$ is the vector of the feedback couplings, ${\bf dW} = (dW_t^{(x)}, dW_t^{(y)})$ is the vector of the Wiener increments describing the measurement, while $A$ and ${\bf B}$ are respectively a $(N\times 2)$-dimensional matrix and a $2$-dimensional vector describing the feedback strategy. The optimization in this scenario could be done analytically if some conditions are fulfilled, as described in \cite{martin2020quantum} (see Appendix \ref{app:singlefeedH} for further details on this method). However for our problem we verify that these conditions are never satisfied and thus a numerical optimization with unbounded couplings has to be performed.  
\begin{figure*}
\centering
\includegraphics[width=\textwidth]{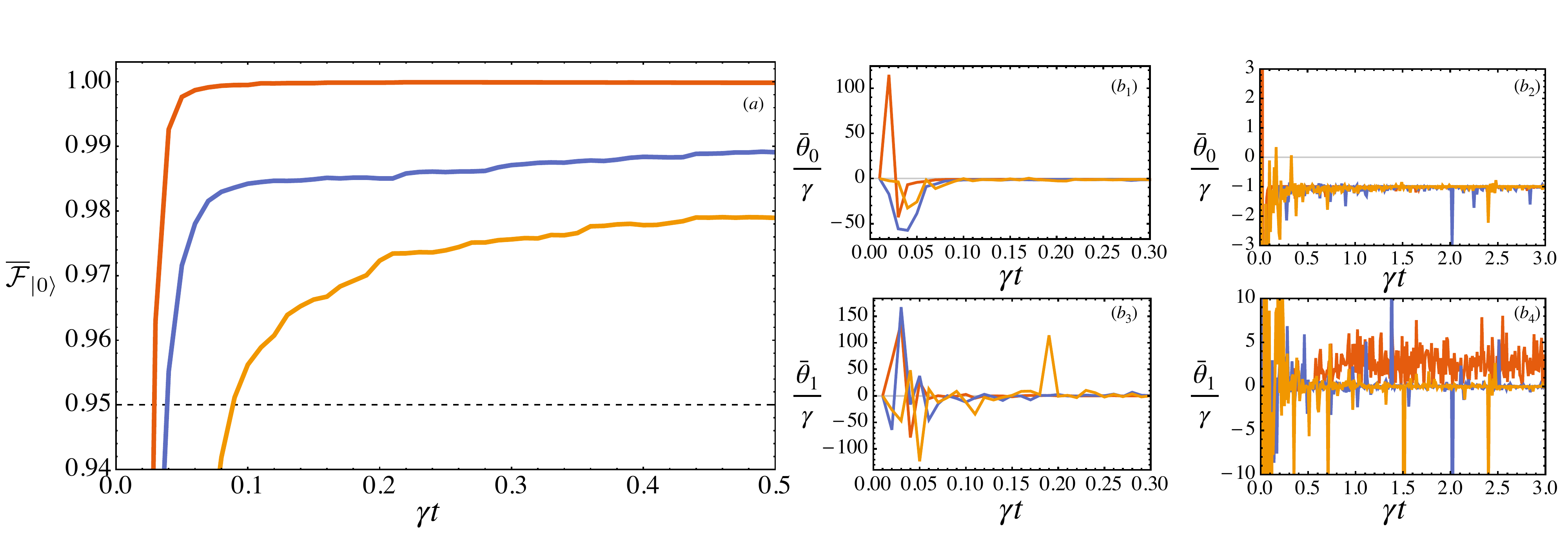}
\caption{Results for multiple feedback control Hamiltonians $\hat{h}_{k}^{(hop)}$ defined in Eq. \eqref{eq:hopping}. $(a)$: average reward function $\overline{\mathcal{F}}_{\vert 0 \rangle}$ as a function of time (black dashed line: threshold $\mathcal{F}^{th} = 0.95$). $(b_{1})-(b_{4})$: averaged feedback couplings $\bar{\theta}_{k}$, corresponding to the $k$th Hamiltonian in \eqref{eq:feedbHamiltoniansOK}: $(b_{1})$ and $(b_{3})$: short time behavior; $(b_{2})$ and $(b_{4})$ time-asymptotic behavior. 
Red line: $N=5$; blue line: $N=11$; orange line: $N=15$. Other parameters: $\eta =1$, $dt=0.01$, number of stochastic trajectory $N_{tj}=5000$.
}
\label{fig:multFeedUnbound}
\end{figure*}
\par
The results are depicted in Fig. \ref{fig:multFeedUnbound} for $N=5,11,15$, where we observe that the protocol is particularly efficient in reaching the target state.
\CB{In order to quantify the efficiency of our protocol, we fix a threshold value for the average fidelity $\mathcal{F}^{th}=0.95$, such that whenever the reward function is larger than this threshold value, the target search is considered successful. }
This efficiency decreases as the size of the graph increases, although the threshold value is reached on rather small time scales $\gamma t <1$. \CB{This is a great improvement with respect  to the performance of the standard quantum spatial search algorithm, reported in Appendix \ref{app:standard-algorithm-cycle}, where the success probability never reaches the threshold value (see Fig. \ref{fig:child-algo} for a comparison).  }

In Fig. \ref{fig:multFeedUnbound} we also consider the average feedback coupling $\bar{\theta}_0$ between  nodes $\ket{0}$ and $\ket{1}$ and the coupling $\bar{\theta}_1$ between  nodes $\ket{1}$ and $\ket{2}$ (see Fig. \ref{fig:cycleembedded} for reference): after  an initial transient, the average value of the feedback couplings reaches an asymptotic value, meaning that the feedback operation is stable after having reached the target node. We notice that the average coupling $\bar{\theta}_0$, 
between nodes $\ket{0}$ and $\ket{1}$, tends to the asymptotic value $\bar{\theta}_0 \to -\gamma$. 
In fact,  when the protocol has  {\em almost} localized the walker in the desired node, the role of the feedback is  to try to stop the dynamics by nullifying the corresponding couplings in the Hamiltonian. We have also numerical evidence that, within numerical noise, the average feedback couplings are symmetric with respect to the $x$-axis ,i.e. $\bar{\theta}_0 = \bar{\theta}_{N-1}$, $\bar{\theta}_1 = \bar{\theta}_{N-2}$, etc., in the configuration in which the target node is placed in $(1,0)$, see Fig. \ref{fig:cycleembedded}.
\CB{The average feedback couplings reported in Fig. \ref{fig:multFeedUnbound} show however large fluctuations, especially $\bar{\theta}_1$, suggesting that in some trajectories larger values of the optimal couplings are chosen by the optimization algorithm. We propose two possible justifications for this behaviour: i) the first one is based on the stochasticity of the single random trajectory, i.e. there might be a time-step in which the measurement project the state far away from the target node, and thus a large correction is needed; ii) the second one is based on the shape of the landscape functions of the feedback couplings for a single trajectory. One can indeed observe that these landscape functions are periodic and thus have many local and equal maxima that can be reached by different values of the feedback couplings. The large fluctuations thus may arise from the fact that the algorithm does not always choose the maximum in the neighbourhood of the maximum found at the previous step (further details on the couplings' landscape functions are given in Appendix \ref{app:landscape}).}
\subsection{Numerical optimization with bounded controls}
\label{sec:multfeedbHnumOpt}
As we have seen in the previous Section, not only numerical optimization of the feedback strategy is necessary, but also large absolute values of the feedback couplings might be needed to perform an efficient search. 
Hence, to test the limits of our protocol, we consider the case where the feedback couplings $\boldsymbol{\theta}$ belong to a bounded domain. In this case  each $\theta_{k}$ can take values from the interval $[-\xi \gamma, \xi \gamma]$, where we introduced the bounding (dimensionless) parameter $\xi$ that quantifies the range of  values admitted for  the feedback couplings $\boldsymbol{\theta}$. 
We  consider the bounding parameter $\xi\geq 1$. We have indeed numerical evidence that for $\xi<1$, that is for feedback couplings smaller than the Laplacian parameter $\gamma$, the protocol fails. As we noticed in the example above, once the walker has been localized over the target the role of the feedback is to stop the dynamics and this effect cannot be achieved efficiently if in general $|\theta_0| < \gamma$.
\par
\begin{figure*}
\centering
\includegraphics[width=\textwidth]{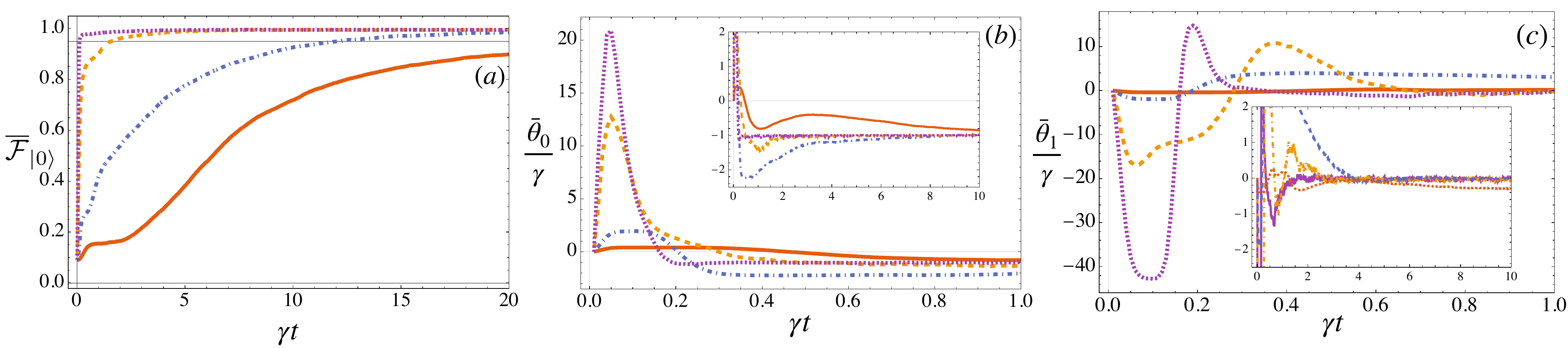}
\caption{{Results for multiple feedback Hamiltonians $\hat{h}_k^{(hop)}$ with bounded control parameters $\boldsymbol\theta$ and for a graph with $N=11$ nodes. $(a)$: average reward function $\overline{\mathcal{F}}_{\vert 0 \rangle}$ as a function of time (black thick line: threshold $\mathcal{F}^{th} = 0.95)$; 
$(b)$ and $(c)$: average feedback coupling $\bar{\theta}_{0}$ and $\bar{\theta}_{1}$ respectively, as a function of time (short time behavior). Insets: average feedback coupling $\bar{\theta}_{0}$ and $\bar{\theta}_{1}$ for a larger time $t$ (asymptotic behavior). 
In all  plots we used  $\xi=1$ (red thick line),  $\xi=5$ (blue dotdashed line)  $\xi=50$ (orange dashed line),  $\xi=100$ (purple dotted line),  $\eta=1$, $dt=0.01$, number of stochastic trajectories $N_{tj}=5000$.}
}
\label{fig:multipleFeedbackBound}
\end{figure*}
\begin{figure}
\centering
\includegraphics[width=0.48\textwidth]{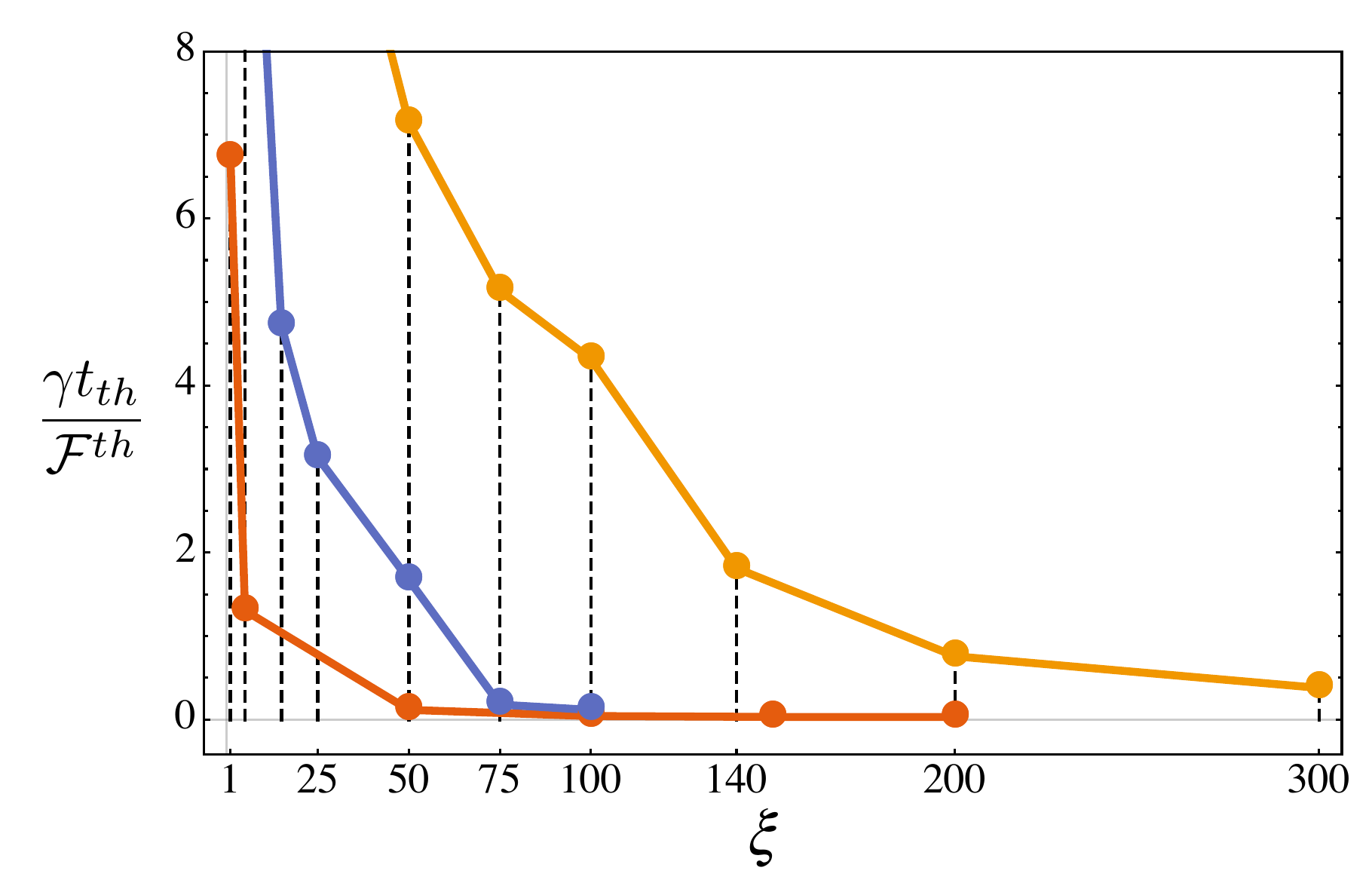}
\caption{Plot of $\gamma t_{th}/\mathcal{F}^{th}$ as a function of the bounding parameter $\xi$ for different graph sizes: $N=5$ (red line)  $N=11$ (blue line) and $N=15$ (orange line). The value $t_{th}$ is the time at which the $\mathcal{F}^{th}=0.95$ is reached on average. By changing the value of the  threshold parameter $\mathcal{F}^{th}$ the qualitative behavior of the curves does not change and the values obtained for different $\xi$ have the same order of magnitude.    We used  the same data and  the same set of parameters of Fig. \ref{fig:multipleFeedbackBound}.
}
\label{fig:tthfuncxi}
\end{figure}

The numerical results are provided in Fig. \ref{fig:multipleFeedbackBound} for $N=11$ (the results are qualitatively similar also for $N=5$ and $N=15$), and for different values of $\xi$ ranging between $\xi=1$ and $\xi=100$. 
As expected, as $\xi$ grows, the efficiency of the protocol improves, i.e. the minimum time $t_{th}$ required
to reach the
 threshold $\mathcal{F}^{th}$  decreases on average. \CB{Moreover, even for small values of $\xi$, this protocol is able to identify the target node with higher probability than the standard quantum algorithm in the same time interval, see Appendix \ref{app:standard-algorithm-cycle}.}
 \\
%
The decreasing of the optimal time $\gamma t_{th}$  for increasing values of $\xi$ can also be seen 
from Fig. \ref{fig:tthfuncxi}, where we plot the ratio of $\gamma t_{th}/\mathcal{F}^{th}$, a quantity that  corresponds to the effective time necessary to reach the target on average \cite{cattaneo18}.
We observe that this quantity in general  quantitatively depends on the chosen threshold value $\mathcal{F}^{th}$  but  it gives  always the same qualitative behaviour. 
We  see that above a certain value of $\xi$, the ratio reaches a minimum asymptotic value. 
Larger values of $\xi$ are necessary to reduce the effective time $\gamma t_{th}/\mathcal{F}_{th}$  when the size $N$ is increased. 
From Fig. \ref{fig:multipleFeedbackBound} and \ref{fig:tthfuncxi},  we notice that the order of magnitude of the $t_{th}$ is  larger if compared with the one obtained for the unbounded feedback (see Fig. \ref{fig:multFeedUnbound}).
\par
In Fig. \ref{fig:multipleFeedbackBound} we also report the average values of the feedback couplings $\bar{\theta}_{0}$ and $\bar{\theta}_{1}$.
\AC{Differently from the unbounded controls scenario, here the noisy fluctuations are much smaller (again, for a more detailed discussion, we refer the reader to Appendix \ref{app:landscape}), and} we have numerical evidence that the time behaviour of $\bar{\theta}_{0}$ and $\bar{\theta}_{1}$  is equal to their symmetric counterpart $\bar{\theta}_{10}$ and $\bar{\theta}_{9}$. 
The qualitative behavior of $\bar{\theta}_{0}$ is the same also for the other values of $N$ and $\xi$ considered: after a first positive peak, it follows a minimum and then a second maximum, which is smaller than the first, and eventually it tends to the finite asymptotically value $\bar{\theta}_{0}\to - \gamma$, confirming our previous intuition. The situation is slightly different for $\bar{\theta}_{1}$, where for $N=11$ and $N=15$ there is a sequence of minima and maxima which asymptotically tend to a value close to $0$  while for $N=5$ the asymptotic values remains positive. %
This behaviour leads to the observation that, for smaller graphs, the couplings  $\theta_1$ and $\theta_{N-2}$ are more relevant with respect to graphs with a larger size.  The other feedback couplings are not particularly interesting, since their average is approximately $0$ everywhere, so we decide to not report them.
\begin{figure*}
\centering
\includegraphics[width=\textwidth]{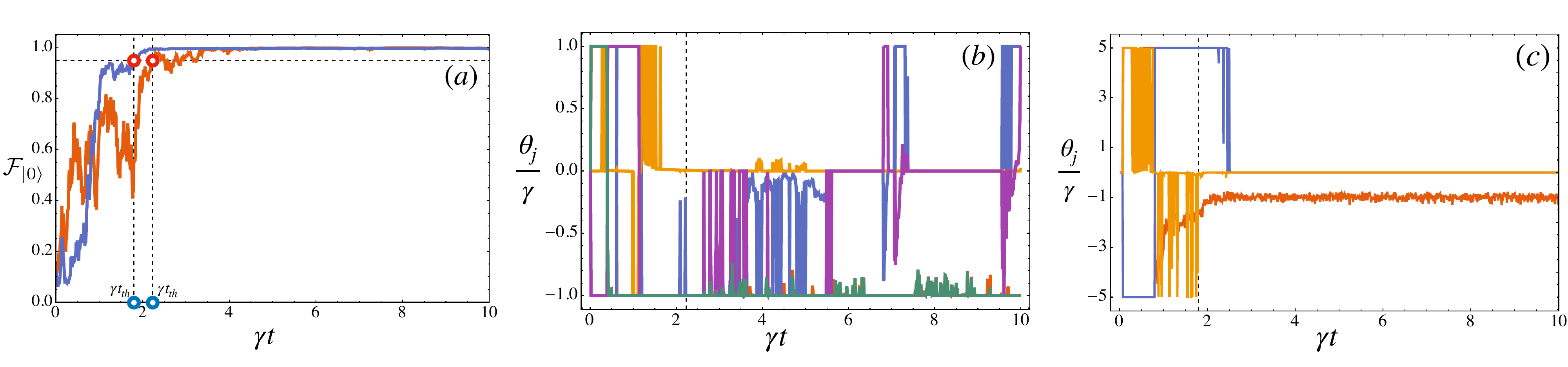}
\caption{Single trajectory for the multiple-feedback Hamiltonian protocol with bounded domain. $(a)$: reward function $\mathcal{F}_{\vert 0\rangle}(\bullet)$ as a function of time for $N=5$, $\xi =1$ (red line) and $N=11$, $\xi=5$ (blue line). In this case, we have respectively $\gamma t_{th}=1.80$ and $\gamma t_{th} = 2.23$, highlighted by the red circles. 
$(b)$: feedback couplings $\boldsymbol{\theta}\,dt$ as a function of time for the $N=5$ trajectory: Red line: $\theta_{0}$; Blue line: $\theta_{1}$; Orange line: $\theta_{2}$; Purple line: $\theta_{3}$; Green line: $\theta_{4}$. $(c)$: feedback couplings $\boldsymbol{\theta} \,dt$ as a function of time for the $N=11$ trajectory: Red line: $\theta_{0}$; Blue line: $\theta_{1}$; Orange line: $\theta_{2}$. All the other feedback couplings are null (apart for some numerical noise of order $10^{-8}$) and we do not report them here. In $(b)$ and $(c)$ the vertical dashed line corresponds to the threshold time $\gamma t_{th}$ of the single trajectory. The parameters considered are $\eta=1$ and $dt=0.01$.\\
}
\label{fig:singtrjboundedxi}
\end{figure*}
\par
So far we have discussed only averaged results on a large number of trajectories. 
To understand in detail the behavior of the protocol, in Fig. \ref{fig:singtrjboundedxi} we report the results for a single stochastic trajectory with bounded domain for $N=5$  with $\xi=1$ and $N=11$  with $\xi=5$. During the transient evolution, when the feedback is driving the walker to the target node $\vert 0 \rangle$,  the value of the reward function is particularly affected by the measurement. After reaching the threshold value, the feedback operation tries to keep the walker into the target vertex. 
However, the stabilization procedure is not given by constant values of the feedback couplings. Similar to what we have discussed in the previous Section, this means that corrections are necessary also after having reached the target state with high fidelity.  These corrections are responsible for the noise we see in the averaged feedback couplings reported in Fig. \ref{fig:multipleFeedbackBound}.  
By inspecting  the single trajectories, we also notice that most of the values taken by $\boldsymbol{\theta}$ during the evolution are either equal to $0$ or to  $\pm \gamma$.  
So it is worth exploring a scenario where the
 feedback couplings $\boldsymbol{\theta}$ belong to  discrete set of possible values.
%

\begin{figure*}
\centering
\includegraphics[width=\textwidth]{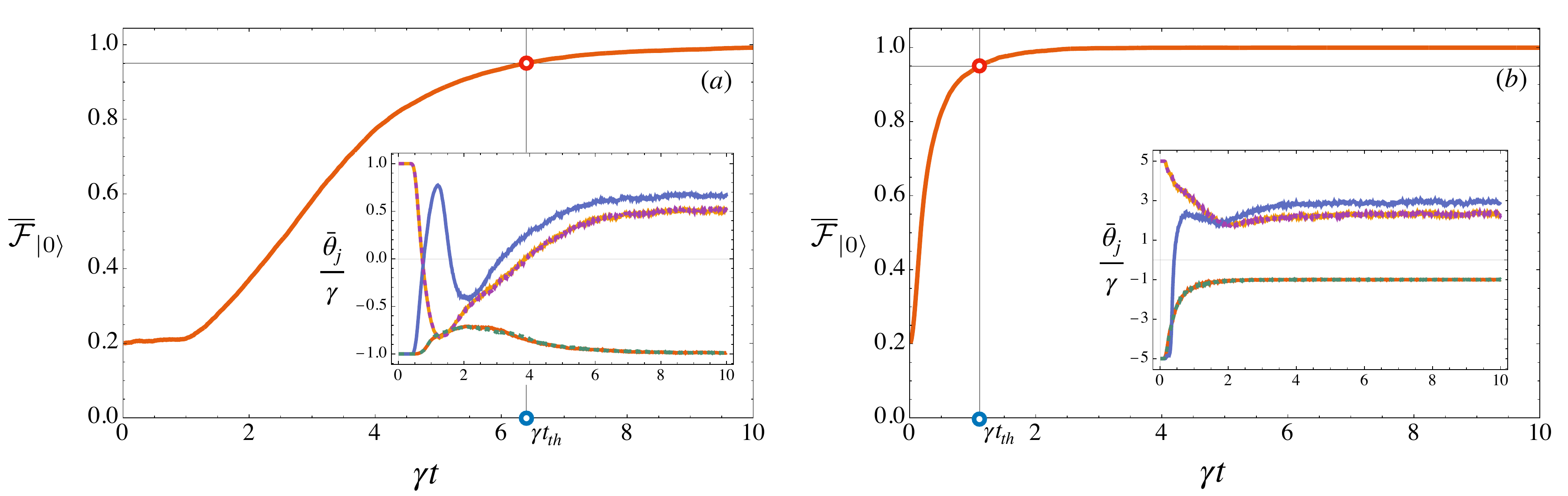}
\caption{Digital feedback control for $N=5$. Left panel: three possible value of the control parameters: $\{0,\pm \gamma \}$. Right panel: five possible value of the control parameters: $\{0,\pm \gamma,\pm \xi\gamma \}$ with $\xi=5$. Main plot: averaged fidelity $\overline{\mathcal{F}}_{\vert 0\rangle}$ as a function time. Inset: averaged feedback couplings $\overline{\boldsymbol{\theta}}$ as a function of time: {Red line: $\theta_{0}$; Orange line: $\theta_{1}$; Blue line: $\theta_{2}$; Purple dashed line: $\theta_{3}$; Green dashed line: $\theta_{4}$. The red and the green line are superposed, as well as the the orange and the purple}. The number of trajectories is $N_{tj}=5000$, and the parameters considered are $\eta=1$ and $dt=0.01$.
}
\label{fig:stepfeedback}
\end{figure*}
\subsection{Numerical optimization with digital feedback control}
\label{sec:multfeedbHswitch}
 We now explore a digital feedback protocol, where the feedback control parameters $\boldsymbol{\theta}$ are picked, at each time step, from a discrete number of values. We study this strategy only for a cycle graph of order $N=5$ since the numerical algorithm we employed is particularly demanding. \\
First, we consider only three possible values for the couplings $\theta_{k}$, belonging to the set $\{ 0, \pm \gamma \}$. 
In this case the optimization algorithm explores all the values of the reward function for all the possible combinations that the five feedback couplings $\boldsymbol{\theta}$ may realize, i.e. $5^{3}=125$ possible combination, and select the one that realizes the maximum $\mathcal{F}_{\vert 0 \rangle}(\bullet)$. In the left panel of Fig. \ref{fig:stepfeedback} we report the results obtained by repeating this algorithm at each step and averaging over $N_{tj}=5000$ trajectories. We see that the threshold value of $\mathcal{F}_{th}$ is reached for a time $\gamma t_{th}=6.40$, which is slightly larger than the value obtained via the continuous bounded protocol $\gamma t_{th} = 6.39$. Regarding the feedback couplings $\boldsymbol{\theta}$, their average values oscillate in the transient time, and after the threshold time $t_{th}$, they stabilize around asymptotic values. We found that approximately  $\bar{\theta}_0 = \bar{\theta}_4 = -\gamma$, 
while   $\bar{\theta}_{1},\bar{\theta}_{2}$ and $\bar{\theta}_{3}$ correspond in general to positive values. 
\par
Then, we consider five possible values for the feedback couplings $\theta_{k}$, belonging to  the set $\{0, \pm \gamma, \pm \xi \gamma\}$, with $\xi$ playing the same role of the bounding factor we introduced before. Here, we may consider the two extra switchers as a boosted feedback operation, i.e. a larger coupling strength than the standard $\gamma$. In this case, the number of possible combination increase and it is equal to $5^{5}=3125$. The averaged results are reported in Fig. \ref{fig:stepfeedback}(b) for $\xi=5$ and for $N_{tj}=5000$ trajectories. 
The threshold value is reached, on average, by the time $\gamma t_{th}=1.12$, which in this case is slightly smaller than the continuous-bounded protocol threshold time $\gamma t_{th}=1.24$. This result is indeed unexpected since the digital feedback is an instance of  the strategies allowed by the continuous bounded domain. 
This suggests that all the algorithms we employed for the optimization of the continuous bounded domain are not particularly efficient in finding the optimal values of the parameters $\boldsymbol{\theta}$, while the brute-force spanning algorithm we considered in this Section cannot fail, as all the possible combinations are tested. 
\\

The averaged values of the feedback couplings reported in the inset of Fig. \ref{fig:stepfeedback} show that the extra switchers are considerably used in the initial stage of the evolution. As time increases, also in this case the values of $\bar{\theta}_{0}$ and $\bar{\theta}_{4}$  reach the  asymptotic value of $\bar{\theta}_0 = \bar{\theta}_4 = - \gamma$. The other couplings, instead, reach an asymptotic value larger than one and are particularly noisy. This is a sign that, in each trajectory, the walker dynamics must  often be corrected by the boosted positive feedback couplings.
\section{Conclusions}
\label{sec:conclusions}
The ability to control or manipulate the dynamics of a quantum walker over a network is important for the development of quantum computation, quantum algorithms and simulations. In this work we proposed a new protocol for searching a target node over a cycle graph by means of a continuous-time quantum walk. The CTQW  interacts with environmental bosonic modes that are continuously monitored and then a proper feedback operation is applied to drive the walker toward the target state. 
The feedback thus plays the role of a dynamic oracle, able to recognize the marked vertex and to change the values of the couplings between the nodes. In this work we analyzed and compared the performances of three different feedback strategies. In the first one, we optimized the feedback couplings without posing any bound on their values; then we considered the case of bounded control, by introducing a bounding parameter $\xi$; finally, we studied the case of digital feedback, where the optimal couplings were picked from a discrete set of values. 
We show how all the three strategies are able to localize the walker on the target node,
\CB{with higher probability with respect to the  quantum spatial search algorithm  with a projective oracle}. In particular, as expected, the minimum time  necessary to reach a threshold  target fidelity is lower in the unbounded case, while the continuous bounded control and the digital feedback strategies achieve similar results. Furthermore, for all considered strategies, we show that once the target vertex is reached, the feedback operates to keep the walker in this position. This is an important difference with respect to standard spatial search protocols \cite{childs04,portugal2013quantum}, where the target is found, with higher probability, at a specific time or in a very narrow time window. The implications are relevant, especially at the experimental and operational level, as in our protocol one does not need to perform the final position measurement at a specific time, but rather at any time larger than the known threshold.
\par 
Different physical realizations of quantum walks have been proposed in recent years. Among others, photonic realizations in integrated optical waveguide \cite{PhysRevLett.100.170506,peruzzo2010quantum,grafe2016integrated, jiao2021two}, single optically trapped atoms in a one-dimensional optical lattice \cite{karski2009quantum,genske2013electric}, or even with trapped ions \cite{PhysRevLett.104.100503,PhysRevLett.103.090504,PhysRevLett.103.183602}. Concerning possible implementations of our scheme, we specifically mention cold-atom platforms \cite{schneider2012fermionic,preiss2015strongly,PhysRevA.93.051602}. Recently it has been demonstrated how one can also achieve  rapid reconfigurability of the network parameters by combination with optical tweezers \cite{doi:10.1126/science.abo0608}. Moreover promising steps towards continuous monitoring of observables in this framework have been put forward \cite{PhysRevA.95.043843}. Instead of using an homodyne detection scheme, an alternative monitoring approach could also be modeled on the spontaneous emission observed in a Bose-Einstein condensate \cite{clark2021quantum}.
Although in this paper we focus on the cycle graph, our scheme can be, in principle, generalized to more general topologies with appropriate adaptations both in the feedback operations and system dynamics, i.e. respectively by changing the feedback Hamiltonian and the system-environment coupling.

\acknowledgments
The authors acknowledge useful discussion with Francesco Albarelli. MGAP is member of INdAM-GNFM.
\appendix

\section{The standard search algorithm in a cycle graph}
\label{app:standard-algorithm-cycle}
\begin{figure}
    \centering
    \includegraphics[width=0.48\textwidth]{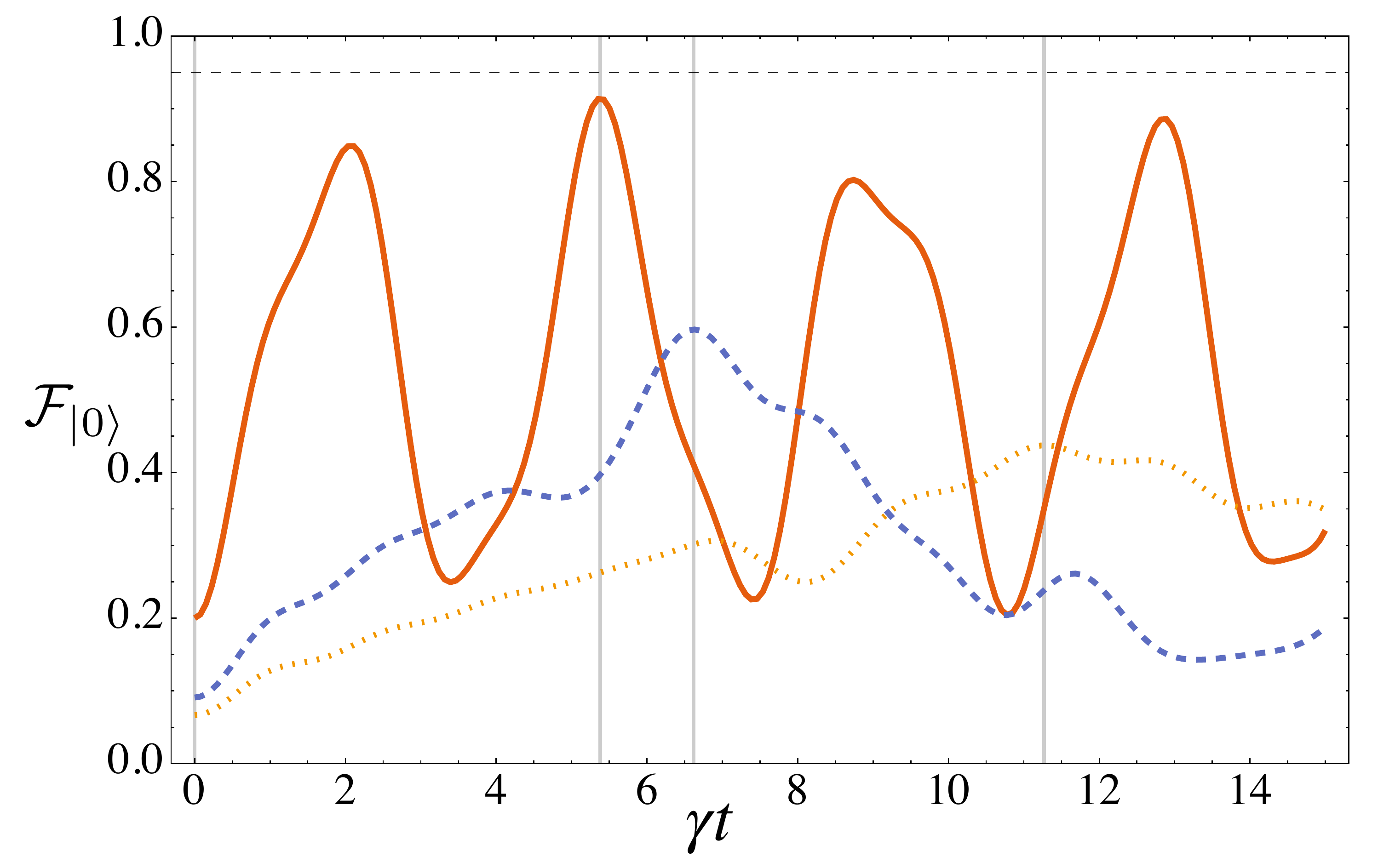}
    \caption{\AC{{Reward function $\mathcal{F}_{\vert 0 \rangle}$ for the optimized standard algorithm with the projective oracle for $N=5$ (red line), $N=11$ (blue dashed line) and $N=15$ (orange dotted line)}. For each size $N$, we found the optimal value for the $\beta$ parameter that gives the maximum value of  $\mathcal{F}_{\vert 0 \rangle}$ in the considered time interval. We obtain the following: for $N=5$, $\gamma t_{opt}=5.38,\beta_{opt}/\gamma=2.24$ and the maximum is $\mathcal{F}_{\vert 0\rangle}(t_{opt},\beta_{opt})=0.91$; for $N=11$, $\gamma t_{opt}=6.23,\beta_{opt}/\gamma=1.19$ and the maximum is $\mathcal{F}_{\vert 0\rangle}(t_{opt},\beta_{opt})=0.60$; for $N=15$, $\gamma t_{opt}=11.28,\beta_{opt}/\gamma=0.92$ and the maximum is $\mathcal{F}_{\vert 0\rangle}(t_{opt},\beta_{opt})=0.44$.}}
    \label{fig:child-algo}
\end{figure}

\CB{
The standard approach of quantum spatial search, using quantum walks, relies on the Hamiltonian in Eq. \eqref{stardardSearch}, which can be rewritten as:
\begin{equation}
    H_S=\gamma L -\beta \ketbra{w}{w},
    \end{equation}
where $\gamma$ scales the time (or alternatively we can set $\gamma=1$) and $\beta/\gamma$ is the oracle parameter that we need to optimize to improve the search. In this way, we can compare the different search strategies, by showing the dynamical quantities of interest in terms of the re-scaled time $\gamma t$.
If we apply this algorithm to one-dimensional lattices, such as the cycle graph, the success probability of finding the target does not scale well with the size of the graph. In figure \ref{fig:child-algo}, we show how the spatial search algorithm performs in the case of a cycle graph. We compare three different sizes: $N=5,11,15$. For each graph, we numerically optimize the oracle parameter and then compute the
probability of finding the target, also called reward function and defined in Eq. \eqref{eq:rewardfun}, as a function of time $\gamma t$. 
We  see that for already $N=15$ the maximum probability of finding the target is less than 0.5 in the time interval considered. 
}

\section{Unconditional master equation and the monitoring operators $\hat{c}_j$}
\label{app:uncondME}

\begin{figure}
    \centering
    \includegraphics[width=0.48\textwidth]{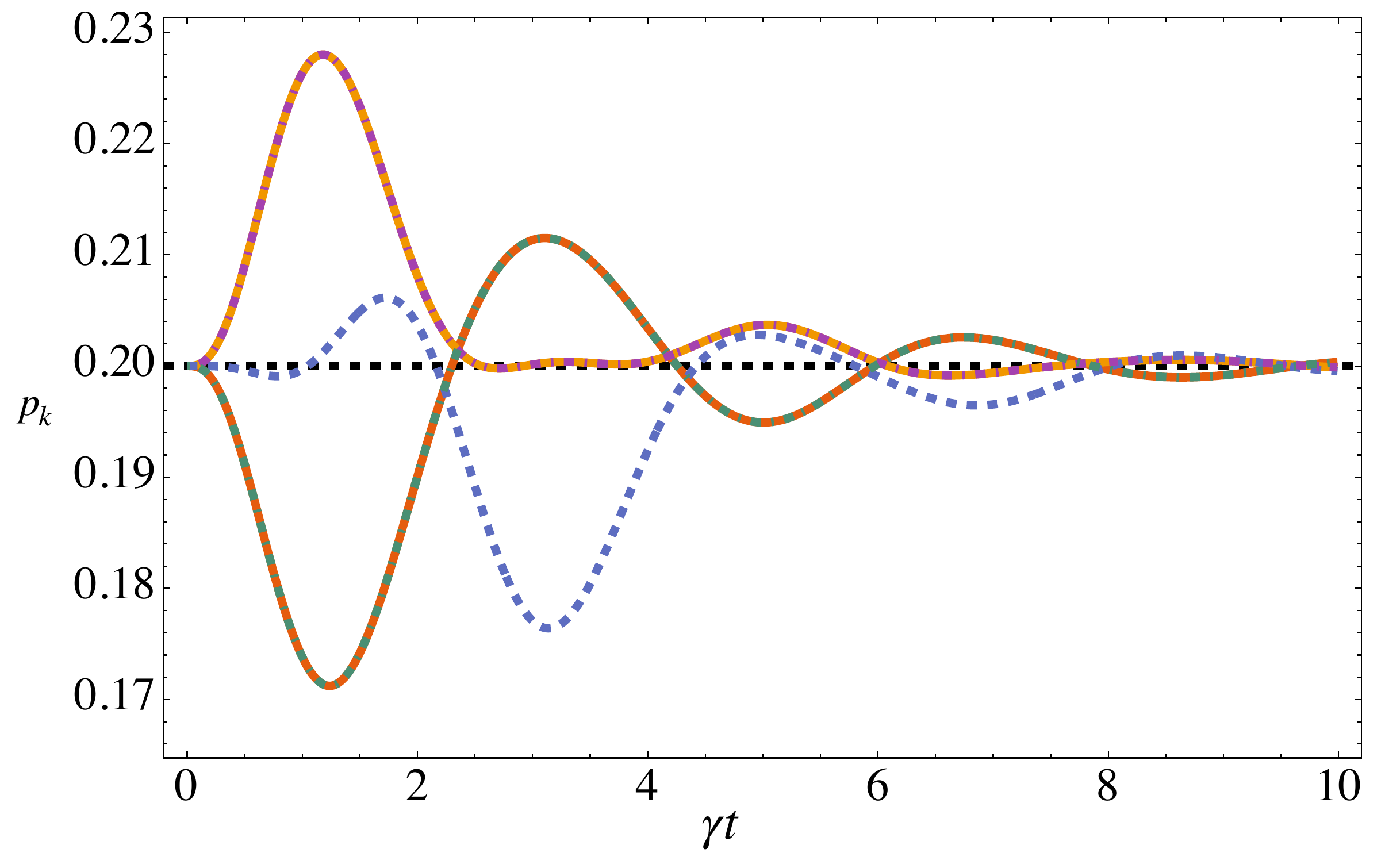}
    \caption{\AC{Time-evolution of diagonal elements of the unconditioned density matrix $p_k=\bra{k}\rho^u(t)\ket{k}$ in the position basis for the unconditioned dynamics \eqref{eq:umeqw} in a cycle graph with $N=5$ and $\kappa=\gamma$. {Pair of jump operators $\hat{c}_1$ and $\hat{c}_2$ in Eqs. \eqref{eq:jumpoperators1} and \eqref{eq:jumpoperators2} or single non-Hermitian jump operator $\hat{c}_0$ in Eq. \eqref{eq:jumpopcomplex}: black dashed line $p_0=p_1=p_2=p_3=p_4=1/5$. Single jump operator $\hat{c}_K$ in Eq. \eqref{eq:singlemonitoring}: green dot dashed line: $p_0(t)$; red line $p_4(t)$; purple dot dashed line: $p_1(t)$; orange line $p_3(t)$; blue dashed line: $p_2(t)$. The initial state is the uniform superposition of all nodes of the graph as in Eq. \eqref{psi0}}.}}
    \label{fig:app-uncond}
\end{figure}

In this appendix, we provide some details regarding the choice of the jump operators in Eq.s \eqref{eq:jumpoperators1}\eqref{eq:jumpoperators2}. We recall that the evolution of an unconditional state is described by the following master equation 
\begin{align}
  d\varrho^u &= -i \gamma [{L},\varrho^{u}(t)]dt +\kappa \sum_j\mathcal{D}[\hat{c}_j] \varrho^{u}(t) dt     \,,
      \label{eq:umeqw}
\end{align}
where $\{\hat{c}_j\}$ is the set of jump operators describing the coupling of the system's degrees of freedom with the surrounding environment, and ${L}$ denotes the Laplacian operator characterizing the quantum walk  defined in Sec. \ref{qwalks}. 
Since the cycle graph is symmetric under translations of the node's index and all nodes are equivalent, we expect an unconditioned dynamics to reflect this invariance. We will \AC{show} how this requirement sets some constraints in the choice of $\hat{c}_j$.
Since our goal is to monitor the position of the walker, one may consider any operator diagonal in the position basis $\{\vert k\rangle\}$, such as for example
\begin{equation}
    \hat{c}_K = \hat{K}=\sum_{k=0}^{N-1}k \vert k \rangle \langle k \vert \,.
    \label{eq:singlemonitoring}
\end{equation}
In this case one has that physically each node couples with the bosonic operator of the external field and this coupling is proportional to the index of the node itself. The unconditioned dynamics in \eqref{eq:umeqw} for an initially equally superposed state given in Eq. \eqref{psi0} would eventually lead to a maximally mixed state at long times, as one expects. However, as we show in Fig. \ref{fig:app-uncond}, during the time evolution, one observes that the symmetry of the graph is lost, as the different probabilities $p_k(t) = \langle k |\varrho^{u}(t)|k\rangle$ have different behaviours  in time. The reason behind the broken symmetry is that one has to fix the node having eigenvalue $k=0$.
Indeed, with a single \AC{real} jump operator diagonal in the position basis, it is not possible to have an unconditioned dynamics that preserve the cycle symmetry. \AC{There are two possible ways to circumvent this problem: the first is to consider a non-Hermitian jump operator}
\begin{equation}
    \hat{c}_0 = \sum_{k=0}^{N-1}e^{i2\pi k/N} \vert k \rangle \langle k \vert.
    \label{eq:jumpopcomplex}
\end{equation}
\AC{The second way is to use two jump operators, each diagonal in the position basis, like the one given in Eqs. \eqref{eq:jumpoperators1}--\eqref{eq:jumpoperators2}}. As remarked in the main text,  the eigenvalues of these operators correspond to the coordinates of the nodes in the $(x,y)$ plane.

\par
\AC{We now discuss the evolution corresponding to the unconditional dynamics for the three choices of jump operators. In Fig. \ref{fig:app-uncond} we report the probabilities $p_k(t) = \langle k \vert \varrho^u(t) \vert k \rangle$ of the diagonal element of the density matrix in the position basis under the master equation \eqref{eq:umeqw}. We see that, both with the non-Hermitian jump operator \eqref{eq:jumpopcomplex} or with the pair of jump operators \eqref{eq:jumpoperators1} -- \eqref{eq:jumpoperators2}, the probabilities $p_k(t)$ are constant in time, and thus describe a proper pure dephasing evolution, keeping the nodes populations constant and preserving the node symmetry, which instead is lost with the single jump operator in \eqref{eq:singlemonitoring}.}
\par
\begin{figure}
    \centering
    \includegraphics[width=0.48\textwidth]{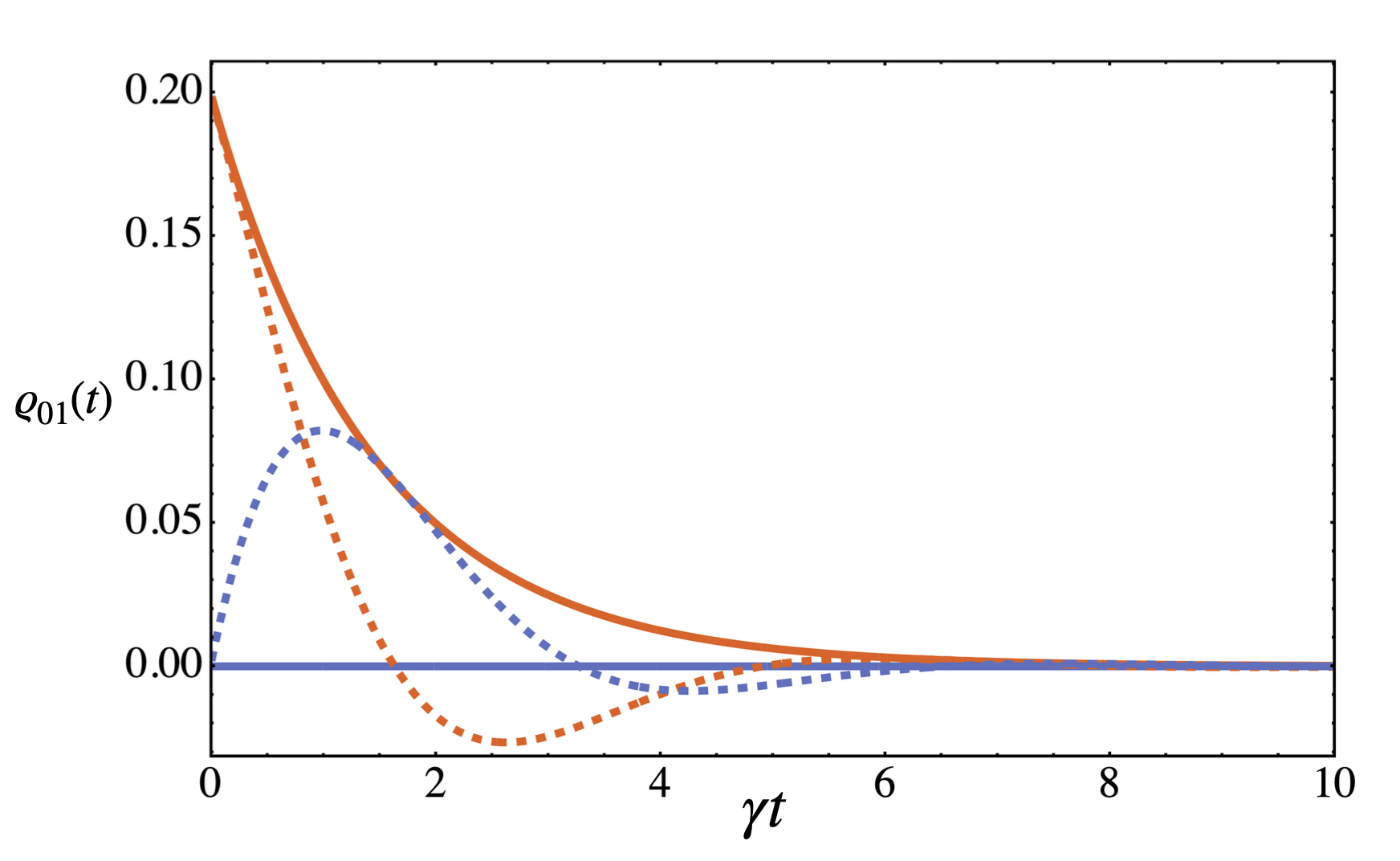}
    \caption{\AC{Real (red lines) and imaginary (blue lines) parts of the off-diagonal element $\varrho_{01}(t)$ for the unconditioned dynamics \eqref{eq:umeqw} in a cycle graph with $N=5$ and $\kappa=\gamma$ for the single jump operator $\hat{c}_0$ (dashed lines) and for the pair of jump operators ($\hat{c}_1$,$\hat{c}_2$) (solid lines). The initial state is prepared in the uniform superposition of all nodes of the graph as in Eq. \eqref{psi0}.}}
    \label{fig:app-uncond-offdiag}
\end{figure}

\AC{One can also show that the two unconditioned dynamics that preserve this symmetry are not equivalent. This can be seen by looking at the off diagonal elements $\varrho_{ij}(t)=\langle i \vert \varrho^u(t) \vert j\rangle$. In fact we first observe that the absolute values of these off-diagonal elements have an identical behaviour as a function of time for the same choice of the coupling constant $\kappa$, leading to the same mixed steady-state diagonal in the position basis. However a different behaviour is observed if we focus on the imaginary and real parts of these quantities. Just as an example, in Fig.  \ref{fig:app-uncond-offdiag} we report their evolution for the element $\varrho_{01}(t)$. While for the pair of jump operators $(\hat{c}_1,\hat{c}_2)$ the imaginary part is always equal to zero, and the real part decreases exponentially to zero, for the single jump operator $\hat{c}_0$, one observes damped oscillations for both quantities.}
\par
\AC{
In the main text we have focused on the evolution due to the pair of jump operators $\hat{c}_1$ and $\hat{c}_2$, as their eigenvalues directly correspond to the coordinates of the walker position. In this sense one could think to be able to couple the walker to two independent environments via quantum non demolition-like interactions, in order to perform continuous monitoring of these observables. In this sense this choice is the one that, in our opinion, better fits the description in terms of continuous monitoring. However in the next Appendix we  show that similar results are obtained  by considering the dynamics with a single jump operator $\hat{c}_0$ and with continuous heterodyne detection.}

\section{Results for the single complex jump operator $\hat{c}_0$}
\label{app:complexjumpresult}
\AC{In this appendix, we report the results of the bounded feedback protocol with the non-Hermitian jump operator $\hat{c}_0$ given in Eq. \eqref{eq:jumpopcomplex}, analysing the reward function \eqref{eq:rewardfun} for a single case of study, analogously to Fig. \ref{fig:multipleFeedbackBound}.}
\par
\AC{In this scenario, we can  still obtain two distinct photocurrents yielding information on the position operators $(\hat{x},\hat{y})$ by performing heterodyne detection instead of homodyne. In fact, in this case, one has two photocurrents given by \cite{WisemanMilburn,GeneralDino}
\begin{align}
    d y_t^{(1)} & = \sqrt{\frac{\eta \kappa}{2}} \text{Tr}[(\hat{c}_0+\hat{c}_0^\dagger)\varrho^c]dt + dW^{(1)}_t \nonumber \\
    &= \sqrt{2 \eta \kappa} \text{Tr}[\hat{x}\varrho^c]dt + dW^{(1)}_t \\
    d y_t^{(2)} & = \sqrt{\frac{\eta \kappa}{2}} \text{Tr}[(-i\hat{c}_0+i\hat{c}_0^\dagger)\varrho^c]dt + dW^{(2)}_t \nonumber 
\\
&= \sqrt{2\eta \kappa} \text{Tr}[\hat{y}\varrho^c]dt + dW^{(2)}_t
\end{align}}
\AC{where $dW^{(1)}_t$ and $dW^{(2)}_t$ denotes two independent Wiener increments. As explained in Sec. \eqref{sec:qcontrol}, the evolution of the conditional state $\varrho^c(t+dt)$, described by Eq. \eqref{eq:evcond}, is determined by a single Kraus operator, that for an heterodyne detection can be written as}
\begin{equation}
    \hat{M}_{{\bf dy}_t} = \mathbb{I}-\frac{\kappa}{2}\hat{c}_0^\dagger \hat{c}_0 dt + \sqrt{\frac{\eta \kappa}{2}}\hat{c}_0(d y_t^{(1)}-idy_t^{(2)}).
\end{equation}
\AC{By looking at the two photocurrents, we can use the same algorithm we used in the main text to drive the walker to the target node.} 
\par
\begin{figure}[!th]
\centering
\includegraphics[width=0.48\textwidth]{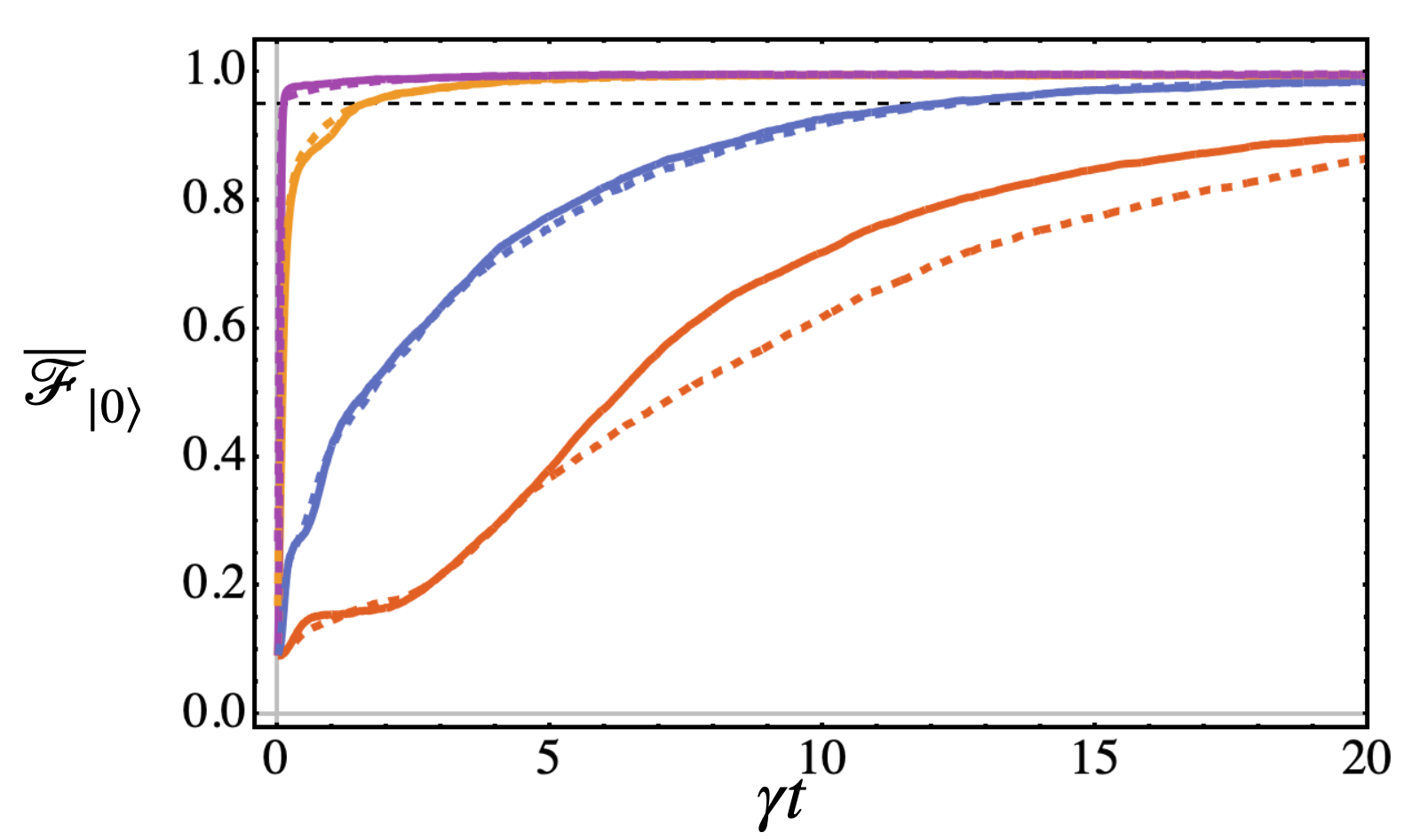}
\caption{\AC{Comparison between the average reward function $\overline{\mathcal{F}}_{\vert 0 \rangle}$ for multiple feedback Hamiltonians $\hat{h}_k^{(hop)}$ with the single non-Hermitian jump operator $\hat{c}_0$ (solid) and the protocol with two jump operator (thick) already reported in Fig. \ref{fig:multipleFeedbackBound}. We consider bounded control parameters $\boldsymbol\theta$ and a graph with $N=11$ nodes. Black dashed line: threshold $\mathcal{F}^{th} = 0.95)$. We used  $\xi=1$ (red line),  $\xi=5$ (blue line)  $\xi=50$ (orange line),  $\xi=100$ (purple line), $\kappa=\gamma$, $\eta=1$, $\gamma dt=0.01$, number of stochastic trajectories $N_{tj}=5000$.}}
\label{fig:fidsingjumpcom}
\end{figure}

\AC{In Fig. \ref{fig:fidsingjumpcom} we report the behavior of the average reward function for the multiple feedback Hamiltonians protocol described in Sec. \ref{sec:qtargeting} and with bounded domains. The results reported are comparable with the one obtained in Fig. \ref{fig:multipleFeedbackBound}, meaning that the two protocols have similar performances. In particular we observe that the curves at fixed $\xi$ are almost coincident, except for the case of $\xi=1$. We may thus conjecture that in general the two strategies corresponding to the different jump operators  lead to the similar results, but for small $\xi$ and for the jump operator $\hat{c}_0$ the algorithm implemented does not find the optimal feedback strategy, leading to smaller values of the average fidelity.}

\section{Shape of the landscape of the reward function $\mathcal{F}_{\vert 0\rangle}$}
\label{app:landscape}

\AC{In this section, we discuss some details regarding the landscape of the reward function $\mathcal{F}_{\vert 0 \rangle}$ for the protocol used in the main text, i.e. with the pair of jump operators given in Eqs. \eqref{eq:jumpoperators1}--\eqref{eq:jumpoperators2}. The aim is to provide an heuristic argument for the large  fluctuations we observe in the average feedback couplings in the unbounded domain, (see Fig. \ref{fig:multFeedUnbound}).}
\par
\AC{The landscape is, by definition, a function of the domain of the reward function (which is $\mathbb{R}^N$, with $N$ is the size of the cycle graph, i.e. the number of feedback couplings); for this reason it is not possible to plot the full landscape even for the smallest case we considered, i.e. $N=5$. Nonetheless, driven by the numerical results obtained in the main text and by symmetry considerations, we can restrict the domain in our analysis. Considering the notation of Fig. \ref{fig:cycleembedded}, we can assume that $\theta_0=\theta_4$ and $\theta_1=\theta_3$, as confirmed by the results shown in Fig. \ref{fig:multFeedUnbound}, where we can clearly see this symmetry, up to some fluctuations. We further assume that $\theta_2=0$, which is supported by numerical results in both the unbounded and bounded case, even though we did not reported them explicitly in the main text. In this way, we can now picture the landscape as a 3D function with two free feedback couplings, $\theta_0$ and $\theta_1$.}
\par
\begin{figure*}[!th]
\centering
\includegraphics[width=0.98\textwidth]{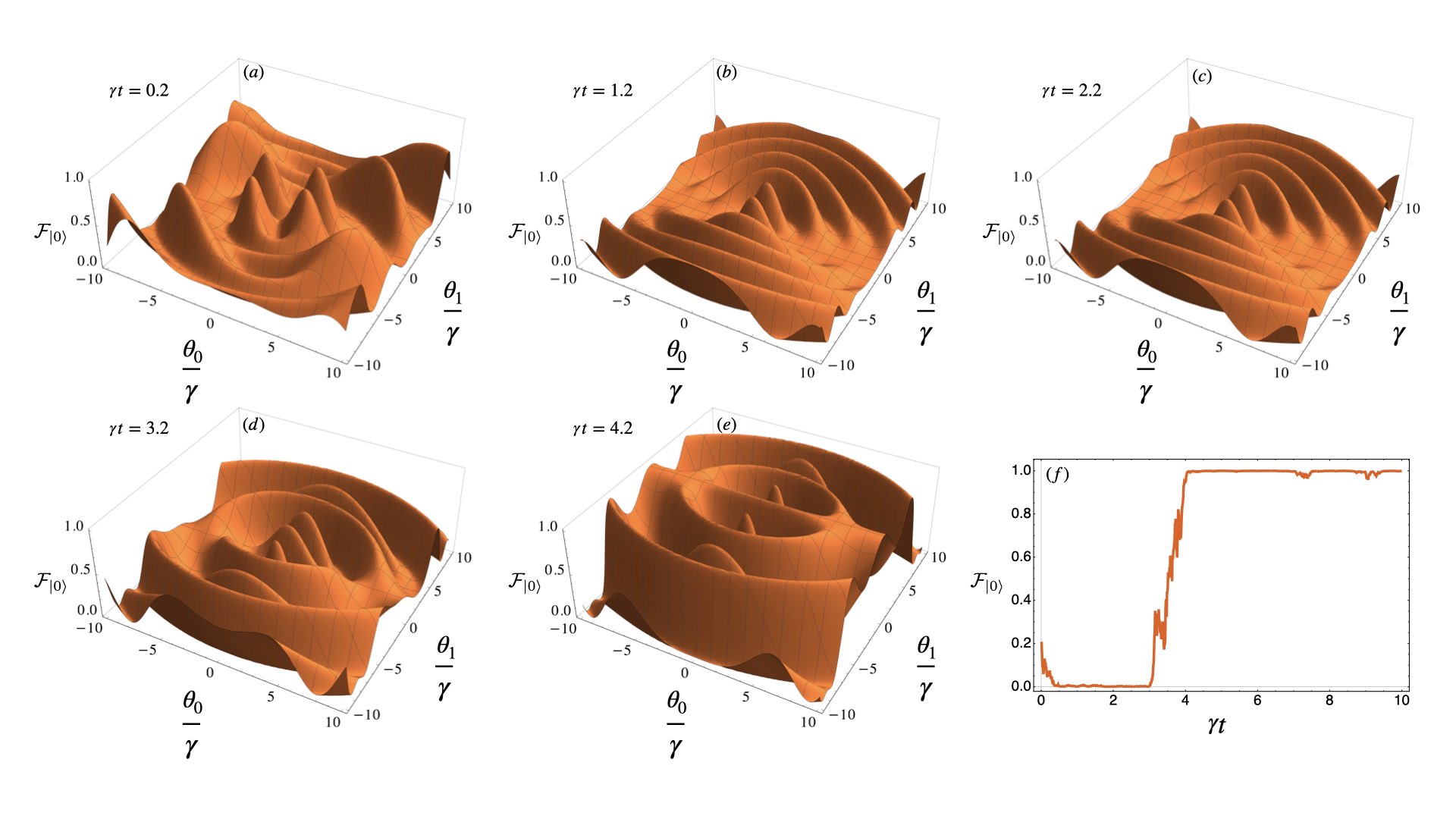}
\caption{\AC{Landscapes for the reward function $\mathcal{F}_{\vert 0 \rangle}$ in a single trajectory at different time step. Panels $(a)-(e)$ are the landscapes, panel $(f)$ is the evolution of the reward function as a function of $\gamma t$. Comparing the two, we see the changes in the landscape as soon as the reward function increase. The parameters considered are $N=5$, $\xi=1$, $\kappa=\gamma$, $\gamma dt=0.01$, $\eta=1$. Please notice that the algorithm see only a smaller square of the values reported, as explained in Appendix \ref{app:landscape}.}}
\label{fig:landscape}
\end{figure*}

\AC{The landscapes for a single trajectory and at different times  are reported in Fig. \ref{fig:landscape}, with $N=5$ and $\xi=1$, which is the smallest domain we have studied. Please, notice that the landscape in the figure is plotted for a larger domain, i.e. $\{\theta_{0}/\gamma,\theta_{1}/\gamma\} \in \{[-10,10],[-10,10]\}$. This means that the bounded algorithm is going to pick values in a smaller square centered in the $3D$ plots we have reported. We show this larger domain to illustrate why smaller domains (i.e. smaller $\xi$) leads to a slower increase of the average fidelity: one indeed observes that the maxima of the reward function are not accessible if $\xi$ is smaller than a certain threshold. We also stress that the algorithm that drives the walker does not assume that the angles are equal, i.e. $\theta_0=\theta_4$ and $\theta_1=\theta_3$ and $\theta_2=0$.}
\par
\AC{The landscapes reported in Fig. \ref{fig:landscape} show an oscillatory shape without a single global maximum but with many local maxima with the same height. This means that when the domain is enlarged (or even unbounded), at each step the algorithm might found a maximum outside the neighbourhood of the maximum found at the previous step. 
In the case of an unbounded domain, the periodicity of the landscape allows the algorithm to find maxima at very large  values of the feedback coupling, which gives rise to the large fluctuations  in the average feedback couplings in Fig. \ref{fig:multFeedUnbound}. To circumvent the problem, one could consider an algorithm in which, at each step, the feedback couplings are allowed to change in a small neighbourhood around the optimal value found at the previous step. We leave this to future investigations but we believe that this protocol should not substantially change the performance of the one used but should solve the large fluctuations problem we have observed.}

\section{Unitary Feedback: analytic expression of the feedback couplings $\boldsymbol{\theta}$}
\label{app:singlefeedH}

In this appendix we provide a detailed and analytical derivation for the expression of the feedback couplings in the case of a single feedback Hamiltonian, assuming that they are of the order of $dt$ and $dW_i$, by following the results presented in \cite{martin2020quantum}. In the interaction picture, the conditioned evolution for the density matrix due to $d_M$ measurements is given by Eq. \eqref{eq:sme}, and can be recasted as
\begin{align}
    \varrho^c(t+dt) & s= \varrho^c(t) + \sqrt{\kappa} \sum_{i=1}^{d_M}\mathcal{H}[\hat{c}_i]\varrho^c(t) dW_i + \nonumber \\
    & \quad + \sum_{i=1}^{d_{M}}\kappa \mathcal{D}[\hat{c}_i]\varrho^c(t) dt,
\end{align}
where the superoperator $\mathcal{D}[\hat{c}_i]\bullet$  and $ \mathcal{H}[\hat{c}_i]\bullet$ are defined in Eq. \eqref{eq:superop1} and \eqref{eq:superop2} respectively. 
\par
As explained in the Sec. \ref{sec:qtargeting}, our feedback Hamiltonian is identified with the adjacency matrix \eqref{eq:adjmat}, i.e.
\begin{equation}
\label{eq:singFeedHami}
    \hat{H}_{fb}(\theta) = \theta \sum_{i=0}^{N-1}\hat{h}^{(hop)}_i,
\end{equation}
with $\hat{h}^{(hop)}_i$ defined in \eqref{eq:hopping}. The single-feedback Hamiltonian protocol proposed here is nothing but the multiple-feedback protocol in which the single couplings change synchronously and with equal strength. Here we assume that the expression for the feedback coupling $\theta$ for the step $t+dt$ can be expanded as
\begin{figure*}[!th]
\centering
\includegraphics[width=\textwidth]{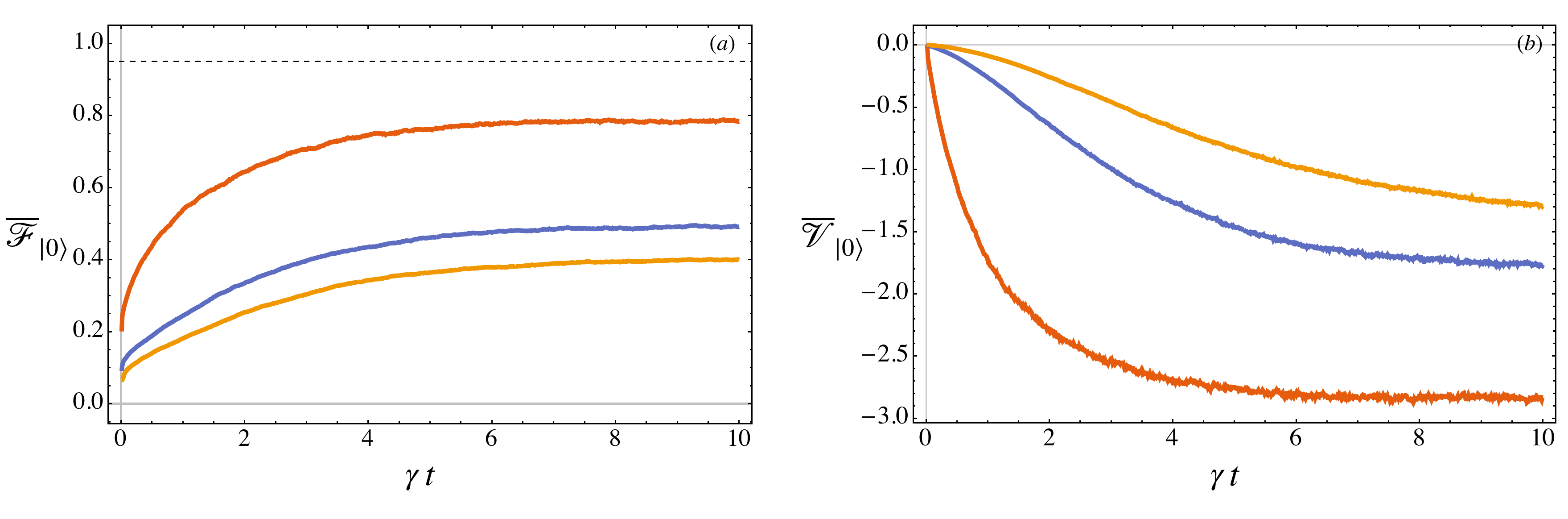}
\caption{Single-feedback Hamiltonian protocol with $\hat{H}_{F}$ defined in Eq. \eqref{eq:singFeedHami} for $N_{tj}=5000$ trajectories and parameters $\eta=\kappa=1$ and $dt=0.01$. $(a)$: average reward function $\bar{\mathcal{F}}_{\vert 0\rangle}$ with respect to the target state as a function of time $t$. Dashed line: threshold fidelity $\mathcal{F}_{th}$. $(b)$: average of the second derivative $\bar{\mathcal{V}}$ defined in Eq. \eqref{eq:secDersingFeed} as a function of time $t$. Red line $N=5$; Blue line $N=11$; Orange line $N=15$.}
\label{fig:singleFeed}
\end{figure*}
\begin{equation}
\label{eq:infanglSingFeed}
   \tilde\theta= \theta \,dt = \sum_{k=1}^{2}A_k dW_t^{(k)} + Bdt\,,
\end{equation}
 with $\boldsymbol{A} =\{A_1,A_2\}$ and $B$ are respectively a $2$-dimensional real vector and a real number. Then, the unitary evolution due to the feedback can be written as
\begin{align}
\hat{U} & = \exp\left\{-i \hat{H}_{fb}(\theta) \,dt \right\} = \\
 & = \exp\left\{-i \, \hat{h}_{fb} \sum_{k=1}^{2}A_{k} dW_{t}^{(k)}-i \hat{h}_{fb}\,B dt \right\} \,,
\end{align}
where we have introduced the operator $\hat{h}_{fb} =\sum_{i=0}^{N-1}\hat{h}^{(hop)}_i$. 
The Taylor expansion of such operator up to first order in $dt$ is obtained 
\begin{align}
\hat{U} & = \mathbb{I} - i\, \hat{h}_{fb} \sum_{k=1}^{2}A_{k} dW_{t}^{(k)} -i \hat{h}_{fb} \, B dt + \\
& \quad - \frac{1}{2} \hat{h}_{fb}^2 \sum_{k=1}^{2} A_{i}^{2} dt.
\end{align}
From the latter expression, we can derive the infinitesimal evolution after the feedback (remember that $dW_{t}^{(i)}dW_{t}^{(j)} = dt \delta_{ij}$)
\begin{widetext}
\begin{align}
    \varrho^f_{{\tilde\theta}}(t+dt) = \hat{U} \varrho^c(t+dt) \hat{U}^\dagger & = \varrho^f_{{\tilde\theta}}(t) + \sqrt{\kappa} \sum_{i=1}^{2}  \mathcal{H}[\hat{c}_i]\varrho^f_{{\tilde\theta}}(t) dW_t^{(i)} + \kappa \sum_{i=1}^{2}\mathcal{D}[\hat{c}_i]\varrho^f_{{\tilde\theta}}(t) dt - i \, [\hat{h}_{fb},\varrho^f_{{\tilde\theta}}(t)]\sum_{i=1}^{2} A_{i} dW_t^{(i)} + \\
    & \quad -i \sqrt{\kappa} [\hat{h}_{fb},\mathcal{H}[\hat{c}_i]\varrho^f_{{\tilde\theta}}(t)] \sum_{i=1}^{2}A_{i} dt -i\, [\hat{h}_{fb},\varrho^f_{{\tilde\theta}}(t)] B dt + \sum_{i=1}^{2} A_{i}^{2}\mathcal{D}[\hat{h}_{fb}]\varrho^f_{{\tilde\theta}}(t) dt = \\
    & = \varrho^f_{{\tilde\theta}}(t) + \sum_{i=1}^{2} \hat{\mathcal{W}}_i \,dW_t^{(i)} + \left(\sum_{i=1}^{2} \hat{\mathcal{T}}_i - i [\hat{h}_{fb},\varrho^f_{{\tilde\theta}}(t)]\right) B dt,
\end{align}
\end{widetext}
where we have grouped the differential factors together, i.e.
\begin{align}
\hat{\mathcal{W}}_{i} & = \sqrt{\kappa} \mathcal{H}[\hat{c}_{i}]\varrho^f_{{\tilde\theta}}(t) -i A_{i}[\hat{h}_{fb},\varrho^f_{{\tilde\theta}}(t)] \\
\hat{\mathcal{T}}_{i} & = \kappa \mathcal{D}[\hat{c}_{i}]\varrho^f_{{\tilde\theta}}(t) - i \sqrt{\kappa} A_{i} [\hat{h}_{fb},\mathcal{H}[\hat{c}_{i}]\varrho^f_{{\tilde\theta}}(t)] + \nonumber \\
& \quad + A_{i}^{2} \mathcal{D}[\hat{h}_{fb}]\varrho^f_{{\tilde\theta}}(t)
\end{align}
To obtain the value of $\boldsymbol{A}$ and $B$ which determines the feedback operation at each time-step, we require that the derivative of the linear reward function $\Lambda(\varrho(t))$ with respect to $\tilde\theta$ at the following time step of the evolution
\begin{align}
    \mathcal{G}(t+dt) & = \frac{\partial}{\partial \tilde\theta} \left(\Lambda\left(\varrho^f_{{\tilde\theta}}(t+dt)\right)\right)\bigg\vert_{{\tilde\theta} =  \tilde\theta_{opt}} =\\
    &=\Lambda \left(\frac{\partial}{\partial \tilde\theta} \varrho^f_{{\tilde\theta}}(t+dt)\bigg\vert_{{\tilde\theta}={\tilde\theta}_{opt}} \right)
\end{align} 
satisfy the extremality condition
\begin{equation}
\label{eq:condext}
	\mathcal{G}(t+dt) = 0.
\end{equation}
In addition, since we are interested in maximizing $\Lambda(\bullet)$, we ask also that the second derivative of the reward function is negative, i.e.
\begin{equation}
\label{eq:secDersingFeed}
	\mathcal{V} = \Lambda\left(\frac{\partial^{2}\varrho^{f}_{\tilde\theta}(t+dt)}{\partial \tilde\theta^{2}}\bigg\vert_{\tilde\theta = \tilde\theta_{opt}}\right) < 0 
\end{equation}
which ensures that the feedback operation maximize the reward function. To find the solution, we first evaluate 
\begin{align}
    \partial_{\tilde\theta} \varrho^f_{{\tilde\theta}}(t+dt) = i [\hat{H}_{fb}(\theta),\varrho^f_{{\tilde\theta}}(t+dt)].
\end{align}
Then, the condition \eqref{eq:condext} can be expanded as follows
\begin{widetext}
\begin{align}
	\mathcal{G}(t+dt) & = -i \Lambda([\hat{h}_{fb},\varrho^f_{{\tilde\theta}}(t)]) - i \sum_{j=1}^{2}\Lambda([\hat{h}_{fb},\hat{\mathcal{W}}_{j}]) dW_{t}^{(j)} - i \left(\sum_{j=1}^{2}\Lambda([\hat{h}_{fb},\hat{\mathcal{T}}_{j}])-i B\Lambda([\hat{h}_{fb},[\hat{h}_{fb},\varrho^f_{{\tilde\theta}}(t)]])\right)dt
\end{align}
\end{widetext}
Now, the first term in the latter equation is null since we assume that at the previous time step the reward function satisfy the extremality condition. Then, considering the terms proportional to $dW_{t}^{j}$ we have that for $j=1,2$ $\Lambda([\hat{h}_{fb},\hat{\mathcal{W}}_{j}]) = 0$, which is nothing but
\begin{equation}
\label{eq:AsingleFeed}
	A_{j} = -i \sqrt{\kappa} \frac{\langle 0 \vert [\hat{h}_{fb},\mathcal{H}[\hat{c}_{j}]\varrho^f_{{\tilde\theta}}(t)]\vert 0 \rangle}{\langle 0 \vert [\hat{h}_{fb},[\hat{h}_{fb},\varrho^f_{{\tilde\theta}}(t)]]\vert 0\rangle} 
\end{equation} 
where we have considered as reward function the one defined in \eqref{eq:rewardfun}, i.e. $\Lambda(\bullet) = \langle 0 \vert \bullet \vert 0 \rangle$. With the same line of reasoning, taking the term proportional to $dt$ we can obtain the scalar function
\begin{align}
\label{eq:BsingleFeed}
B = -i \frac{\sum_{j=1}^{2}\langle 0\vert [\hat{h}_{fb},\hat{\mathcal{T}}_{j}] \vert 0 \rangle}{\langle 0 \vert [\hat{h}_{fb},[\hat{h}_{fb},\varrho^f_{{\tilde\theta}}(t)]]\vert 0 \rangle},
\end{align}
We notice that these equations are valid if $\langle 0 \vert [\hat{h}_{fb},[\hat{h}_{fb},\varrho^f_{{\tilde\theta}}(t)]]\vert 0\rangle\neq 0$, a condition that must be checked at each step of the feedback operation. 
\par
In addition, the condition for maximizing the reward function at each time step certain time can be simply written as
\begin{equation}
\mathcal{V}_{\vert 0 \rangle}(t+dt) = -i [\hat{h}_{fb},[\hat{h}_{fb}, \varrho^{f}_{\theta}(t+dt)]] < 0.
\end{equation}
If this condition fails, we chose not to act with the feedback operation and skip to the next time-step, even though numerical evidence shows that this situation rarely occurs.
\par
The numerical results of this protocol are reported in Fig. \ref{fig:singleFeed}, left panel. The average fidelity $\overline{\mathcal{F}}_{\vert 0 \rangle}$ for $N_{tj} =5000$ trajectories for three different values of $N=5,11,15$. As the size increases, the efficiency of the protocol worsens. Moreover, it never reaches the threshold value $\overline{\mathcal{F}}^{th}_{\vert 0 \rangle}$. 
\par
In the right panel of \ref{fig:singleFeed} we report the average value of the second derivative, i.e. $\overline{\mathcal{V}}_{\vert 0 \rangle} = \mathbb{E}_{traj}[\mathcal{V}_{\vert 0 \rangle} (t+dt)]$. The results obtained show that the feedback operation is always optimal \emph{on average} at each time step. However, since the threshold value $\overline{\mathcal{F}}^{th}_{\vert 0 \rangle}$ is never reached, we conclude that the single-feedback Hamiltonian is inefficient in achieving the targeting goal, even though the $\theta$ found according to Eq. and \eqref{eq:AsingleFeed} and \eqref{eq:BsingleFeed}  to be the optimal one. In addition, the absolute values of $\mathcal{V}$ decrease as $N$ increases, showing that the efficiency of the protocol worsens as the size increases, as we have already observed in the main text for the multi-coupling feedback Hamiltonian.


\bibliography{bibliography}

\begin{thebibliography}{99}%
\makeatletter
\providecommand \@ifxundefined [1]{%
 \@ifx{#1\undefined}
}%
\providecommand \@ifnum [1]{%
 \ifnum #1\expandafter \@firstoftwo
 \else \expandafter \@secondoftwo
 \fi
}%
\providecommand \@ifx [1]{%
 \ifx #1\expandafter \@firstoftwo
 \else \expandafter \@secondoftwo
 \fi
}%
\providecommand \natexlab [1]{#1}%
\providecommand \enquote  [1]{``#1''}%
\providecommand \bibnamefont  [1]{#1}%
\providecommand \bibfnamefont [1]{#1}%
\providecommand \citenamefont [1]{#1}%
\providecommand \href@noop [0]{\@secondoftwo}%
\providecommand \href [0]{\begingroup \@sanitize@url \@href}%
\providecommand \@href[1]{\@@startlink{#1}\@@href}%
\providecommand \@@href[1]{\endgroup#1\@@endlink}%
\providecommand \@sanitize@url [0]{\catcode `\\12\catcode `\$12\catcode
  `\&12\catcode `\#12\catcode `\^12\catcode `\_12\catcode `\%12\relax}%
\providecommand \@@startlink[1]{}%
\providecommand \@@endlink[0]{}%
\providecommand \url  [0]{\begingroup\@sanitize@url \@url }%
\providecommand \@url [1]{\endgroup\@href {#1}{\urlprefix }}%
\providecommand \urlprefix  [0]{URL }%
\providecommand \Eprint [0]{\href }%
\providecommand \doibase [0]{http://dx.doi.org/}%
\providecommand \selectlanguage [0]{\@gobble}%
\providecommand \bibinfo  [0]{\@secondoftwo}%
\providecommand \bibfield  [0]{\@secondoftwo}%
\providecommand \translation [1]{[#1]}%
\providecommand \BibitemOpen [0]{}%
\providecommand \bibitemStop [0]{}%
\providecommand \bibitemNoStop [0]{.\EOS\space}%
\providecommand \EOS [0]{\spacefactor3000\relax}%
\providecommand \BibitemShut  [1]{\csname bibitem#1\endcsname}%
\let\auto@bib@innerbib\@empty
\bibitem [{\citenamefont {Farhi}\ and\ \citenamefont
  {Gutmann}(1998{\natexlab{a}})}]{farhi981}%
  \BibitemOpen
  \bibfield  {author} {\bibinfo {author} {\bibfnamefont {E.}~\bibnamefont
  {Farhi}}\ and\ \bibinfo {author} {\bibfnamefont {S.}~\bibnamefont
  {Gutmann}},\ }\bibfield  {title} {\enquote {\bibinfo {title} {Quantum
  computation and decision trees},}\ }\href {\doibase 10.1103/PhysRevA.58.915}
  {\bibfield  {journal} {\bibinfo  {journal} {Phys. Rev. A}\ }\textbf {\bibinfo
  {volume} {58}},\ \bibinfo {pages} {915--928} (\bibinfo {year}
  {1998}{\natexlab{a}})}\BibitemShut {NoStop}%
\bibitem [{\citenamefont {Kempe}(2003)}]{kempe003}%
  \BibitemOpen
  \bibfield  {author} {\bibinfo {author} {\bibfnamefont {J.}~\bibnamefont
  {Kempe}},\ }\bibfield  {title} {\enquote {\bibinfo {title} {Quantum random
  walks: An introductory overview},}\ }\href {\doibase
  10.1080/00107151031000110776} {\bibfield  {journal} {\bibinfo  {journal}
  {Contemp. Phys.}\ }\textbf {\bibinfo {volume} {44}},\ \bibinfo {pages}
  {307--327} (\bibinfo {year} {2003})}\BibitemShut {NoStop}%
\bibitem [{\citenamefont {Venegas-Andraca}(2012)}]{venegas12}%
  \BibitemOpen
  \bibfield  {author} {\bibinfo {author} {\bibfnamefont {S.~E.}\ \bibnamefont
  {Venegas-Andraca}},\ }\bibfield  {title} {\enquote {\bibinfo {title} {Quantum
  walks: a comprehensive review},}\ }\href {\doibase
  https://doi.org/10.1007/s11128-012-0432-5} {\bibfield  {journal} {\bibinfo
  {journal} {Quant. Inf. Process.}\ }\textbf {\bibinfo {volume} {11}},\
  \bibinfo {pages} {1015--1106} (\bibinfo {year} {2012})}\BibitemShut {NoStop}%
\bibitem [{\citenamefont {Childs}(2009)}]{childs09}%
  \BibitemOpen
  \bibfield  {author} {\bibinfo {author} {\bibfnamefont {A.~M.}\ \bibnamefont
  {Childs}},\ }\bibfield  {title} {\enquote {\bibinfo {title} {Universal
  computation by quantum walk},}\ }\href {\doibase
  10.1103/PhysRevLett.102.180501} {\bibfield  {journal} {\bibinfo  {journal}
  {Phys. Rev. Lett.}\ }\textbf {\bibinfo {volume} {102}},\ \bibinfo {pages}
  {180501} (\bibinfo {year} {2009})}\BibitemShut {NoStop}%
\bibitem [{\citenamefont {Kendon}(2014)}]{kendon14comp}%
  \BibitemOpen
  \bibfield  {author} {\bibinfo {author} {\bibfnamefont {V.}~\bibnamefont
  {Kendon}},\ }\bibfield  {title} {\enquote {\bibinfo {title} {Quantum walk
  computation},}\ }\href {\doibase 10.1063/1.4903129} {\bibfield  {journal}
  {\bibinfo  {journal} {AIP Conf. Proc.}\ }\textbf {\bibinfo {volume} {1633}},\
  \bibinfo {pages} {177--179} (\bibinfo {year} {2014})}\BibitemShut {NoStop}%
\bibitem [{\citenamefont {{A. P. Hines, and P. C. E. Stamp}}(2007)}]{hines07}%
  \BibitemOpen
  \bibfield  {author} {\bibinfo {author} {\bibnamefont {{A. P. Hines, and P. C.
  E. Stamp}}},\ }\bibfield  {title} {\enquote {\bibinfo {title} {Quantum walks,
  quantum gates, and quantum computers},}\ }\href {\doibase
  10.1103/PhysRevA.75.062321} {\bibfield  {journal} {\bibinfo  {journal} {Phys.
  Rev. A}\ }\textbf {\bibinfo {volume} {75}},\ \bibinfo {pages} {062321}
  (\bibinfo {year} {2007})}\BibitemShut {NoStop}%
\bibitem [{\citenamefont {{A. M. Childs, and E. Farhi, and Gutmann,
  S.}}(2002)}]{chil02}%
  \BibitemOpen
  \bibfield  {author} {\bibinfo {author} {\bibnamefont {{A. M. Childs, and E.
  Farhi, and Gutmann, S.}}},\ }\bibfield  {title} {\enquote {\bibinfo {title}
  {An example of the difference between quantum and classical random walks},}\
  }\href {\doibase 10.1023/A:1019609420309} {\bibfield  {journal} {\bibinfo
  {journal} {Quantum Inf. Proc.}\ }\textbf {\bibinfo {volume} {1}},\ \bibinfo
  {pages} {35--43} (\bibinfo {year} {2002})}\BibitemShut {NoStop}%
\bibitem [{\citenamefont {{Childs, A. M. and Cleve, R. and Deotto, E. and
  Farhi, E. and Gutmann, S. and Spielman, D. A.}}(2003)}]{childs_cleve_03}%
  \BibitemOpen
  \bibfield  {author} {\bibinfo {author} {\bibnamefont {{Childs, A. M. and
  Cleve, R. and Deotto, E. and Farhi, E. and Gutmann, S. and Spielman, D.
  A.}}},\ }\bibfield  {title} {\enquote {\bibinfo {title} {Exponential
  algorithmic speedup by a quantum walk},}\ }in\ \href {\doibase
  10.1145/780542.780552} {\emph {\bibinfo {booktitle} {Proceedings of the
  Thirty-Fifth Annual ACM Symposium on Theory of Computing}}},\ \bibinfo
  {series and number} {STOC '03}\ (\bibinfo  {publisher} {Association for
  Computing Machinery},\ \bibinfo {year} {2003})\ pp.\ \bibinfo {pages}
  {59--68}\BibitemShut {NoStop}%
\bibitem [{\citenamefont {Gamble}\ \emph {et~al.}(2010)\citenamefont {Gamble},
  \citenamefont {Friesen}, \citenamefont {Zhou}, \citenamefont {Joynt},\ and\
  \citenamefont {Coppersmith}}]{gamble10}%
  \BibitemOpen
  \bibfield  {author} {\bibinfo {author} {\bibfnamefont {J.~K.}\ \bibnamefont
  {Gamble}}, \bibinfo {author} {\bibfnamefont {M.}~\bibnamefont {Friesen}},
  \bibinfo {author} {\bibfnamefont {D.}~\bibnamefont {Zhou}}, \bibinfo {author}
  {\bibfnamefont {R.}~\bibnamefont {Joynt}}, \ and\ \bibinfo {author}
  {\bibfnamefont {S.~N.}\ \bibnamefont {Coppersmith}},\ }\bibfield  {title}
  {\enquote {\bibinfo {title} {Two-particle quantum walks applied to the graph
  isomorphism problem},}\ }\href {\doibase 10.1103/PhysRevA.81.052313}
  {\bibfield  {journal} {\bibinfo  {journal} {Phys. Rev. A}\ }\textbf {\bibinfo
  {volume} {81}},\ \bibinfo {pages} {052313} (\bibinfo {year}
  {2010})}\BibitemShut {NoStop}%
\bibitem [{\citenamefont {{Rossi, L. and Torsello, A. and Hancock, E.
  R.}}(2015)}]{rossi15}%
  \BibitemOpen
  \bibfield  {author} {\bibinfo {author} {\bibnamefont {{Rossi, L. and
  Torsello, A. and Hancock, E. R.}}},\ }\bibfield  {title} {\enquote {\bibinfo
  {title} {Measuring graph similarity through continuous-time quantum walks and
  the quantum jensen-shannon divergence},}\ }\href {\doibase
  10.1103/PhysRevE.91.022815} {\bibfield  {journal} {\bibinfo  {journal} {Phys.
  Rev. E}\ }\textbf {\bibinfo {volume} {91}},\ \bibinfo {pages} {022815}
  (\bibinfo {year} {2015})}\BibitemShut {NoStop}%
\bibitem [{\citenamefont {{Cade, C. and Montanaro, A. and Belovs,
  A.}}(2018)}]{cade18}%
  \BibitemOpen
  \bibfield  {author} {\bibinfo {author} {\bibnamefont {{Cade, C. and
  Montanaro, A. and Belovs, A.}}},\ }\bibfield  {title} {\enquote {\bibinfo
  {title} {Time and space efficient quantum algorithms for detecting cycles and
  testing bipartiteness},}\ }\href@noop {} {\bibfield  {journal} {\bibinfo
  {journal} {Quantum Info. Comput.}\ }\textbf {\bibinfo {volume} {18}},\
  \bibinfo {pages} {18--50} (\bibinfo {year} {2018})}\BibitemShut {NoStop}%
\bibitem [{\citenamefont {{A. Callison and N. Chancellor and F. Mintert and V.
  Kendon}}(2019)}]{Callison_2019}%
  \BibitemOpen
  \bibfield  {author} {\bibinfo {author} {\bibnamefont {{A. Callison and N.
  Chancellor and F. Mintert and V. Kendon}}},\ }\bibfield  {title} {\enquote
  {\bibinfo {title} {Finding spin glass ground states using quantum walks},}\
  }\href {\doibase 10.1088/1367-2630/ab5ca2} {\bibfield  {journal} {\bibinfo
  {journal} {New J. Phys.}\ }\textbf {\bibinfo {volume} {21}},\ \bibinfo
  {pages} {123022} (\bibinfo {year} {2019})}\BibitemShut {NoStop}%
\bibitem [{\citenamefont {{Marsh, S. and Wang, J. B.}}(2020)}]{marsh20}%
  \BibitemOpen
  \bibfield  {author} {\bibinfo {author} {\bibnamefont {{Marsh, S. and Wang, J.
  B.}}},\ }\bibfield  {title} {\enquote {\bibinfo {title} {Combinatorial
  optimization via highly efficient quantum walks},}\ }\href {\doibase
  10.1103/PhysRevResearch.2.023302} {\bibfield  {journal} {\bibinfo  {journal}
  {Phys. Rev. Research}\ }\textbf {\bibinfo {volume} {2}},\ \bibinfo {pages}
  {023302} (\bibinfo {year} {2020})}\BibitemShut {NoStop}%
\bibitem [{\citenamefont {{Kryukov, A. and Abramov, R. and Fedichkin, L. E. and
  Alodjants, A. and Melnikov, A. A.}}(2022)}]{kryukov22}%
  \BibitemOpen
  \bibfield  {author} {\bibinfo {author} {\bibnamefont {{Kryukov, A. and
  Abramov, R. and Fedichkin, L. E. and Alodjants, A. and Melnikov, A. A.}}},\
  }\bibfield  {title} {\enquote {\bibinfo {title} {Supervised graph
  classification for chiral quantum walks},}\ }\href {\doibase
  10.1103/PhysRevA.105.022208} {\bibfield  {journal} {\bibinfo  {journal}
  {Phys. Rev. A}\ }\textbf {\bibinfo {volume} {105}},\ \bibinfo {pages}
  {022208} (\bibinfo {year} {2022})}\BibitemShut {NoStop}%
\bibitem [{\citenamefont {Ambainis}(2003)}]{ambainis2003}%
  \BibitemOpen
  \bibfield  {author} {\bibinfo {author} {\bibfnamefont {A.}~\bibnamefont
  {Ambainis}},\ }\bibfield  {title} {\enquote {\bibinfo {title} {Quantum walks
  and their algorithmic applications},}\ }\href {\doibase
  10.1142/S0219749903000383} {\bibfield  {journal} {\bibinfo  {journal} {Int.
  J. Quantum Inf.}\ }\textbf {\bibinfo {volume} {1}},\ \bibinfo {pages}
  {507--518} (\bibinfo {year} {2003})}\BibitemShut {NoStop}%
\bibitem [{\citenamefont {Kadian}\ \emph {et~al.}(2021)\citenamefont {Kadian},
  \citenamefont {Garhwal},\ and\ \citenamefont {Kumar}}]{kadian21}%
  \BibitemOpen
  \bibfield  {author} {\bibinfo {author} {\bibfnamefont {K.}~\bibnamefont
  {Kadian}}, \bibinfo {author} {\bibfnamefont {S.}~\bibnamefont {Garhwal}}, \
  and\ \bibinfo {author} {\bibfnamefont {A.}~\bibnamefont {Kumar}},\ }\bibfield
   {title} {\enquote {\bibinfo {title} {Quantum walk and its application
  domains: A systematic review},}\ }\href {\doibase
  https://doi.org/10.1016/j.cosrev.2021.100419} {\bibfield  {journal} {\bibinfo
   {journal} {Comput. Sci. Rev.}\ }\textbf {\bibinfo {volume} {41}},\ \bibinfo
  {pages} {100419} (\bibinfo {year} {2021})}\BibitemShut {NoStop}%
\bibitem [{\citenamefont {Venegas-Andraca}(2008)}]{venegas2008quantum}%
  \BibitemOpen
  \bibfield  {author} {\bibinfo {author} {\bibfnamefont {S.~E.}\ \bibnamefont
  {Venegas-Andraca}},\ }\href@noop {} {\emph {\bibinfo {title} {Quantum Walks
  for Computer Scientists}}},\ Synthesis Lectures on Quantum Computing\
  (\bibinfo  {publisher} {Morgan \& Claypool Publishers},\ \bibinfo {address}
  {San Rafael},\ \bibinfo {year} {2008})\BibitemShut {NoStop}%
\bibitem [{\citenamefont {Childs}\ and\ \citenamefont
  {Goldstone}(2004)}]{childs04}%
  \BibitemOpen
  \bibfield  {author} {\bibinfo {author} {\bibfnamefont {A.~M.}\ \bibnamefont
  {Childs}}\ and\ \bibinfo {author} {\bibfnamefont {J.}~\bibnamefont
  {Goldstone}},\ }\bibfield  {title} {\enquote {\bibinfo {title} {Spatial
  search by quantum walk},}\ }\href {\doibase 10.1103/PhysRevA.70.022314}
  {\bibfield  {journal} {\bibinfo  {journal} {Phys. Rev. A}\ }\textbf {\bibinfo
  {volume} {70}},\ \bibinfo {pages} {022314} (\bibinfo {year}
  {2004})}\BibitemShut {NoStop}%
\bibitem [{\citenamefont {Janmark}\ \emph {et~al.}(2014)\citenamefont
  {Janmark}, \citenamefont {Meyer},\ and\ \citenamefont {Wong}}]{Janmark14}%
  \BibitemOpen
  \bibfield  {author} {\bibinfo {author} {\bibfnamefont {J.}~\bibnamefont
  {Janmark}}, \bibinfo {author} {\bibfnamefont {D.~A.}\ \bibnamefont {Meyer}},
  \ and\ \bibinfo {author} {\bibfnamefont {T.~G.}\ \bibnamefont {Wong}},\
  }\bibfield  {title} {\enquote {\bibinfo {title} {Global symmetry is
  unnecessary for fast quantum search},}\ }\href {\doibase
  10.1103/PhysRevLett.112.210502} {\bibfield  {journal} {\bibinfo  {journal}
  {Phys. Rev. Lett.}\ }\textbf {\bibinfo {volume} {112}},\ \bibinfo {pages}
  {210502} (\bibinfo {year} {2014})}\BibitemShut {NoStop}%
\bibitem [{\citenamefont {Childs}\ and\ \citenamefont {Ge}(2014)}]{childs2014}%
  \BibitemOpen
  \bibfield  {author} {\bibinfo {author} {\bibfnamefont {A.~M.}\ \bibnamefont
  {Childs}}\ and\ \bibinfo {author} {\bibfnamefont {Y.}~\bibnamefont {Ge}},\
  }\bibfield  {title} {\enquote {\bibinfo {title} {Spatial search by
  continuous-time quantum walks on crystal lattices},}\ }\href {\doibase
  10.1103/PhysRevA.89.052337} {\bibfield  {journal} {\bibinfo  {journal} {Phys.
  Rev. A}\ }\textbf {\bibinfo {volume} {89}},\ \bibinfo {pages} {052337}
  (\bibinfo {year} {2014})}\BibitemShut {NoStop}%
\bibitem [{\citenamefont {Chakraborty}\ \emph
  {et~al.}(2020{\natexlab{a}})\citenamefont {Chakraborty}, \citenamefont
  {Novo},\ and\ \citenamefont {Roland}}]{Chakraborty20_anyGraph}%
  \BibitemOpen
  \bibfield  {author} {\bibinfo {author} {\bibfnamefont {S.}~\bibnamefont
  {Chakraborty}}, \bibinfo {author} {\bibfnamefont {L.}~\bibnamefont {Novo}}, \
  and\ \bibinfo {author} {\bibfnamefont {J.}~\bibnamefont {Roland}},\
  }\bibfield  {title} {\enquote {\bibinfo {title} {Finding a marked node on any
  graph via continuous-time quantum walks},}\ }\href {\doibase
  10.1103/PhysRevA.102.022227} {\bibfield  {journal} {\bibinfo  {journal}
  {Phys. Rev. A}\ }\textbf {\bibinfo {volume} {102}},\ \bibinfo {pages}
  {022227} (\bibinfo {year} {2020}{\natexlab{a}})}\BibitemShut {NoStop}%
\bibitem [{\citenamefont {Paris}\ \emph {et~al.}(2021)\citenamefont {Paris},
  \citenamefont {Benedetti},\ and\ \citenamefont {Olivares}}]{paris-oracle-21}%
  \BibitemOpen
  \bibfield  {author} {\bibinfo {author} {\bibfnamefont {M.~G.~A.}\
  \bibnamefont {Paris}}, \bibinfo {author} {\bibfnamefont {C.}~\bibnamefont
  {Benedetti}}, \ and\ \bibinfo {author} {\bibfnamefont {S.}~\bibnamefont
  {Olivares}},\ }\bibfield  {title} {\enquote {\bibinfo {title} {Improving
  quantum search on simple graphs by pretty good structured oracles},}\ }\href
  {\doibase 10.3390/sym13010096} {\bibfield  {journal} {\bibinfo  {journal}
  {Symmetry}\ }\textbf {\bibinfo {volume} {13}} (\bibinfo {year} {2021}),\
  10.3390/sym13010096}\BibitemShut {NoStop}%
\bibitem [{\citenamefont {Benedetti}\ \emph {et~al.}(2021)\citenamefont
  {Benedetti}, \citenamefont {Tamascelli}, \citenamefont {Paris},\ and\
  \citenamefont {Crespi}}]{benedetti21}%
  \BibitemOpen
  \bibfield  {author} {\bibinfo {author} {\bibfnamefont {C.}~\bibnamefont
  {Benedetti}}, \bibinfo {author} {\bibfnamefont {D.}~\bibnamefont
  {Tamascelli}}, \bibinfo {author} {\bibfnamefont {M.~G.~A.}\ \bibnamefont
  {Paris}}, \ and\ \bibinfo {author} {\bibfnamefont {A.}~\bibnamefont
  {Crespi}},\ }\bibfield  {title} {\enquote {\bibinfo {title} {Quantum spatial
  search in two-dimensional waveguide arrays},}\ }\href {\doibase
  10.1103/PhysRevApplied.16.054036} {\bibfield  {journal} {\bibinfo  {journal}
  {Phys. Rev. Applied}\ }\textbf {\bibinfo {volume} {16}},\ \bibinfo {pages}
  {054036} (\bibinfo {year} {2021})}\BibitemShut {NoStop}%
\bibitem [{\citenamefont {Portugal}(2018)}]{portugal2013quantum}%
  \BibitemOpen
  \bibfield  {author} {\bibinfo {author} {\bibfnamefont {R.}~\bibnamefont
  {Portugal}},\ }\href@noop {} {\emph {\bibinfo {title} {Quantum walks and
  search algorithms}}}\ (\bibinfo  {publisher} {Springer Nature Switzerland},\
  \bibinfo {address} {AG},\ \bibinfo {year} {2018})\BibitemShut {NoStop}%
\bibitem [{\citenamefont {Pemberton-Ross}\ and\ \citenamefont
  {Kay}(2011)}]{alastair}%
  \BibitemOpen
  \bibfield  {author} {\bibinfo {author} {\bibfnamefont {P.~J.}\ \bibnamefont
  {Pemberton-Ross}}\ and\ \bibinfo {author} {\bibfnamefont {A.}~\bibnamefont
  {Kay}},\ }\bibfield  {title} {\enquote {\bibinfo {title} {Perfect quantum
  routing in regular spin networks},}\ }\href {\doibase
  10.1103/PhysRevLett.106.020503} {\bibfield  {journal} {\bibinfo  {journal}
  {Phys. Rev. Lett.}\ }\textbf {\bibinfo {volume} {106}},\ \bibinfo {pages}
  {020503} (\bibinfo {year} {2011})}\BibitemShut {NoStop}%
\bibitem [{\citenamefont {Chudzicki}\ and\ \citenamefont
  {Strauch}(2010)}]{Chudzicki10}%
  \BibitemOpen
  \bibfield  {author} {\bibinfo {author} {\bibfnamefont {C.}~\bibnamefont
  {Chudzicki}}\ and\ \bibinfo {author} {\bibfnamefont {F.~W.}\ \bibnamefont
  {Strauch}},\ }\bibfield  {title} {\enquote {\bibinfo {title} {Parallel state
  transfer and efficient quantum routing on quantum networks},}\ }\href
  {\doibase 10.1103/PhysRevLett.105.260501} {\bibfield  {journal} {\bibinfo
  {journal} {Phys. Rev. Lett.}\ }\textbf {\bibinfo {volume} {105}},\ \bibinfo
  {pages} {260501} (\bibinfo {year} {2010})}\BibitemShut {NoStop}%
\bibitem [{\citenamefont {Paganelli}\ \emph {et~al.}(2013)\citenamefont
  {Paganelli}, \citenamefont {Lorenzo}, \citenamefont {Apollaro}, \citenamefont
  {Plastina},\ and\ \citenamefont {Giorgi}}]{paganelli13}%
  \BibitemOpen
  \bibfield  {author} {\bibinfo {author} {\bibfnamefont {S.}~\bibnamefont
  {Paganelli}}, \bibinfo {author} {\bibfnamefont {S.}~\bibnamefont {Lorenzo}},
  \bibinfo {author} {\bibfnamefont {T.~J.~G.}\ \bibnamefont {Apollaro}},
  \bibinfo {author} {\bibfnamefont {F.}~\bibnamefont {Plastina}}, \ and\
  \bibinfo {author} {\bibfnamefont {G.~L.}\ \bibnamefont {Giorgi}},\ }\bibfield
   {title} {\enquote {\bibinfo {title} {Routing quantum information in spin
  chains},}\ }\href {\doibase 10.1103/PhysRevA.87.062309} {\bibfield  {journal}
  {\bibinfo  {journal} {Phys. Rev. A}\ }\textbf {\bibinfo {volume} {87}},\
  \bibinfo {pages} {062309} (\bibinfo {year} {2013})}\BibitemShut {NoStop}%
\bibitem [{\citenamefont {Christandl}\ \emph {et~al.}(2005)\citenamefont
  {Christandl}, \citenamefont {Datta}, \citenamefont {Dorlas}, \citenamefont
  {Ekert}, \citenamefont {Kay},\ and\ \citenamefont {Landahl}}]{Christandl05}%
  \BibitemOpen
  \bibfield  {author} {\bibinfo {author} {\bibfnamefont {M.}~\bibnamefont
  {Christandl}}, \bibinfo {author} {\bibfnamefont {N.}~\bibnamefont {Datta}},
  \bibinfo {author} {\bibfnamefont {T.~C.}\ \bibnamefont {Dorlas}}, \bibinfo
  {author} {\bibfnamefont {A.}~\bibnamefont {Ekert}}, \bibinfo {author}
  {\bibfnamefont {A.}~\bibnamefont {Kay}}, \ and\ \bibinfo {author}
  {\bibfnamefont {A.~J.}\ \bibnamefont {Landahl}},\ }\bibfield  {title}
  {\enquote {\bibinfo {title} {Perfect transfer of arbitrary states in quantum
  spin networks},}\ }\href {\doibase 10.1103/PhysRevA.71.032312} {\bibfield
  {journal} {\bibinfo  {journal} {Phys. Rev. A}\ }\textbf {\bibinfo {volume}
  {71}},\ \bibinfo {pages} {032312} (\bibinfo {year} {2005})}\BibitemShut
  {NoStop}%
\bibitem [{\citenamefont {Kay}(2010)}]{kay10}%
  \BibitemOpen
  \bibfield  {author} {\bibinfo {author} {\bibfnamefont {A.}~\bibnamefont
  {Kay}},\ }\bibfield  {title} {\enquote {\bibinfo {title} {Perfect, efficient,
  state transfer and its application as a constructive tool},}\ }\href
  {\doibase 10.1142/S0219749910006514} {\bibfield  {journal} {\bibinfo
  {journal} {Int. J. Quantum Inf.}\ }\textbf {\bibinfo {volume} {08}},\
  \bibinfo {pages} {641--676} (\bibinfo {year} {2010})}\BibitemShut {NoStop}%
\bibitem [{\citenamefont {Tamascelli}\ \emph {et~al.}(2016)\citenamefont
  {Tamascelli}, \citenamefont {Olivares}, \citenamefont {Rossotti},
  \citenamefont {Osellame},\ and\ \citenamefont {Paris}}]{tama16}%
  \BibitemOpen
  \bibfield  {author} {\bibinfo {author} {\bibfnamefont {D.}~\bibnamefont
  {Tamascelli}}, \bibinfo {author} {\bibfnamefont {S.}~\bibnamefont
  {Olivares}}, \bibinfo {author} {\bibfnamefont {S.}~\bibnamefont {Rossotti}},
  \bibinfo {author} {\bibfnamefont {R.}~\bibnamefont {Osellame}}, \ and\
  \bibinfo {author} {\bibfnamefont {M.~G.~A.}\ \bibnamefont {Paris}},\
  }\bibfield  {title} {\enquote {\bibinfo {title} {Quantum state transfer via
  {Bloch} oscillations},}\ }\href {https://doi.org/10.1038/srep26054}
  {\bibfield  {journal} {\bibinfo  {journal} {Sci. Rep.}\ }\textbf {\bibinfo
  {volume} {6}},\ \bibinfo {pages} {26054} (\bibinfo {year}
  {2016})}\BibitemShut {NoStop}%
\bibitem [{\citenamefont {Wiseman}\ and\ \citenamefont
  {Milburn}(2010)}]{WisemanMilburn}%
  \BibitemOpen
  \bibfield  {author} {\bibinfo {author} {\bibfnamefont {H.~M.}\ \bibnamefont
  {Wiseman}}\ and\ \bibinfo {author} {\bibfnamefont {G.~J.}\ \bibnamefont
  {Milburn}},\ }\href@noop {} {\emph {\bibinfo {title} {{Quantum Measurement
  and Control}}}}\ (\bibinfo  {publisher} {Cambridge University Press},\
  \bibinfo {address} {New York},\ \bibinfo {year} {2010})\BibitemShut {NoStop}%
\bibitem [{\citenamefont {Wiseman}\ and\ \citenamefont
  {Milburn}(1993)}]{Wiseman1993}%
  \BibitemOpen
  \bibfield  {author} {\bibinfo {author} {\bibfnamefont {H.~M.}\ \bibnamefont
  {Wiseman}}\ and\ \bibinfo {author} {\bibfnamefont {G.~J.}\ \bibnamefont
  {Milburn}},\ }\bibfield  {title} {\enquote {\bibinfo {title} {Quantum theory
  of optical feedback via homodyne detection},}\ }\href {\doibase
  10.1103/PhysRevLett.70.548} {\bibfield  {journal} {\bibinfo  {journal} {Phys.
  Rev. Lett.}\ }\textbf {\bibinfo {volume} {70}},\ \bibinfo {pages} {548--551}
  (\bibinfo {year} {1993})}\BibitemShut {NoStop}%
\bibitem [{\citenamefont {Wiseman}(1994)}]{Wiseman1994}%
  \BibitemOpen
  \bibfield  {author} {\bibinfo {author} {\bibfnamefont {H.~M.}\ \bibnamefont
  {Wiseman}},\ }\bibfield  {title} {\enquote {\bibinfo {title} {Quantum theory
  of continuous feedback},}\ }\href {\doibase 10.1103/PhysRevA.49.2133}
  {\bibfield  {journal} {\bibinfo  {journal} {Phys. Rev. A}\ }\textbf {\bibinfo
  {volume} {49}},\ \bibinfo {pages} {2133--2150} (\bibinfo {year}
  {1994})}\BibitemShut {NoStop}%
\bibitem [{\citenamefont {Doherty}\ and\ \citenamefont
  {Jacobs}(1999)}]{Doherty1999}%
  \BibitemOpen
  \bibfield  {author} {\bibinfo {author} {\bibfnamefont {A.~C.}\ \bibnamefont
  {Doherty}}\ and\ \bibinfo {author} {\bibfnamefont {K.}~\bibnamefont
  {Jacobs}},\ }\bibfield  {title} {\enquote {\bibinfo {title} {Feedback control
  of quantum systems using continuous state estimation},}\ }\href {\doibase
  10.1103/PhysRevA.60.2700} {\bibfield  {journal} {\bibinfo  {journal} {Phys.
  Rev. A}\ }\textbf {\bibinfo {volume} {60}},\ \bibinfo {pages} {2700--2711}
  (\bibinfo {year} {1999})}\BibitemShut {NoStop}%
\bibitem [{\citenamefont {Thomsen}\ \emph {et~al.}(2002)\citenamefont
  {Thomsen}, \citenamefont {Mancini},\ and\ \citenamefont
  {Wiseman}}]{Thomsen2002}%
  \BibitemOpen
  \bibfield  {author} {\bibinfo {author} {\bibfnamefont {L.~K.}\ \bibnamefont
  {Thomsen}}, \bibinfo {author} {\bibfnamefont {Stefano}\ \bibnamefont
  {Mancini}}, \ and\ \bibinfo {author} {\bibfnamefont {Howard~M.}\ \bibnamefont
  {Wiseman}},\ }\bibfield  {title} {\enquote {\bibinfo {title} {{Spin squeezing
  via quantum feedback}},}\ }\href {\doibase 10.1103/PhysRevA.65.061801}
  {\bibfield  {journal} {\bibinfo  {journal} {Phys. Rev. A}\ }\textbf {\bibinfo
  {volume} {65}},\ \bibinfo {pages} {061801} (\bibinfo {year} {2002})},\
  \Eprint {http://arxiv.org/abs/quant-ph/0202028} {quant-ph/0202028}
  \BibitemShut {NoStop}%
\bibitem [{\citenamefont {Serafini}\ and\ \citenamefont
  {Mancini}(2010)}]{SerafozziMancini}%
  \BibitemOpen
  \bibfield  {author} {\bibinfo {author} {\bibfnamefont {Alessio}\ \bibnamefont
  {Serafini}}\ and\ \bibinfo {author} {\bibfnamefont {Stefano}\ \bibnamefont
  {Mancini}},\ }\bibfield  {title} {\enquote {\bibinfo {title} {{Determination
  of Maximal Gaussian Entanglement Achievable by Feedback-Controlled
  Dynamics}},}\ }\href {\doibase 10.1103/PhysRevLett.104.220501} {\bibfield
  {journal} {\bibinfo  {journal} {Phys. Rev. Lett.}\ }\textbf {\bibinfo
  {volume} {104}},\ \bibinfo {pages} {220501} (\bibinfo {year}
  {2010})}\BibitemShut {NoStop}%
\bibitem [{\citenamefont {Szorkovszky}\ \emph {et~al.}(2011)\citenamefont
  {Szorkovszky}, \citenamefont {Doherty}, \citenamefont {Harris},\ and\
  \citenamefont {Bowen}}]{Szorkovszky2011}%
  \BibitemOpen
  \bibfield  {author} {\bibinfo {author} {\bibfnamefont {A.}~\bibnamefont
  {Szorkovszky}}, \bibinfo {author} {\bibfnamefont {A.~C.}\ \bibnamefont
  {Doherty}}, \bibinfo {author} {\bibfnamefont {G.~I.}\ \bibnamefont {Harris}},
  \ and\ \bibinfo {author} {\bibfnamefont {W.~P.}\ \bibnamefont {Bowen}},\
  }\bibfield  {title} {\enquote {\bibinfo {title} {Mechanical squeezing via
  parametric amplification and weak measurement},}\ }\href {\doibase
  10.1103/PhysRevLett.107.213603} {\bibfield  {journal} {\bibinfo  {journal}
  {Phys. Rev. Lett.}\ }\textbf {\bibinfo {volume} {107}},\ \bibinfo {pages}
  {213603} (\bibinfo {year} {2011})}\BibitemShut {NoStop}%
\bibitem [{\citenamefont {Genoni}\ \emph {et~al.}(2013)\citenamefont {Genoni},
  \citenamefont {Mancini},\ and\ \citenamefont {Serafini}}]{Genoni2013PRA}%
  \BibitemOpen
  \bibfield  {author} {\bibinfo {author} {\bibfnamefont {Marco~G.}\
  \bibnamefont {Genoni}}, \bibinfo {author} {\bibfnamefont {Stefano}\
  \bibnamefont {Mancini}}, \ and\ \bibinfo {author} {\bibfnamefont {Alessio}\
  \bibnamefont {Serafini}},\ }\bibfield  {title} {\enquote {\bibinfo {title}
  {Optimal feedback control of linear quantum systems in the presence of
  thermal noise},}\ }\href {\doibase 10.1103/PhysRevA.87.042333} {\bibfield
  {journal} {\bibinfo  {journal} {Phys. Rev. A}\ }\textbf {\bibinfo {volume}
  {87}},\ \bibinfo {pages} {042333} (\bibinfo {year} {2013})}\BibitemShut
  {NoStop}%
\bibitem [{\citenamefont {Genoni}\ \emph {et~al.}(2015)\citenamefont {Genoni},
  \citenamefont {Zhang}, \citenamefont {Millen}, \citenamefont {Barker},\ and\
  \citenamefont {Serafini}}]{Genoni2015NJP}%
  \BibitemOpen
  \bibfield  {author} {\bibinfo {author} {\bibfnamefont {Marco~G}\ \bibnamefont
  {Genoni}}, \bibinfo {author} {\bibfnamefont {Jinglei}\ \bibnamefont {Zhang}},
  \bibinfo {author} {\bibfnamefont {James}\ \bibnamefont {Millen}}, \bibinfo
  {author} {\bibfnamefont {Peter~F}\ \bibnamefont {Barker}}, \ and\ \bibinfo
  {author} {\bibfnamefont {Alessio}\ \bibnamefont {Serafini}},\ }\bibfield
  {title} {\enquote {\bibinfo {title} {Quantum cooling and squeezing of a
  levitating nanosphere via time-continuous measurements},}\ }\href {\doibase
  10.1088/1367-2630/17/7/073019} {\bibfield  {journal} {\bibinfo  {journal}
  {New Journal of Physics}\ }\textbf {\bibinfo {volume} {17}},\ \bibinfo
  {pages} {073019} (\bibinfo {year} {2015})}\BibitemShut {NoStop}%
\bibitem [{\citenamefont {Hofer}\ and\ \citenamefont
  {Hammerer}(2015)}]{Hofer2015}%
  \BibitemOpen
  \bibfield  {author} {\bibinfo {author} {\bibfnamefont {Sebastian~G.}\
  \bibnamefont {Hofer}}\ and\ \bibinfo {author} {\bibfnamefont {Klemens}\
  \bibnamefont {Hammerer}},\ }\bibfield  {title} {\enquote {\bibinfo {title}
  {{Entanglement-enhanced time-continuous quantum control in optomechanics}},}\
  }\href {\doibase 10.1103/PhysRevA.91.033822} {\bibfield  {journal} {\bibinfo
  {journal} {Phys. Rev. A}\ }\textbf {\bibinfo {volume} {91}},\ \bibinfo
  {pages} {033822} (\bibinfo {year} {2015})},\ \Eprint
  {http://arxiv.org/abs/1411.1337} {1411.1337} \BibitemShut {NoStop}%
\bibitem [{\citenamefont {Martin}\ \emph {et~al.}(2015)\citenamefont {Martin},
  \citenamefont {Motzoi}, \citenamefont {Li}, \citenamefont {Sarovar},\ and\
  \citenamefont {Whaley}}]{Martin2015}%
  \BibitemOpen
  \bibfield  {author} {\bibinfo {author} {\bibfnamefont {L.}~\bibnamefont
  {Martin}}, \bibinfo {author} {\bibfnamefont {F.}~\bibnamefont {Motzoi}},
  \bibinfo {author} {\bibfnamefont {H.}~\bibnamefont {Li}}, \bibinfo {author}
  {\bibfnamefont {M.}~\bibnamefont {Sarovar}}, \ and\ \bibinfo {author}
  {\bibfnamefont {K.~B.}\ \bibnamefont {Whaley}},\ }\bibfield  {title}
  {\enquote {\bibinfo {title} {Deterministic generation of remote entanglement
  with active quantum feedback},}\ }\href {\doibase 10.1103/PhysRevA.92.062321}
  {\bibfield  {journal} {\bibinfo  {journal} {Phys. Rev. A}\ }\textbf {\bibinfo
  {volume} {92}},\ \bibinfo {pages} {062321} (\bibinfo {year}
  {2015})}\BibitemShut {NoStop}%
\bibitem [{\citenamefont {Martin}\ \emph
  {et~al.}(2017{\natexlab{a}})\citenamefont {Martin}, \citenamefont {Sayrafi},\
  and\ \citenamefont {Whaley}}]{martin2017optimal}%
  \BibitemOpen
  \bibfield  {author} {\bibinfo {author} {\bibfnamefont {L.}~\bibnamefont
  {Martin}}, \bibinfo {author} {\bibfnamefont {M.}~\bibnamefont {Sayrafi}}, \
  and\ \bibinfo {author} {\bibfnamefont {K~B.}\ \bibnamefont {Whaley}},\
  }\bibfield  {title} {\enquote {\bibinfo {title} {What is the optimal way to
  prepare a bell state using measurement and feedback?}}\ }\href {\doibase
  10.1088/2058-9565/aa804c} {\bibfield  {journal} {\bibinfo  {journal} {Quantum
  Sci. Technol.}\ }\textbf {\bibinfo {volume} {2}},\ \bibinfo {pages} {044006}
  (\bibinfo {year} {2017}{\natexlab{a}})}\BibitemShut {NoStop}%
\bibitem [{\citenamefont {Martin}\ \emph
  {et~al.}(2017{\natexlab{b}})\citenamefont {Martin}, \citenamefont {Sayrafi},\
  and\ \citenamefont {Whaley}}]{Martin:17}%
  \BibitemOpen
  \bibfield  {author} {\bibinfo {author} {\bibfnamefont {L.}~\bibnamefont
  {Martin}}, \bibinfo {author} {\bibfnamefont {M.}~\bibnamefont {Sayrafi}}, \
  and\ \bibinfo {author} {\bibfnamefont {B}~\bibnamefont {Whaley}},\ }\bibfield
   {title} {\enquote {\bibinfo {title} {Optimal protocols for remote
  entanglement generation},}\ }in\ \href {\doibase 10.1364/QIM.2017.QF6C.3}
  {\emph {\bibinfo {booktitle} {Quantum Information and Measurement (QIM)
  2017}}}\ (\bibinfo  {publisher} {Optical Society of America},\ \bibinfo
  {year} {2017})\ p.\ \bibinfo {pages} {QF6C.3}\BibitemShut {NoStop}%
\bibitem [{\citenamefont {Brunelli}\ \emph {et~al.}(2019)\citenamefont
  {Brunelli}, \citenamefont {Malz},\ and\ \citenamefont
  {Nunnenkamp}}]{Brunelli2019PRL}%
  \BibitemOpen
  \bibfield  {author} {\bibinfo {author} {\bibfnamefont {Matteo}\ \bibnamefont
  {Brunelli}}, \bibinfo {author} {\bibfnamefont {Daniel}\ \bibnamefont {Malz}},
  \ and\ \bibinfo {author} {\bibfnamefont {Andreas}\ \bibnamefont
  {Nunnenkamp}},\ }\bibfield  {title} {\enquote {\bibinfo {title} {Conditional
  dynamics of optomechanical two-tone backaction-evading measurements},}\
  }\href {\doibase 10.1103/PhysRevLett.123.093602} {\bibfield  {journal}
  {\bibinfo  {journal} {Phys. Rev. Lett.}\ }\textbf {\bibinfo {volume} {123}},\
  \bibinfo {pages} {093602} (\bibinfo {year} {2019})}\BibitemShut {NoStop}%
\bibitem [{\citenamefont {Martin}(2020)}]{martin2020quantum}%
  \BibitemOpen
  \bibfield  {author} {\bibinfo {author} {\bibfnamefont {Leigh~S.}\
  \bibnamefont {Martin}},\ }\href@noop {} {\enquote {\bibinfo {title} {Quantum
  feedback for measurement and control},}\ } (\bibinfo {year} {2020}),\ \Eprint
  {http://arxiv.org/abs/2004.09766} {arXiv:2004.09766 [quant-ph]} \BibitemShut
  {NoStop}%
\bibitem [{\citenamefont {Jiang}\ \emph {et~al.}(2020)\citenamefont {Jiang},
  \citenamefont {Wang}, \citenamefont {Martin},\ and\ \citenamefont
  {Whaley}}]{Jiang2020}%
  \BibitemOpen
  \bibfield  {author} {\bibinfo {author} {\bibfnamefont {Y.}~\bibnamefont
  {Jiang}}, \bibinfo {author} {\bibfnamefont {X.}~\bibnamefont {Wang}},
  \bibinfo {author} {\bibfnamefont {L.}~\bibnamefont {Martin}}, \ and\ \bibinfo
  {author} {\bibfnamefont {K.~B.}\ \bibnamefont {Whaley}},\ }\bibfield  {title}
  {\enquote {\bibinfo {title} {Optimality of feedback control for qubit
  purification under inefficient measurement},}\ }\href {\doibase
  10.1103/PhysRevA.102.022612} {\bibfield  {journal} {\bibinfo  {journal}
  {Phys. Rev. A}\ }\textbf {\bibinfo {volume} {102}},\ \bibinfo {pages}
  {022612} (\bibinfo {year} {2020})}\BibitemShut {NoStop}%
\bibitem [{\citenamefont {Zhang}\ \emph {et~al.}(2020)\citenamefont {Zhang},
  \citenamefont {Martin},\ and\ \citenamefont {Whaley}}]{Zhang2020}%
  \BibitemOpen
  \bibfield  {author} {\bibinfo {author} {\bibfnamefont {S.}~\bibnamefont
  {Zhang}}, \bibinfo {author} {\bibfnamefont {L.~S.}\ \bibnamefont {Martin}}, \
  and\ \bibinfo {author} {\bibfnamefont {K.~B.}\ \bibnamefont {Whaley}},\
  }\bibfield  {title} {\enquote {\bibinfo {title} {Locally optimal
  measurement-based quantum feedback with application to multiqubit
  entanglement generation},}\ }\href {\doibase 10.1103/PhysRevA.102.062418}
  {\bibfield  {journal} {\bibinfo  {journal} {Phys. Rev. A}\ }\textbf {\bibinfo
  {volume} {102}},\ \bibinfo {pages} {062418} (\bibinfo {year}
  {2020})}\BibitemShut {NoStop}%
\bibitem [{\citenamefont {Rossi}\ \emph {et~al.}(2020)\citenamefont {Rossi},
  \citenamefont {Albarelli}, \citenamefont {Tamascelli},\ and\ \citenamefont
  {Genoni}}]{Rossi2020PRL}%
  \BibitemOpen
  \bibfield  {author} {\bibinfo {author} {\bibfnamefont {Matteo A.~C.}\
  \bibnamefont {Rossi}}, \bibinfo {author} {\bibfnamefont {Francesco}\
  \bibnamefont {Albarelli}}, \bibinfo {author} {\bibfnamefont {Dario}\
  \bibnamefont {Tamascelli}}, \ and\ \bibinfo {author} {\bibfnamefont
  {Marco~G.}\ \bibnamefont {Genoni}},\ }\bibfield  {title} {\enquote {\bibinfo
  {title} {Noisy quantum metrology enhanced by continuous nondemolition
  measurement},}\ }\href {\doibase 10.1103/PhysRevLett.125.200505} {\bibfield
  {journal} {\bibinfo  {journal} {Phys. Rev. Lett.}\ }\textbf {\bibinfo
  {volume} {125}},\ \bibinfo {pages} {200505} (\bibinfo {year}
  {2020})}\BibitemShut {NoStop}%
\bibitem [{\citenamefont {Di~Giovanni}\ \emph {et~al.}(2021)\citenamefont
  {Di~Giovanni}, \citenamefont {Brunelli},\ and\ \citenamefont
  {Genoni}}]{DiGiovanni2021}%
  \BibitemOpen
  \bibfield  {author} {\bibinfo {author} {\bibfnamefont {Antonio}\ \bibnamefont
  {Di~Giovanni}}, \bibinfo {author} {\bibfnamefont {Matteo}\ \bibnamefont
  {Brunelli}}, \ and\ \bibinfo {author} {\bibfnamefont {Marco~G.}\ \bibnamefont
  {Genoni}},\ }\bibfield  {title} {\enquote {\bibinfo {title} {Unconditional
  mechanical squeezing via backaction-evading measurements and nonoptimal
  feedback control},}\ }\href {\doibase 10.1103/PhysRevA.103.022614} {\bibfield
   {journal} {\bibinfo  {journal} {Phys. Rev. A}\ }\textbf {\bibinfo {volume}
  {103}},\ \bibinfo {pages} {022614} (\bibinfo {year} {2021})}\BibitemShut
  {NoStop}%
\bibitem [{\citenamefont {Rossi}\ \emph
  {et~al.}(2018{\natexlab{a}})\citenamefont {Rossi}, \citenamefont {Mason},
  \citenamefont {Chen}, \citenamefont {Tsaturyan},\ and\ \citenamefont
  {Schliesser}}]{Rossi2018}%
  \BibitemOpen
  \bibfield  {author} {\bibinfo {author} {\bibfnamefont {Massimiliano}\
  \bibnamefont {Rossi}}, \bibinfo {author} {\bibfnamefont {David}\ \bibnamefont
  {Mason}}, \bibinfo {author} {\bibfnamefont {Junxin}\ \bibnamefont {Chen}},
  \bibinfo {author} {\bibfnamefont {Yeghishe}\ \bibnamefont {Tsaturyan}}, \
  and\ \bibinfo {author} {\bibfnamefont {Albert}\ \bibnamefont {Schliesser}},\
  }\bibfield  {title} {\enquote {\bibinfo {title} {Measurement-based quantum
  control of mechanical motion},}\ }\href {\doibase 10.1038/s41586-018-0643-8}
  {\bibfield  {journal} {\bibinfo  {journal} {Nature}\ }\textbf {\bibinfo
  {volume} {563}},\ \bibinfo {pages} {53--58} (\bibinfo {year}
  {2018}{\natexlab{a}})}\BibitemShut {NoStop}%
\bibitem [{\citenamefont {Magrini}\ \emph
  {et~al.}(2021{\natexlab{a}})\citenamefont {Magrini}, \citenamefont
  {Rosenzweig}, \citenamefont {Bach}, \citenamefont {Deutschmann-Olek},
  \citenamefont {Hofer}, \citenamefont {Hong}, \citenamefont {Kiesel},
  \citenamefont {Kugi},\ and\ \citenamefont {Aspelmeyer}}]{Magrini2021}%
  \BibitemOpen
  \bibfield  {author} {\bibinfo {author} {\bibfnamefont {Lorenzo}\ \bibnamefont
  {Magrini}}, \bibinfo {author} {\bibfnamefont {Philipp}\ \bibnamefont
  {Rosenzweig}}, \bibinfo {author} {\bibfnamefont {Constanze}\ \bibnamefont
  {Bach}}, \bibinfo {author} {\bibfnamefont {Andreas}\ \bibnamefont
  {Deutschmann-Olek}}, \bibinfo {author} {\bibfnamefont {Sebastian~G.}\
  \bibnamefont {Hofer}}, \bibinfo {author} {\bibfnamefont {Sungkun}\
  \bibnamefont {Hong}}, \bibinfo {author} {\bibfnamefont {Nikolai}\
  \bibnamefont {Kiesel}}, \bibinfo {author} {\bibfnamefont {Andreas}\
  \bibnamefont {Kugi}}, \ and\ \bibinfo {author} {\bibfnamefont {Markus}\
  \bibnamefont {Aspelmeyer}},\ }\bibfield  {title} {\enquote {\bibinfo {title}
  {Real-time optimal quantum control of mechanical motion at room
  temperature},}\ }\href {\doibase 10.1038/s41586-021-03602-3} {\bibfield
  {journal} {\bibinfo  {journal} {Nature}\ }\textbf {\bibinfo {volume} {595}},\
  \bibinfo {pages} {373--377} (\bibinfo {year}
  {2021}{\natexlab{a}})}\BibitemShut {NoStop}%
\bibitem [{\citenamefont {Tebbenjohanns}\ \emph
  {et~al.}(2021{\natexlab{a}})\citenamefont {Tebbenjohanns}, \citenamefont
  {Mattana}, \citenamefont {Rossi}, \citenamefont {Frimmer},\ and\
  \citenamefont {Novotny}}]{Tebbenjohanns2021}%
  \BibitemOpen
  \bibfield  {author} {\bibinfo {author} {\bibfnamefont {Felix}\ \bibnamefont
  {Tebbenjohanns}}, \bibinfo {author} {\bibfnamefont {M.~Luisa}\ \bibnamefont
  {Mattana}}, \bibinfo {author} {\bibfnamefont {Massimiliano}\ \bibnamefont
  {Rossi}}, \bibinfo {author} {\bibfnamefont {Martin}\ \bibnamefont {Frimmer}},
  \ and\ \bibinfo {author} {\bibfnamefont {Lukas}\ \bibnamefont {Novotny}},\
  }\bibfield  {title} {\enquote {\bibinfo {title} {Quantum control of a
  nanoparticle optically levitated in cryogenic free space},}\ }\href {\doibase
  10.1038/s41586-021-03617-w} {\bibfield  {journal} {\bibinfo  {journal}
  {Nature}\ }\textbf {\bibinfo {volume} {595}},\ \bibinfo {pages} {378--382}
  (\bibinfo {year} {2021}{\natexlab{a}})}\BibitemShut {NoStop}%
\bibitem [{\citenamefont {Wong}\ \emph {et~al.}(2016)\citenamefont {Wong},
  \citenamefont {Tarrataca},\ and\ \citenamefont
  {Nahimov}}]{wong2016laplacian}%
  \BibitemOpen
  \bibfield  {author} {\bibinfo {author} {\bibfnamefont {T.~G}\ \bibnamefont
  {Wong}}, \bibinfo {author} {\bibfnamefont {L.}~\bibnamefont {Tarrataca}}, \
  and\ \bibinfo {author} {\bibfnamefont {N.}~\bibnamefont {Nahimov}},\
  }\bibfield  {title} {\enquote {\bibinfo {title} {Laplacian versus adjacency
  matrix in quantum walk search},}\ }\href {\doibase
  https://doi.org/10.1007/s11128-016-1373-1} {\bibfield  {journal} {\bibinfo
  {journal} {Quantum Inf. Process.}\ }\textbf {\bibinfo {volume} {15}},\
  \bibinfo {pages} {4029} (\bibinfo {year} {2016})}\BibitemShut {NoStop}%
\bibitem [{\citenamefont {{Lu, D. and Biamonte, J. D. and Li, J. and Li, H. and
  Johnson, T. H. and Bergholm, V. and Faccin, M. and Zimbor\'as, Z. and
  Laflamme, R. and Baugh, J. and Lloyd, S.}}(2016)}]{cqw2}%
  \BibitemOpen
  \bibfield  {author} {\bibinfo {author} {\bibnamefont {{Lu, D. and Biamonte,
  J. D. and Li, J. and Li, H. and Johnson, T. H. and Bergholm, V. and Faccin,
  M. and Zimbor\'as, Z. and Laflamme, R. and Baugh, J. and Lloyd, S.}}},\
  }\bibfield  {title} {\enquote {\bibinfo {title} {Chiral quantum walks},}\
  }\href {\doibase 10.1103/PhysRevA.93.042302} {\bibfield  {journal} {\bibinfo
  {journal} {Phys. Rev. A}\ }\textbf {\bibinfo {volume} {93}},\ \bibinfo
  {pages} {042302} (\bibinfo {year} {2016})}\BibitemShut {NoStop}%
\bibitem [{\citenamefont {{J. Turner and J. Biamonte}}(2021)}]{Turner_2021}%
  \BibitemOpen
  \bibfield  {author} {\bibinfo {author} {\bibnamefont {{J. Turner and J.
  Biamonte}}},\ }\bibfield  {title} {\enquote {\bibinfo {title} {Topological
  classification of time-asymmetry in unitary quantum processes},}\ }\href
  {\doibase 10.1088/1751-8121/abf9d0} {\bibfield  {journal} {\bibinfo
  {journal} {J. Phys. A: Math. Theo.}\ }\textbf {\bibinfo {volume} {54}},\
  \bibinfo {pages} {235301} (\bibinfo {year} {2021})}\BibitemShut {NoStop}%
\bibitem [{\citenamefont {Frigerio}\ \emph {et~al.}(2021)\citenamefont
  {Frigerio}, \citenamefont {Benedetti}, \citenamefont {Olivares},\ and\
  \citenamefont {Paris}}]{frigerio21}%
  \BibitemOpen
  \bibfield  {author} {\bibinfo {author} {\bibfnamefont {M.}~\bibnamefont
  {Frigerio}}, \bibinfo {author} {\bibfnamefont {C.}~\bibnamefont {Benedetti}},
  \bibinfo {author} {\bibfnamefont {S.}~\bibnamefont {Olivares}}, \ and\
  \bibinfo {author} {\bibfnamefont {M.~G.~A.}\ \bibnamefont {Paris}},\
  }\bibfield  {title} {\enquote {\bibinfo {title} {Generalized
  quantum-classical correspondence for random walks on graphs},}\ }\href
  {\doibase 10.1103/PhysRevA.104.L030201} {\bibfield  {journal} {\bibinfo
  {journal} {Phys. Rev. A}\ }\textbf {\bibinfo {volume} {104}},\ \bibinfo
  {pages} {L030201} (\bibinfo {year} {2021})}\BibitemShut {NoStop}%
\bibitem [{\citenamefont {Frigerio}\ \emph {et~al.}(2022)\citenamefont
  {Frigerio}, \citenamefont {Benedetti}, \citenamefont {Olivares},\ and\
  \citenamefont {Paris}}]{frigerio22}%
  \BibitemOpen
  \bibfield  {author} {\bibinfo {author} {\bibfnamefont {Massimo}\ \bibnamefont
  {Frigerio}}, \bibinfo {author} {\bibfnamefont {Claudia}\ \bibnamefont
  {Benedetti}}, \bibinfo {author} {\bibfnamefont {Stefano}\ \bibnamefont
  {Olivares}}, \ and\ \bibinfo {author} {\bibfnamefont {Matteo G.~A.}\
  \bibnamefont {Paris}},\ }\bibfield  {title} {\enquote {\bibinfo {title}
  {Quantum-classical distance as a tool to design optimal chiral quantum
  walks},}\ }\href {\doibase 10.1103/PhysRevA.105.032425} {\bibfield  {journal}
  {\bibinfo  {journal} {Phys. Rev. A}\ }\textbf {\bibinfo {volume} {105}},\
  \bibinfo {pages} {032425} (\bibinfo {year} {2022})}\BibitemShut {NoStop}%
\bibitem [{\citenamefont {Farhi}\ and\ \citenamefont
  {Gutmann}(1998{\natexlab{b}})}]{farhi98}%
  \BibitemOpen
  \bibfield  {author} {\bibinfo {author} {\bibfnamefont {E.}~\bibnamefont
  {Farhi}}\ and\ \bibinfo {author} {\bibfnamefont {S.}~\bibnamefont
  {Gutmann}},\ }\bibfield  {title} {\enquote {\bibinfo {title} {Analog analogue
  of a digital quantum computation},}\ }\href {\doibase
  10.1103/PhysRevA.57.2403} {\bibfield  {journal} {\bibinfo  {journal} {Phys.
  Rev. A}\ }\textbf {\bibinfo {volume} {57}},\ \bibinfo {pages} {2403--2406}
  (\bibinfo {year} {1998}{\natexlab{b}})}\BibitemShut {NoStop}%
\bibitem [{\citenamefont {Chakraborty}\ \emph {et~al.}(2016)\citenamefont
  {Chakraborty}, \citenamefont {Novo}, \citenamefont {Ambainis},\ and\
  \citenamefont {Omar}}]{Chakraborty16}%
  \BibitemOpen
  \bibfield  {author} {\bibinfo {author} {\bibfnamefont {S.}~\bibnamefont
  {Chakraborty}}, \bibinfo {author} {\bibfnamefont {L.}~\bibnamefont {Novo}},
  \bibinfo {author} {\bibfnamefont {A.}~\bibnamefont {Ambainis}}, \ and\
  \bibinfo {author} {\bibfnamefont {Y.}~\bibnamefont {Omar}},\ }\bibfield
  {title} {\enquote {\bibinfo {title} {Spatial search by quantum walk is
  optimal for almost all graphs},}\ }\href {\doibase
  10.1103/PhysRevLett.116.100501} {\bibfield  {journal} {\bibinfo  {journal}
  {Phys. Rev. Lett.}\ }\textbf {\bibinfo {volume} {116}},\ \bibinfo {pages}
  {100501} (\bibinfo {year} {2016})}\BibitemShut {NoStop}%
\bibitem [{\citenamefont {Philipp}\ \emph {et~al.}(2016)\citenamefont
  {Philipp}, \citenamefont {Tarrataca},\ and\ \citenamefont
  {Boettcher}}]{philipp16}%
  \BibitemOpen
  \bibfield  {author} {\bibinfo {author} {\bibfnamefont {P.}~\bibnamefont
  {Philipp}}, \bibinfo {author} {\bibfnamefont {L.}~\bibnamefont {Tarrataca}},
  \ and\ \bibinfo {author} {\bibfnamefont {S.}~\bibnamefont {Boettcher}},\
  }\bibfield  {title} {\enquote {\bibinfo {title} {Continuous-time quantum
  search on balanced trees},}\ }\href {\doibase 10.1103/PhysRevA.93.032305}
  {\bibfield  {journal} {\bibinfo  {journal} {Phys. Rev. A}\ }\textbf {\bibinfo
  {volume} {93}},\ \bibinfo {pages} {032305} (\bibinfo {year}
  {2016})}\BibitemShut {NoStop}%
\bibitem [{\citenamefont {T.~G}(2016)}]{wong16}%
  \BibitemOpen
  \bibfield  {author} {\bibinfo {author} {\bibfnamefont {Wong}\ \bibnamefont
  {T.~G}},\ }\bibfield  {title} {\enquote {\bibinfo {title} {Quantum walk
  search on johnson graphs},}\ }\href {\doibase 10.1088/1751-8113/49/19/195303}
  {\bibfield  {journal} {\bibinfo  {journal} {J. Phys. A: Math. Theor.}\
  }\textbf {\bibinfo {volume} {49}},\ \bibinfo {pages} {195303} (\bibinfo
  {year} {2016})}\BibitemShut {NoStop}%
\bibitem [{\citenamefont {{Chakraborty, S. and Novo, L. and Di Giorgio, S. and
  Omar, Y.}}(2017)}]{omar17}%
  \BibitemOpen
  \bibfield  {author} {\bibinfo {author} {\bibnamefont {{Chakraborty, S. and
  Novo, L. and Di Giorgio, S. and Omar, Y.}}},\ }\bibfield  {title} {\enquote
  {\bibinfo {title} {Optimal quantum spatial search on random temporal
  networks},}\ }\href {\doibase 10.1103/PhysRevLett.119.220503} {\bibfield
  {journal} {\bibinfo  {journal} {Phys. Rev. Lett.}\ }\textbf {\bibinfo
  {volume} {119}},\ \bibinfo {pages} {220503} (\bibinfo {year}
  {2017})}\BibitemShut {NoStop}%
\bibitem [{\citenamefont {Wong}\ \emph {et~al.}(2018)\citenamefont {Wong},
  \citenamefont {W\"unscher}, \citenamefont {Lockhart},\ and\ \citenamefont
  {Severini}}]{wong18}%
  \BibitemOpen
  \bibfield  {author} {\bibinfo {author} {\bibfnamefont {T.~G.}\ \bibnamefont
  {Wong}}, \bibinfo {author} {\bibfnamefont {K.}~\bibnamefont {W\"unscher}},
  \bibinfo {author} {\bibfnamefont {J.}~\bibnamefont {Lockhart}}, \ and\
  \bibinfo {author} {\bibfnamefont {S.}~\bibnamefont {Severini}},\ }\bibfield
  {title} {\enquote {\bibinfo {title} {Quantum walk search on kronecker
  graphs},}\ }\href {\doibase 10.1103/PhysRevA.98.012338} {\bibfield  {journal}
  {\bibinfo  {journal} {Phys. Rev. A}\ }\textbf {\bibinfo {volume} {98}},\
  \bibinfo {pages} {012338} (\bibinfo {year} {2018})}\BibitemShut {NoStop}%
\bibitem [{\citenamefont {{Wang, Y. and Wu, S. and Wang, W.}}(2020)}]{wang20}%
  \BibitemOpen
  \bibfield  {author} {\bibinfo {author} {\bibnamefont {{Wang, Y. and Wu, S.
  and Wang, W.}}},\ }\bibfield  {title} {\enquote {\bibinfo {title} {Optimal
  quantum search on truncated simplex lattices},}\ }\href {\doibase
  10.1103/PhysRevA.101.062333} {\bibfield  {journal} {\bibinfo  {journal}
  {Phys. Rev. A}\ }\textbf {\bibinfo {volume} {101}},\ \bibinfo {pages}
  {062333} (\bibinfo {year} {2020})}\BibitemShut {NoStop}%
\bibitem [{\citenamefont {Chakraborty}\ \emph
  {et~al.}(2020{\natexlab{b}})\citenamefont {Chakraborty}, \citenamefont
  {Novo},\ and\ \citenamefont {Roland}}]{chakraborty20}%
  \BibitemOpen
  \bibfield  {author} {\bibinfo {author} {\bibfnamefont {S.}~\bibnamefont
  {Chakraborty}}, \bibinfo {author} {\bibfnamefont {L.}~\bibnamefont {Novo}}, \
  and\ \bibinfo {author} {\bibfnamefont {J.}~\bibnamefont {Roland}},\
  }\bibfield  {title} {\enquote {\bibinfo {title} {Optimality of spatial search
  via continuous-time quantum walks},}\ }\href {\doibase
  10.1103/PhysRevA.102.032214} {\bibfield  {journal} {\bibinfo  {journal}
  {Phys. Rev. A}\ }\textbf {\bibinfo {volume} {102}},\ \bibinfo {pages}
  {032214} (\bibinfo {year} {2020}{\natexlab{b}})}\BibitemShut {NoStop}%
\bibitem [{\citenamefont {Foulger}\ \emph {et~al.}(2014)\citenamefont
  {Foulger}, \citenamefont {Gnutzmann},\ and\ \citenamefont
  {Tanner}}]{tanner14}%
  \BibitemOpen
  \bibfield  {author} {\bibinfo {author} {\bibfnamefont {I.}~\bibnamefont
  {Foulger}}, \bibinfo {author} {\bibfnamefont {S.}~\bibnamefont {Gnutzmann}},
  \ and\ \bibinfo {author} {\bibfnamefont {G.}~\bibnamefont {Tanner}},\
  }\bibfield  {title} {\enquote {\bibinfo {title} {Quantum search on graphene
  lattices},}\ }\href {\doibase 10.1103/PhysRevLett.112.070504} {\bibfield
  {journal} {\bibinfo  {journal} {Phys. Rev. Lett.}\ }\textbf {\bibinfo
  {volume} {112}},\ \bibinfo {pages} {070504} (\bibinfo {year}
  {2014})}\BibitemShut {NoStop}%
\bibitem [{\citenamefont {Jacobs}\ and\ \citenamefont
  {Steck}(2006)}]{SteckJacobs}%
  \BibitemOpen
  \bibfield  {author} {\bibinfo {author} {\bibfnamefont {K.}~\bibnamefont
  {Jacobs}}\ and\ \bibinfo {author} {\bibfnamefont {D.~A.}\ \bibnamefont
  {Steck}},\ }\bibfield  {title} {\enquote {\bibinfo {title} {{A
  straightforward introduction to continuous quantum measurement}},}\ }\href
  {\doibase 10.1080/00107510601101934} {\bibfield  {journal} {\bibinfo
  {journal} {Contemp. Phys.}\ }\textbf {\bibinfo {volume} {47}},\ \bibinfo
  {pages} {279} (\bibinfo {year} {2006})},\ \Eprint
  {http://arxiv.org/abs/quant-ph/0611067} {quant-ph/0611067} \BibitemShut
  {NoStop}%
\bibitem [{\citenamefont {Brun}(2002)}]{Brun2002}%
  \BibitemOpen
  \bibfield  {author} {\bibinfo {author} {\bibfnamefont {T.~A.}\ \bibnamefont
  {Brun}},\ }\bibfield  {title} {\enquote {\bibinfo {title} {{A simple model of
  quantum trajectories}},}\ }\href {\doibase 10.1119/1.1475328} {\bibfield
  {journal} {\bibinfo  {journal} {Am. J. Phys.}\ }\textbf {\bibinfo {volume}
  {70}},\ \bibinfo {pages} {719--737} (\bibinfo {year} {2002})}\BibitemShut
  {NoStop}%
\bibitem [{\citenamefont {Murch}\ \emph {et~al.}(2013)\citenamefont {Murch},
  \citenamefont {Weber}, \citenamefont {Macklin},\ and\ \citenamefont
  {Siddiqi}}]{Murch:2013aa}%
  \BibitemOpen
  \bibfield  {author} {\bibinfo {author} {\bibfnamefont {K.~W.}\ \bibnamefont
  {Murch}}, \bibinfo {author} {\bibfnamefont {S.~J.}\ \bibnamefont {Weber}},
  \bibinfo {author} {\bibfnamefont {C.}~\bibnamefont {Macklin}}, \ and\
  \bibinfo {author} {\bibfnamefont {I.}~\bibnamefont {Siddiqi}},\ }\bibfield
  {title} {\enquote {\bibinfo {title} {Observing single quantum trajectories of
  a superconducting quantum bit},}\ }\href {\doibase 10.1038/nature12539}
  {\bibfield  {journal} {\bibinfo  {journal} {Nature}\ }\textbf {\bibinfo
  {volume} {502}},\ \bibinfo {pages} {211--214} (\bibinfo {year}
  {2013})}\BibitemShut {NoStop}%
\bibitem [{\citenamefont {Naghiloo}\ \emph {et~al.}(2016)\citenamefont
  {Naghiloo}, \citenamefont {Foroozani}, \citenamefont {Tan}, \citenamefont
  {Jadbabaie},\ and\ \citenamefont {Murch}}]{Naghiloo:2016aa}%
  \BibitemOpen
  \bibfield  {author} {\bibinfo {author} {\bibfnamefont {M.}~\bibnamefont
  {Naghiloo}}, \bibinfo {author} {\bibfnamefont {N.}~\bibnamefont {Foroozani}},
  \bibinfo {author} {\bibfnamefont {D.}~\bibnamefont {Tan}}, \bibinfo {author}
  {\bibfnamefont {A.}~\bibnamefont {Jadbabaie}}, \ and\ \bibinfo {author}
  {\bibfnamefont {K.~W.}\ \bibnamefont {Murch}},\ }\bibfield  {title} {\enquote
  {\bibinfo {title} {Mapping quantum state dynamics in spontaneous emission},}\
  }\href {\doibase 10.1038/ncomms11527} {\bibfield  {journal} {\bibinfo
  {journal} {Nat. Commun.}\ }\textbf {\bibinfo {volume} {7}},\ \bibinfo {pages}
  {11527} (\bibinfo {year} {2016})}\BibitemShut {NoStop}%
\bibitem [{\citenamefont {Hacohen-Gourgy}\ \emph {et~al.}(2016)\citenamefont
  {Hacohen-Gourgy}, \citenamefont {Martin}, \citenamefont {Flurin},
  \citenamefont {Ramasesh}, \citenamefont {Whaley},\ and\ \citenamefont
  {Siddiqi}}]{Hacohen-Gourgy2016}%
  \BibitemOpen
  \bibfield  {author} {\bibinfo {author} {\bibfnamefont {S.}~\bibnamefont
  {Hacohen-Gourgy}}, \bibinfo {author} {\bibfnamefont {L.~S.}\ \bibnamefont
  {Martin}}, \bibinfo {author} {\bibfnamefont {E.}~\bibnamefont {Flurin}},
  \bibinfo {author} {\bibfnamefont {V.~V.}\ \bibnamefont {Ramasesh}}, \bibinfo
  {author} {\bibfnamefont {K.~B.}\ \bibnamefont {Whaley}}, \ and\ \bibinfo
  {author} {\bibfnamefont {I.}~\bibnamefont {Siddiqi}},\ }\bibfield  {title}
  {\enquote {\bibinfo {title} {Quantum dynamics of simultaneously measured
  non-commuting observables},}\ }\href {\doibase 10.1038/nature19762}
  {\bibfield  {journal} {\bibinfo  {journal} {Nature}\ }\textbf {\bibinfo
  {volume} {538}},\ \bibinfo {pages} {491--494} (\bibinfo {year}
  {2016})}\BibitemShut {NoStop}%
\bibitem [{\citenamefont {Ficheux}\ \emph {et~al.}(2018)\citenamefont
  {Ficheux}, \citenamefont {Jezouin}, \citenamefont {Leghtas},\ and\
  \citenamefont {Huard}}]{Ficheux2018}%
  \BibitemOpen
  \bibfield  {author} {\bibinfo {author} {\bibfnamefont {Q.}~\bibnamefont
  {Ficheux}}, \bibinfo {author} {\bibfnamefont {S.}~\bibnamefont {Jezouin}},
  \bibinfo {author} {\bibfnamefont {Z.}~\bibnamefont {Leghtas}}, \ and\
  \bibinfo {author} {\bibfnamefont {B.}~\bibnamefont {Huard}},\ }\bibfield
  {title} {\enquote {\bibinfo {title} {Dynamics of a qubit while simultaneously
  monitoring its relaxation and dephasing},}\ }\href {\doibase
  10.1038/s41467-018-04372-9} {\bibfield  {journal} {\bibinfo  {journal} {Nat.
  Commun.}\ }\textbf {\bibinfo {volume} {9}},\ \bibinfo {pages} {1926}
  (\bibinfo {year} {2018})}\BibitemShut {NoStop}%
\bibitem [{\citenamefont {Minev}\ \emph {et~al.}(2019)\citenamefont {Minev},
  \citenamefont {Mundhada}, \citenamefont {Shankar}, \citenamefont {Reinhold},
  \citenamefont {Guti{\'e}rrez-J{\'a}uregui}, \citenamefont {Schoelkopf},
  \citenamefont {Mirrahimi}, \citenamefont {Carmichael},\ and\ \citenamefont
  {Devoret}}]{Minev:2019aa}%
  \BibitemOpen
  \bibfield  {author} {\bibinfo {author} {\bibfnamefont {Z.~K.}\ \bibnamefont
  {Minev}}, \bibinfo {author} {\bibfnamefont {S.~O.}\ \bibnamefont {Mundhada}},
  \bibinfo {author} {\bibfnamefont {S.}~\bibnamefont {Shankar}}, \bibinfo
  {author} {\bibfnamefont {P.}~\bibnamefont {Reinhold}}, \bibinfo {author}
  {\bibfnamefont {R.}~\bibnamefont {Guti{\'e}rrez-J{\'a}uregui}}, \bibinfo
  {author} {\bibfnamefont {R.~J.}\ \bibnamefont {Schoelkopf}}, \bibinfo
  {author} {\bibfnamefont {M.}~\bibnamefont {Mirrahimi}}, \bibinfo {author}
  {\bibfnamefont {H.~J.}\ \bibnamefont {Carmichael}}, \ and\ \bibinfo {author}
  {\bibfnamefont {M.~H.}\ \bibnamefont {Devoret}},\ }\bibfield  {title}
  {\enquote {\bibinfo {title} {To catch and reverse a quantum jump
  mid-flight},}\ }\href {\doibase 10.1038/s41586-019-1287-z} {\bibfield
  {journal} {\bibinfo  {journal} {Nature}\ }\textbf {\bibinfo {volume} {570}},\
  \bibinfo {pages} {200--204} (\bibinfo {year} {2019})}\BibitemShut {NoStop}%
\bibitem [{\citenamefont {Wieczorek}\ \emph {et~al.}(2015)\citenamefont
  {Wieczorek}, \citenamefont {Hofer}, \citenamefont {Hoelscher-Obermaier},
  \citenamefont {Riedinger}, \citenamefont {Hammerer},\ and\ \citenamefont
  {Aspelmeyer}}]{Wieczorek2015}%
  \BibitemOpen
  \bibfield  {author} {\bibinfo {author} {\bibfnamefont {W.}~\bibnamefont
  {Wieczorek}}, \bibinfo {author} {\bibfnamefont {S.~G.}\ \bibnamefont
  {Hofer}}, \bibinfo {author} {\bibfnamefont {J.}~\bibnamefont
  {Hoelscher-Obermaier}}, \bibinfo {author} {\bibfnamefont {R.}~\bibnamefont
  {Riedinger}}, \bibinfo {author} {\bibfnamefont {K.}~\bibnamefont {Hammerer}},
  \ and\ \bibinfo {author} {\bibfnamefont {M.}~\bibnamefont {Aspelmeyer}},\
  }\bibfield  {title} {\enquote {\bibinfo {title} {Optimal state estimation for
  cavity optomechanical systems},}\ }\href {\doibase
  10.1103/PhysRevLett.114.223601} {\bibfield  {journal} {\bibinfo  {journal}
  {Phys. Rev. Lett.}\ }\textbf {\bibinfo {volume} {114}},\ \bibinfo {pages}
  {223601} (\bibinfo {year} {2015})}\BibitemShut {NoStop}%
\bibitem [{\citenamefont {Rossi}\ \emph
  {et~al.}(2018{\natexlab{b}})\citenamefont {Rossi}, \citenamefont {Mason},
  \citenamefont {Chen}, \citenamefont {Tsaturyan},\ and\ \citenamefont
  {Schliesser}}]{Rossi:2018aa}%
  \BibitemOpen
  \bibfield  {author} {\bibinfo {author} {\bibfnamefont {M.}~\bibnamefont
  {Rossi}}, \bibinfo {author} {\bibfnamefont {D.}~\bibnamefont {Mason}},
  \bibinfo {author} {\bibfnamefont {J.}~\bibnamefont {Chen}}, \bibinfo {author}
  {\bibfnamefont {Y.}~\bibnamefont {Tsaturyan}}, \ and\ \bibinfo {author}
  {\bibfnamefont {A.}~\bibnamefont {Schliesser}},\ }\bibfield  {title}
  {\enquote {\bibinfo {title} {Measurement-based quantum control of mechanical
  motion},}\ }\href {\doibase 10.1038/s41586-018-0643-8} {\bibfield  {journal}
  {\bibinfo  {journal} {Nature}\ }\textbf {\bibinfo {volume} {563}},\ \bibinfo
  {pages} {53--58} (\bibinfo {year} {2018}{\natexlab{b}})}\BibitemShut
  {NoStop}%
\bibitem [{\citenamefont {Rossi}\ \emph {et~al.}(2019)\citenamefont {Rossi},
  \citenamefont {Mason}, \citenamefont {Chen},\ and\ \citenamefont
  {Schliesser}}]{Rossi2019}%
  \BibitemOpen
  \bibfield  {author} {\bibinfo {author} {\bibfnamefont {M.}~\bibnamefont
  {Rossi}}, \bibinfo {author} {\bibfnamefont {D.}~\bibnamefont {Mason}},
  \bibinfo {author} {\bibfnamefont {J.}~\bibnamefont {Chen}}, \ and\ \bibinfo
  {author} {\bibfnamefont {A.}~\bibnamefont {Schliesser}},\ }\bibfield  {title}
  {\enquote {\bibinfo {title} {Observing and verifying the quantum trajectory
  of a mechanical resonator},}\ }\href {\doibase
  10.1103/PhysRevLett.123.163601} {\bibfield  {journal} {\bibinfo  {journal}
  {Phys. Rev. Lett.}\ }\textbf {\bibinfo {volume} {123}},\ \bibinfo {pages}
  {163601} (\bibinfo {year} {2019})}\BibitemShut {NoStop}%
\bibitem [{\citenamefont {Magrini}\ \emph
  {et~al.}(2021{\natexlab{b}})\citenamefont {Magrini}, \citenamefont
  {Rosenzweig}, \citenamefont {Bach}, \citenamefont {Deutschmann-Olek},
  \citenamefont {Hofer}, \citenamefont {Hong}, \citenamefont {Kiesel},
  \citenamefont {Kugi},\ and\ \citenamefont {Aspelmeyer}}]{Magrini:2021aa}%
  \BibitemOpen
  \bibfield  {author} {\bibinfo {author} {\bibfnamefont {L.}~\bibnamefont
  {Magrini}}, \bibinfo {author} {\bibfnamefont {P.}~\bibnamefont {Rosenzweig}},
  \bibinfo {author} {\bibfnamefont {C.}~\bibnamefont {Bach}}, \bibinfo {author}
  {\bibfnamefont {A.}~\bibnamefont {Deutschmann-Olek}}, \bibinfo {author}
  {\bibfnamefont {S.~G.}\ \bibnamefont {Hofer}}, \bibinfo {author}
  {\bibfnamefont {S.}~\bibnamefont {Hong}}, \bibinfo {author} {\bibfnamefont
  {N.}~\bibnamefont {Kiesel}}, \bibinfo {author} {\bibfnamefont
  {A.}~\bibnamefont {Kugi}}, \ and\ \bibinfo {author} {\bibfnamefont
  {M.}~\bibnamefont {Aspelmeyer}},\ }\bibfield  {title} {\enquote {\bibinfo
  {title} {Real-time optimal quantum control of mechanical motion at room
  temperature},}\ }\href {\doibase 10.1038/s41586-021-03602-3} {\bibfield
  {journal} {\bibinfo  {journal} {Nature}\ }\textbf {\bibinfo {volume} {595}},\
  \bibinfo {pages} {373--377} (\bibinfo {year}
  {2021}{\natexlab{b}})}\BibitemShut {NoStop}%
\bibitem [{\citenamefont {Tebbenjohanns}\ \emph
  {et~al.}(2021{\natexlab{b}})\citenamefont {Tebbenjohanns}, \citenamefont
  {Mattana}, \citenamefont {Rossi}, \citenamefont {Frimmer},\ and\
  \citenamefont {Novotny}}]{Tebbenjohanns:2021aa}%
  \BibitemOpen
  \bibfield  {author} {\bibinfo {author} {\bibfnamefont {F.}~\bibnamefont
  {Tebbenjohanns}}, \bibinfo {author} {\bibfnamefont {M.~L.}\ \bibnamefont
  {Mattana}}, \bibinfo {author} {\bibfnamefont {M.}~\bibnamefont {Rossi}},
  \bibinfo {author} {\bibfnamefont {M.}~\bibnamefont {Frimmer}}, \ and\
  \bibinfo {author} {\bibfnamefont {L.}~\bibnamefont {Novotny}},\ }\bibfield
  {title} {\enquote {\bibinfo {title} {Quantum control of a nanoparticle
  optically levitated in cryogenic free space},}\ }\href {\doibase
  10.1038/s41586-021-03617-w} {\bibfield  {journal} {\bibinfo  {journal}
  {Nature}\ }\textbf {\bibinfo {volume} {595}},\ \bibinfo {pages} {378--382}
  (\bibinfo {year} {2021}{\natexlab{b}})}\BibitemShut {NoStop}%
\bibitem [{\citenamefont {Rouchon}(2015)}]{rouchon2015arxiv}%
  \BibitemOpen
  \bibfield  {author} {\bibinfo {author} {\bibfnamefont {P.}~\bibnamefont
  {Rouchon}},\ }\href@noop {} {\enquote {\bibinfo {title} {Models and feedback
  stabilization of open quantum systems},}\ } (\bibinfo {year} {2015}),\
  \Eprint {http://arxiv.org/abs/1407.7810} {arXiv:1407.7810 [math.OC]}
  \BibitemShut {NoStop}%
\bibitem [{\citenamefont {Rouchon}\ and\ \citenamefont
  {Ralph}(2015)}]{Rouchon2015}%
  \BibitemOpen
  \bibfield  {author} {\bibinfo {author} {\bibfnamefont {P.}~\bibnamefont
  {Rouchon}}\ and\ \bibinfo {author} {\bibfnamefont {J.~F.}\ \bibnamefont
  {Ralph}},\ }\bibfield  {title} {\enquote {\bibinfo {title} {Efficient quantum
  filtering for quantum feedback control},}\ }\href {\doibase
  10.1103/PhysRevA.91.012118} {\bibfield  {journal} {\bibinfo  {journal} {Phys.
  Rev. A}\ }\textbf {\bibinfo {volume} {91}},\ \bibinfo {pages} {012118}
  (\bibinfo {year} {2015})}\BibitemShut {NoStop}%
\bibitem [{\citenamefont {Ristè}\ and\ \citenamefont
  {DiCarlo}(2015)}]{riste2015}%
  \BibitemOpen
  \bibfield  {author} {\bibinfo {author} {\bibfnamefont {D.}~\bibnamefont
  {Ristè}}\ and\ \bibinfo {author} {\bibfnamefont {L.}~\bibnamefont
  {DiCarlo}},\ }\bibfield  {title} {\enquote {\bibinfo {title} {Digital
  feedback in superconducting quantum circuits},}\ }\href
  {http://arxiv.org/abs/1508.01385} {\bibfield  {journal} {\bibinfo  {journal}
  {arXiv:1508.01385 [cond-mat, physics:quant-ph]}\ } (\bibinfo {year}
  {2015})},\ \bibinfo {note} {arXiv: 1508.01385}\BibitemShut {NoStop}%
\bibitem [{sci()}]{scipy.opt.min}%
  \BibitemOpen
  \href@noop {} {}\bibinfo {howpublished}
  {\url{https://docs.scipy.org/doc/scipy/reference/generated/scipy.optimize.minimize.html}}\BibitemShut
  {NoStop}%
\bibitem [{\citenamefont {Cattaneo}\ \emph {et~al.}(2018)\citenamefont
  {Cattaneo}, \citenamefont {Rossi}, \citenamefont {Paris},\ and\ \citenamefont
  {Maniscalco}}]{cattaneo18}%
  \BibitemOpen
  \bibfield  {author} {\bibinfo {author} {\bibfnamefont {M.}~\bibnamefont
  {Cattaneo}}, \bibinfo {author} {\bibfnamefont {M.~A.~C.}\ \bibnamefont
  {Rossi}}, \bibinfo {author} {\bibfnamefont {M.~G.~A.}\ \bibnamefont {Paris}},
  \ and\ \bibinfo {author} {\bibfnamefont {S.}~\bibnamefont {Maniscalco}},\
  }\bibfield  {title} {\enquote {\bibinfo {title} {Quantum spatial search on
  graphs subject to dynamical noise},}\ }\href {\doibase
  10.1103/PhysRevA.98.052347} {\bibfield  {journal} {\bibinfo  {journal} {Phys.
  Rev. A}\ }\textbf {\bibinfo {volume} {98}},\ \bibinfo {pages} {052347}
  (\bibinfo {year} {2018})}\BibitemShut {NoStop}%
\bibitem [{\citenamefont {Perets}\ \emph {et~al.}(2008)\citenamefont {Perets},
  \citenamefont {Lahini}, \citenamefont {Pozzi}, \citenamefont {Sorel},
  \citenamefont {Morandotti},\ and\ \citenamefont
  {Silberberg}}]{PhysRevLett.100.170506}%
  \BibitemOpen
  \bibfield  {author} {\bibinfo {author} {\bibfnamefont {Hagai~B.}\
  \bibnamefont {Perets}}, \bibinfo {author} {\bibfnamefont {Yoav}\ \bibnamefont
  {Lahini}}, \bibinfo {author} {\bibfnamefont {Francesca}\ \bibnamefont
  {Pozzi}}, \bibinfo {author} {\bibfnamefont {Marc}\ \bibnamefont {Sorel}},
  \bibinfo {author} {\bibfnamefont {Roberto}\ \bibnamefont {Morandotti}}, \
  and\ \bibinfo {author} {\bibfnamefont {Yaron}\ \bibnamefont {Silberberg}},\
  }\bibfield  {title} {\enquote {\bibinfo {title} {Realization of quantum walks
  with negligible decoherence in waveguide lattices},}\ }\href {\doibase
  10.1103/PhysRevLett.100.170506} {\bibfield  {journal} {\bibinfo  {journal}
  {Phys. Rev. Lett.}\ }\textbf {\bibinfo {volume} {100}},\ \bibinfo {pages}
  {170506} (\bibinfo {year} {2008})}\BibitemShut {NoStop}%
\bibitem [{\citenamefont {Peruzzo}\ \emph {et~al.}(2010)\citenamefont
  {Peruzzo}, \citenamefont {Lobino}, \citenamefont {Matthews}, \citenamefont
  {Matsuda}, \citenamefont {Politi}, \citenamefont {Poulios}, \citenamefont
  {Zhou}, \citenamefont {Lahini}, \citenamefont {Ismail}, \citenamefont
  {W{\"o}rhoff} \emph {et~al.}}]{peruzzo2010quantum}%
  \BibitemOpen
  \bibfield  {author} {\bibinfo {author} {\bibfnamefont {Alberto}\ \bibnamefont
  {Peruzzo}}, \bibinfo {author} {\bibfnamefont {Mirko}\ \bibnamefont {Lobino}},
  \bibinfo {author} {\bibfnamefont {Jonathan~CF}\ \bibnamefont {Matthews}},
  \bibinfo {author} {\bibfnamefont {Nobuyuki}\ \bibnamefont {Matsuda}},
  \bibinfo {author} {\bibfnamefont {Alberto}\ \bibnamefont {Politi}}, \bibinfo
  {author} {\bibfnamefont {Konstantinos}\ \bibnamefont {Poulios}}, \bibinfo
  {author} {\bibfnamefont {Xiao-Qi}\ \bibnamefont {Zhou}}, \bibinfo {author}
  {\bibfnamefont {Yoav}\ \bibnamefont {Lahini}}, \bibinfo {author}
  {\bibfnamefont {Nur}\ \bibnamefont {Ismail}}, \bibinfo {author}
  {\bibfnamefont {Kerstin}\ \bibnamefont {W{\"o}rhoff}},  \emph {et~al.},\
  }\bibfield  {title} {\enquote {\bibinfo {title} {Quantum walks of correlated
  photons},}\ }\href@noop {} {\bibfield  {journal} {\bibinfo  {journal}
  {Science}\ }\textbf {\bibinfo {volume} {329}},\ \bibinfo {pages} {1500--1503}
  (\bibinfo {year} {2010})}\BibitemShut {NoStop}%
\bibitem [{\citenamefont {Gr{\"a}fe}\ \emph {et~al.}(2016)\citenamefont
  {Gr{\"a}fe}, \citenamefont {Heilmann}, \citenamefont {Lebugle}, \citenamefont
  {Guzman-Silva}, \citenamefont {Perez-Leija},\ and\ \citenamefont
  {Szameit}}]{grafe2016integrated}%
  \BibitemOpen
  \bibfield  {author} {\bibinfo {author} {\bibfnamefont {Markus}\ \bibnamefont
  {Gr{\"a}fe}}, \bibinfo {author} {\bibfnamefont {Ren{\'e}}\ \bibnamefont
  {Heilmann}}, \bibinfo {author} {\bibfnamefont {Maxime}\ \bibnamefont
  {Lebugle}}, \bibinfo {author} {\bibfnamefont {Diego}\ \bibnamefont
  {Guzman-Silva}}, \bibinfo {author} {\bibfnamefont {Armando}\ \bibnamefont
  {Perez-Leija}}, \ and\ \bibinfo {author} {\bibfnamefont {Alexander}\
  \bibnamefont {Szameit}},\ }\bibfield  {title} {\enquote {\bibinfo {title}
  {Integrated photonic quantum walks},}\ }\href@noop {} {\bibfield  {journal}
  {\bibinfo  {journal} {Journal of Optics}\ }\textbf {\bibinfo {volume} {18}},\
  \bibinfo {pages} {103002} (\bibinfo {year} {2016})}\BibitemShut {NoStop}%
\bibitem [{\citenamefont {Jiao}\ \emph {et~al.}(2021)\citenamefont {Jiao},
  \citenamefont {Gao}, \citenamefont {Zhou}, \citenamefont {Wang},
  \citenamefont {Ren}, \citenamefont {Xu}, \citenamefont {Qiao}, \citenamefont
  {Wang},\ and\ \citenamefont {Jin}}]{jiao2021two}%
  \BibitemOpen
  \bibfield  {author} {\bibinfo {author} {\bibfnamefont {Zhi-Qiang}\
  \bibnamefont {Jiao}}, \bibinfo {author} {\bibfnamefont {Jun}\ \bibnamefont
  {Gao}}, \bibinfo {author} {\bibfnamefont {Wen-Hao}\ \bibnamefont {Zhou}},
  \bibinfo {author} {\bibfnamefont {Xiao-Wei}\ \bibnamefont {Wang}}, \bibinfo
  {author} {\bibfnamefont {Ruo-Jing}\ \bibnamefont {Ren}}, \bibinfo {author}
  {\bibfnamefont {Xiao-Yun}\ \bibnamefont {Xu}}, \bibinfo {author}
  {\bibfnamefont {Lu-Feng}\ \bibnamefont {Qiao}}, \bibinfo {author}
  {\bibfnamefont {Yao}\ \bibnamefont {Wang}}, \ and\ \bibinfo {author}
  {\bibfnamefont {Xian-Min}\ \bibnamefont {Jin}},\ }\bibfield  {title}
  {\enquote {\bibinfo {title} {Two-dimensional quantum walks of correlated
  photons},}\ }\href@noop {} {\bibfield  {journal} {\bibinfo  {journal}
  {Optica}\ }\textbf {\bibinfo {volume} {8}},\ \bibinfo {pages} {1129--1135}
  (\bibinfo {year} {2021})}\BibitemShut {NoStop}%
\bibitem [{\citenamefont {Karski}\ \emph {et~al.}(2009)\citenamefont {Karski},
  \citenamefont {F{\"o}rster}, \citenamefont {Choi}, \citenamefont {Steffen},
  \citenamefont {Alt}, \citenamefont {Meschede},\ and\ \citenamefont
  {Widera}}]{karski2009quantum}%
  \BibitemOpen
  \bibfield  {author} {\bibinfo {author} {\bibfnamefont {Michal}\ \bibnamefont
  {Karski}}, \bibinfo {author} {\bibfnamefont {Leonid}\ \bibnamefont
  {F{\"o}rster}}, \bibinfo {author} {\bibfnamefont {Jai-Min}\ \bibnamefont
  {Choi}}, \bibinfo {author} {\bibfnamefont {Andreas}\ \bibnamefont {Steffen}},
  \bibinfo {author} {\bibfnamefont {Wolfgang}\ \bibnamefont {Alt}}, \bibinfo
  {author} {\bibfnamefont {Dieter}\ \bibnamefont {Meschede}}, \ and\ \bibinfo
  {author} {\bibfnamefont {Artur}\ \bibnamefont {Widera}},\ }\bibfield  {title}
  {\enquote {\bibinfo {title} {Quantum walk in position space with single
  optically trapped atoms},}\ }\href@noop {} {\bibfield  {journal} {\bibinfo
  {journal} {Science}\ }\textbf {\bibinfo {volume} {325}},\ \bibinfo {pages}
  {174--177} (\bibinfo {year} {2009})}\BibitemShut {NoStop}%
\bibitem [{\citenamefont {Genske}\ \emph {et~al.}(2013)\citenamefont {Genske},
  \citenamefont {Alt}, \citenamefont {Steffen}, \citenamefont {Werner},
  \citenamefont {Werner}, \citenamefont {Meschede},\ and\ \citenamefont
  {Alberti}}]{genske2013electric}%
  \BibitemOpen
  \bibfield  {author} {\bibinfo {author} {\bibfnamefont {Maximilian}\
  \bibnamefont {Genske}}, \bibinfo {author} {\bibfnamefont {Wolfgang}\
  \bibnamefont {Alt}}, \bibinfo {author} {\bibfnamefont {Andreas}\ \bibnamefont
  {Steffen}}, \bibinfo {author} {\bibfnamefont {Albert~H}\ \bibnamefont
  {Werner}}, \bibinfo {author} {\bibfnamefont {Reinhard~F}\ \bibnamefont
  {Werner}}, \bibinfo {author} {\bibfnamefont {Dieter}\ \bibnamefont
  {Meschede}}, \ and\ \bibinfo {author} {\bibfnamefont {Andrea}\ \bibnamefont
  {Alberti}},\ }\bibfield  {title} {\enquote {\bibinfo {title} {Electric
  quantum walks with individual atoms},}\ }\href@noop {} {\bibfield  {journal}
  {\bibinfo  {journal} {Physical review letters}\ }\textbf {\bibinfo {volume}
  {110}},\ \bibinfo {pages} {190601} (\bibinfo {year} {2013})}\BibitemShut
  {NoStop}%
\bibitem [{\citenamefont {Z\"ahringer}\ \emph {et~al.}(2010)\citenamefont
  {Z\"ahringer}, \citenamefont {Kirchmair}, \citenamefont {Gerritsma},
  \citenamefont {Solano}, \citenamefont {Blatt},\ and\ \citenamefont
  {Roos}}]{PhysRevLett.104.100503}%
  \BibitemOpen
  \bibfield  {author} {\bibinfo {author} {\bibfnamefont {F.}~\bibnamefont
  {Z\"ahringer}}, \bibinfo {author} {\bibfnamefont {G.}~\bibnamefont
  {Kirchmair}}, \bibinfo {author} {\bibfnamefont {R.}~\bibnamefont
  {Gerritsma}}, \bibinfo {author} {\bibfnamefont {E.}~\bibnamefont {Solano}},
  \bibinfo {author} {\bibfnamefont {R.}~\bibnamefont {Blatt}}, \ and\ \bibinfo
  {author} {\bibfnamefont {C.~F.}\ \bibnamefont {Roos}},\ }\bibfield  {title}
  {\enquote {\bibinfo {title} {Realization of a quantum walk with one and two
  trapped ions},}\ }\href {\doibase 10.1103/PhysRevLett.104.100503} {\bibfield
  {journal} {\bibinfo  {journal} {Phys. Rev. Lett.}\ }\textbf {\bibinfo
  {volume} {104}},\ \bibinfo {pages} {100503} (\bibinfo {year}
  {2010})}\BibitemShut {NoStop}%
\bibitem [{\citenamefont {Schmitz}\ \emph {et~al.}(2009)\citenamefont
  {Schmitz}, \citenamefont {Matjeschk}, \citenamefont {Schneider},
  \citenamefont {Glueckert}, \citenamefont {Enderlein}, \citenamefont {Huber},\
  and\ \citenamefont {Schaetz}}]{PhysRevLett.103.090504}%
  \BibitemOpen
  \bibfield  {author} {\bibinfo {author} {\bibfnamefont {H.}~\bibnamefont
  {Schmitz}}, \bibinfo {author} {\bibfnamefont {R.}~\bibnamefont {Matjeschk}},
  \bibinfo {author} {\bibfnamefont {Ch.}\ \bibnamefont {Schneider}}, \bibinfo
  {author} {\bibfnamefont {J.}~\bibnamefont {Glueckert}}, \bibinfo {author}
  {\bibfnamefont {M.}~\bibnamefont {Enderlein}}, \bibinfo {author}
  {\bibfnamefont {T.}~\bibnamefont {Huber}}, \ and\ \bibinfo {author}
  {\bibfnamefont {T.}~\bibnamefont {Schaetz}},\ }\bibfield  {title} {\enquote
  {\bibinfo {title} {Quantum walk of a trapped ion in phase space},}\ }\href
  {\doibase 10.1103/PhysRevLett.103.090504} {\bibfield  {journal} {\bibinfo
  {journal} {Phys. Rev. Lett.}\ }\textbf {\bibinfo {volume} {103}},\ \bibinfo
  {pages} {090504} (\bibinfo {year} {2009})}\BibitemShut {NoStop}%
\bibitem [{\citenamefont {Xue}\ \emph {et~al.}(2009)\citenamefont {Xue},
  \citenamefont {Sanders},\ and\ \citenamefont
  {Leibfried}}]{PhysRevLett.103.183602}%
  \BibitemOpen
  \bibfield  {author} {\bibinfo {author} {\bibfnamefont {Peng}\ \bibnamefont
  {Xue}}, \bibinfo {author} {\bibfnamefont {Barry~C.}\ \bibnamefont {Sanders}},
  \ and\ \bibinfo {author} {\bibfnamefont {Dietrich}\ \bibnamefont
  {Leibfried}},\ }\bibfield  {title} {\enquote {\bibinfo {title} {Quantum walk
  on a line for a trapped ion},}\ }\href {\doibase
  10.1103/PhysRevLett.103.183602} {\bibfield  {journal} {\bibinfo  {journal}
  {Phys. Rev. Lett.}\ }\textbf {\bibinfo {volume} {103}},\ \bibinfo {pages}
  {183602} (\bibinfo {year} {2009})}\BibitemShut {NoStop}%
\bibitem [{\citenamefont {Schneider}\ \emph {et~al.}(2012)\citenamefont
  {Schneider}, \citenamefont {Hackerm{\"u}ller}, \citenamefont {Ronzheimer},
  \citenamefont {Will}, \citenamefont {Braun}, \citenamefont {Best},
  \citenamefont {Bloch}, \citenamefont {Demler}, \citenamefont {Mandt},
  \citenamefont {Rasch} \emph {et~al.}}]{schneider2012fermionic}%
  \BibitemOpen
  \bibfield  {author} {\bibinfo {author} {\bibfnamefont {Ulrich}\ \bibnamefont
  {Schneider}}, \bibinfo {author} {\bibfnamefont {Lucia}\ \bibnamefont
  {Hackerm{\"u}ller}}, \bibinfo {author} {\bibfnamefont {Jens~Philipp}\
  \bibnamefont {Ronzheimer}}, \bibinfo {author} {\bibfnamefont {Sebastian}\
  \bibnamefont {Will}}, \bibinfo {author} {\bibfnamefont {Simon}\ \bibnamefont
  {Braun}}, \bibinfo {author} {\bibfnamefont {Thorsten}\ \bibnamefont {Best}},
  \bibinfo {author} {\bibfnamefont {Immanuel}\ \bibnamefont {Bloch}}, \bibinfo
  {author} {\bibfnamefont {Eugene}\ \bibnamefont {Demler}}, \bibinfo {author}
  {\bibfnamefont {Stephan}\ \bibnamefont {Mandt}}, \bibinfo {author}
  {\bibfnamefont {David}\ \bibnamefont {Rasch}},  \emph {et~al.},\ }\bibfield
  {title} {\enquote {\bibinfo {title} {Fermionic transport and
  out-of-equilibrium dynamics in a homogeneous hubbard model with ultracold
  atoms},}\ }\href@noop {} {\bibfield  {journal} {\bibinfo  {journal} {Nature
  Physics}\ }\textbf {\bibinfo {volume} {8}},\ \bibinfo {pages} {213--218}
  (\bibinfo {year} {2012})}\BibitemShut {NoStop}%
\bibitem [{\citenamefont {Preiss}\ \emph {et~al.}(2015)\citenamefont {Preiss},
  \citenamefont {Ma}, \citenamefont {Tai}, \citenamefont {Lukin}, \citenamefont
  {Rispoli}, \citenamefont {Zupancic}, \citenamefont {Lahini}, \citenamefont
  {Islam},\ and\ \citenamefont {Greiner}}]{preiss2015strongly}%
  \BibitemOpen
  \bibfield  {author} {\bibinfo {author} {\bibfnamefont {Philipp~M}\
  \bibnamefont {Preiss}}, \bibinfo {author} {\bibfnamefont {Ruichao}\
  \bibnamefont {Ma}}, \bibinfo {author} {\bibfnamefont {M~Eric}\ \bibnamefont
  {Tai}}, \bibinfo {author} {\bibfnamefont {Alexander}\ \bibnamefont {Lukin}},
  \bibinfo {author} {\bibfnamefont {Matthew}\ \bibnamefont {Rispoli}}, \bibinfo
  {author} {\bibfnamefont {Philip}\ \bibnamefont {Zupancic}}, \bibinfo {author}
  {\bibfnamefont {Yoav}\ \bibnamefont {Lahini}}, \bibinfo {author}
  {\bibfnamefont {Rajibul}\ \bibnamefont {Islam}}, \ and\ \bibinfo {author}
  {\bibfnamefont {Markus}\ \bibnamefont {Greiner}},\ }\bibfield  {title}
  {\enquote {\bibinfo {title} {Strongly correlated quantum walks in optical
  lattices},}\ }\href@noop {} {\bibfield  {journal} {\bibinfo  {journal}
  {Science}\ }\textbf {\bibinfo {volume} {347}},\ \bibinfo {pages} {1229--1233}
  (\bibinfo {year} {2015})}\BibitemShut {NoStop}%
\bibitem [{\citenamefont {Meier}\ \emph {et~al.}(2016)\citenamefont {Meier},
  \citenamefont {An},\ and\ \citenamefont {Gadway}}]{PhysRevA.93.051602}%
  \BibitemOpen
  \bibfield  {author} {\bibinfo {author} {\bibfnamefont {Eric~J.}\ \bibnamefont
  {Meier}}, \bibinfo {author} {\bibfnamefont {Fangzhao~Alex}\ \bibnamefont
  {An}}, \ and\ \bibinfo {author} {\bibfnamefont {Bryce}\ \bibnamefont
  {Gadway}},\ }\bibfield  {title} {\enquote {\bibinfo {title} {Atom-optics
  simulator of lattice transport phenomena},}\ }\href {\doibase
  10.1103/PhysRevA.93.051602} {\bibfield  {journal} {\bibinfo  {journal} {Phys.
  Rev. A}\ }\textbf {\bibinfo {volume} {93}},\ \bibinfo {pages} {051602}
  (\bibinfo {year} {2016})}\BibitemShut {NoStop}%
\bibitem [{\citenamefont {Young}\ \emph {et~al.}(2022)\citenamefont {Young},
  \citenamefont {Eckner}, \citenamefont {Schine}, \citenamefont {Childs},\ and\
  \citenamefont {Kaufman}}]{doi:10.1126/science.abo0608}%
  \BibitemOpen
  \bibfield  {author} {\bibinfo {author} {\bibfnamefont {Aaron~W.}\
  \bibnamefont {Young}}, \bibinfo {author} {\bibfnamefont {William~J.}\
  \bibnamefont {Eckner}}, \bibinfo {author} {\bibfnamefont {Nathan}\
  \bibnamefont {Schine}}, \bibinfo {author} {\bibfnamefont {Andrew~M.}\
  \bibnamefont {Childs}}, \ and\ \bibinfo {author} {\bibfnamefont {Adam~M.}\
  \bibnamefont {Kaufman}},\ }\bibfield  {title} {\enquote {\bibinfo {title}
  {Tweezer-programmable 2d quantum walks in a hubbard-regime lattice},}\ }\href
  {\doibase 10.1126/science.abo0608} {\bibfield  {journal} {\bibinfo  {journal}
  {Science}\ }\textbf {\bibinfo {volume} {377}},\ \bibinfo {pages} {885--889}
  (\bibinfo {year} {2022})},\ \Eprint
  {http://arxiv.org/abs/https://www.science.org/doi/pdf/10.1126/science.abo0608}
  {https://www.science.org/doi/pdf/10.1126/science.abo0608} \BibitemShut
  {NoStop}%
\bibitem [{\citenamefont {Laflamme}\ \emph {et~al.}(2017)\citenamefont
  {Laflamme}, \citenamefont {Yang},\ and\ \citenamefont
  {Zoller}}]{PhysRevA.95.043843}%
  \BibitemOpen
  \bibfield  {author} {\bibinfo {author} {\bibfnamefont {C.}~\bibnamefont
  {Laflamme}}, \bibinfo {author} {\bibfnamefont {D.}~\bibnamefont {Yang}}, \
  and\ \bibinfo {author} {\bibfnamefont {P.}~\bibnamefont {Zoller}},\
  }\bibfield  {title} {\enquote {\bibinfo {title} {Continuous measurement of an
  atomic current},}\ }\href {\doibase 10.1103/PhysRevA.95.043843} {\bibfield
  {journal} {\bibinfo  {journal} {Phys. Rev. A}\ }\textbf {\bibinfo {volume}
  {95}},\ \bibinfo {pages} {043843} (\bibinfo {year} {2017})}\BibitemShut
  {NoStop}%
\bibitem [{\citenamefont {Clark}\ \emph {et~al.}(2021)\citenamefont {Clark},
  \citenamefont {Groiseau}, \citenamefont {Shaw}, \citenamefont {Dadras},
  \citenamefont {Binegar}, \citenamefont {Wimberger}, \citenamefont {Summy},\
  and\ \citenamefont {Liu}}]{clark2021quantum}%
  \BibitemOpen
  \bibfield  {author} {\bibinfo {author} {\bibfnamefont {JH}~\bibnamefont
  {Clark}}, \bibinfo {author} {\bibfnamefont {C}~\bibnamefont {Groiseau}},
  \bibinfo {author} {\bibfnamefont {ZN}~\bibnamefont {Shaw}}, \bibinfo {author}
  {\bibfnamefont {S}~\bibnamefont {Dadras}}, \bibinfo {author} {\bibfnamefont
  {C}~\bibnamefont {Binegar}}, \bibinfo {author} {\bibfnamefont
  {S}~\bibnamefont {Wimberger}}, \bibinfo {author} {\bibfnamefont
  {GS}~\bibnamefont {Summy}}, \ and\ \bibinfo {author} {\bibfnamefont
  {Y}~\bibnamefont {Liu}},\ }\bibfield  {title} {\enquote {\bibinfo {title}
  {Quantum to classical walk transitions tuned by spontaneous emissions},}\
  }\href@noop {} {\bibfield  {journal} {\bibinfo  {journal} {Physical Review
  Research}\ }\textbf {\bibinfo {volume} {3}},\ \bibinfo {pages} {043062}
  (\bibinfo {year} {2021})}\BibitemShut {NoStop}%
\bibitem [{\citenamefont {Genoni}\ \emph {et~al.}(2014)\citenamefont {Genoni},
  \citenamefont {Mancini},\ and\ \citenamefont {Serafini}}]{GeneralDino}%
  \BibitemOpen
  \bibfield  {author} {\bibinfo {author} {\bibfnamefont {M.~G.}\ \bibnamefont
  {Genoni}}, \bibinfo {author} {\bibfnamefont {S.}~\bibnamefont {Mancini}}, \
  and\ \bibinfo {author} {\bibfnamefont {A.}~\bibnamefont {Serafini}},\
  }\bibfield  {title} {\enquote {\bibinfo {title} {General-dyne unravelling of
  a thermal master equation},}\ }\href {\doibase 10.1134/S1061920814030054}
  {\bibfield  {journal} {\bibinfo  {journal} {Russian Journal of Mathematical
  Physics}\ }\textbf {\bibinfo {volume} {21}},\ \bibinfo {pages} {329--336}
  (\bibinfo {year} {2014})}\BibitemShut {NoStop}%
\end{thebibliography}%

\end{document}